\documentclass[10pt,a4paper]{article}
\pdfoutput=1
\usepackage[english]{babel}
\usepackage[latin1]{inputenc}
\usepackage{amsfonts,amsbsy,bm,euscript,mathrsfs}
\usepackage{amssymb,stmaryrd,faktor}
\usepackage[tbtags]{amsmath}
\usepackage{graphicx}
\usepackage[title,titletoc]{appendix}
\usepackage[bookmarks=true,colorlinks=true,linkcolor=blue,citecolor=blue,urlcolor=blue,bookmarksnumbered]{hyperref}
\usepackage[bulletsep]{collref}

\newcommand{\AdS}{\textup{AdS}}
\newcommand{\CFT}{\textup{CFT}}
\newcommand{\Sphere}{\textup{S}}
\newcommand{\Torus}{\textup{T}}

\newcommand{\HL}{\mbox{\scriptsize HL}}

\DeclareMathOperator{\sech}{sech}

\numberwithin{equation}{section}

\makeatletter
\renewcommand\section{\@startsection {section}{1}{\z@}
{-3.5ex \@plus -1ex \@minus -.2ex}
{2.3ex \@plus.2ex}
{\normalfont\Large\bfseries}}
\renewcommand\subsection{\@startsection{subsection}{2}{\z@}
{-3.25ex\@plus -1ex \@minus -.2ex}
{1.5ex \@plus.2ex}
{\normalfont\large\bfseries}}
\makeatother

\newcommand{\alg}[1]{\mathfrak{#1}}

\newcommand{\beq}{\begin{equation}}
\newcommand{\eeq}{\end{equation}}
\newcommand{\beqa}{\begin{eqnarray}}
\newcommand{\eeqa}{\end{eqnarray}}

\begin{document}

\thispagestyle{empty}
\begin{flushright}\footnotesize\ttfamily
DMUS-MP-18-04
\end{flushright}
\vspace{2em}

\begin{center}

{\Large\bf \vspace{0.2cm}
{\color{black} The low-energy limit of $\AdS_3$/$\CFT_2$ and its TBA}} 
\vspace{1.5cm}

\textrm{\large Diego Bombardelli ${}^{\circ}$\footnote{\texttt{diegobombardelli@gmail.com}} \ \ \ \ Bogdan Stefa\'nski, jr.${}^{*}$\footnote{\texttt{Bogdan.Stefanski.1@city.ac.uk}} \ \ \ \ Alessandro Torrielli${}^{\dagger}$\footnote{\texttt{a.torrielli@surrey.ac.uk}}}

\vspace{2em}

\vspace{1em}
\begingroup\itshape
${}^{\circ}$ Department of Physics and INFN, University of Torino, Via P.\ Giuria 1, 10125 Torino, Italy.\\
${}^{*}$ Centre for Mathematical Science, City, University of London,
Northampton Square, EC1V 0HB London, UK\\
${}^{\dagger}$ Department of Mathematics, University of Surrey, GU2 7XH, Guildford, UK
\par\endgroup

\end{center}

\vspace{2em}

\begin{abstract}\noindent
We investigate low-energy string excitations in $\AdS_3\times \Sphere^3\times\Torus^4$. When the worldsheet is decompactified, the theory has gapless modes whose spectrum at low energies is determined by massless relativistic integrable S matrices of the type introduced by Al.~B.~Zamolodchikov. The S matrices are non-trivial only for excitations with identical worldsheet chirality, indicating that the low-energy theory is a $\CFT_2$. We construct a Thermodynamic Bethe Ansatz (TBA) for these excitations and show how the massless modes' wrapping effects may be incorporated into the $\AdS_3$ spectral problem. Using the TBA and its associated Y-system, we determine the central charge of the low-energy $\CFT_2$ to be $c=6$ from calculating the vacuum energy for antiperiodic fermions - with the vacuum energy being zero for periodic fermions in agreement with a supersymmetric theory - and find the energies of some excited states.
\end{abstract}

\vspace{3cm}

\begin{center}
In honour of Ludvig Faddeev. 
\end{center}

\newpage

\overfullrule=0pt
\parskip=2pt
\parindent=12pt
\headheight=0.0in \headsep=0.0in \topmargin=0.0in \oddsidemargin=0in

\vspace{-3cm}
\thispagestyle{empty}
\vspace{-1cm}

\tableofcontents

\setcounter{footnote}{0}

\section{\label{sec:level1}Introduction}

The closed superstring spectrum on $\AdS_3\times \Sphere^3\times \Torus^4$ and $\AdS_3\times \Sphere^3\times \Sphere^3\times \Sphere^1$ can be found exactly in $\alpha'$, in the large-volume limit, by solving a set of Bethe Equations (BEs)~\cite{Borsato:2016kbm,Borsato:2016xns}, building on earlier integrable results of these backgrounds~\cite{David:2008yk,Babichenko:2009dk,OhlssonSax:2011ms,Ahn:2012hw,Borsato:2012ud,Borsato:2012ss,Borsato:2013qpa,Borsato:2013hoa}. These algebraic equations follow from the exact worldsheet S matrix~\cite{Borsato:2014exa,Borsato:2014hja,Lloyd:2014bsa,Borsato:2015mma} upon making a Bethe Ansatz for the energy eigenstates. The ansatz is consistent since the worldsheet S matrix satisfies the Yang-Baxter equation. In~\cite{Baggio:2017kza} these BEs were used to determine the protected closed string states. Agreement was found with supergravity results~\cite{deBoer:1998kjm,Eberhardt:2017fsi}, the calculation of which was only completed in the case of $\AdS_3\times \Sphere^3\times \Sphere^3\times \Sphere^1$ recently~\cite{Eberhardt:2017fsi}.

A physical state is made up of a number of fundamental excitations, or magnons, each carrying a momentum, whose value is determined by solving the BEs; the energy of such a state is the sum of the energies of the individual magnons. The dispersion relation of a magnon is fixed by a shortening condition and takes the form
\begin{equation}
\label{eq:disp-rel}
E(p)=\sqrt{m^2+4h^2\sin^2\left(\frac{p}{2}\right)}\,,
\end{equation}
in $\AdS_3$ backgrounds with R-R flux. Above, $h=h(\alpha')=\tfrac{R^2_{\AdS}}{2\pi\alpha'}+\dots$ is the coupling constant that enters the BEs, and $m$ is the magnon mass. In $\AdS_3\times \Sphere^3\times \Torus^4$ $m^2=0\,,1$, while for $\AdS_3\times \Sphere^3\times \Sphere^3\times \Sphere^1$ $m^2=0\,,\alpha\,,1-\alpha$, where $\alpha=\frac{R^2_{\AdS}}{R^2_{\Sphere_+}}$.~\footnote{The $m^2=1$ modes on $\AdS_3\times \Sphere^3\times \Sphere^3\times \Sphere^1$  are believed to be composite~\cite{Borsato:2015mma}.} 

In contradistinction to higher-dimensional examples, when the worldsheet theory is decompactified, the $m^2=0$ modes of $\AdS_3$ backgrounds give rise to a gapless spectrum. This has important consequences, notably on the protected spectrum~\cite{Sax:2012jv,Baggio:2017kza}, but also on the Berenstein-Maldacena-Nastase (BMN) limit~\cite{Berenstein:2002jq}. In this limit, the magnon momenta are rescaled as $p\rightarrow \frac{p}{h}$ and $h$ is taken large. The dispersion 
relation~\eqref{eq:disp-rel} becomes relativistic
\begin{equation}
\label{eq:disp-rel-rel}
E(p)\rightarrow\sqrt{m^2+p^2}\,,
\end{equation}
and in higher-dimensional integrable holographic models, the S matrix trivializes. The S matrix has a perturbative expansion
\begin{equation}
S=1+h^{-2}S^{(1)}+h^{-4}S^{(2)}+\dots\,.
\label{eq:pert-Smat-exp}
\end{equation}
The leading-order term is trivial and the corrections can be matched to $\alpha'$-perturbative worldsheet scattering computations. In $\AdS_3$ integrable models, the BMN limit is more subtle. This is because massless magnons can be left- or right-moving
{\em relativistic} massless modes in this limit.
~\footnote{Away from the BMN limit, the dispersion relation is non-relativistic and periodic. Therefore by increasing the momentum of a left-moving magnon it becomes a right-moving one.}
As a result, at small momenta the all-loop massless/massless S matrix reduces to {\em four} S matrices, depending on what worldsheet chirality the scattering excitations have.~\footnote{It is straightforward to check that the S matrices for massive/massive, massive/massless scattering has the conventional expansion given in equation~\eqref{eq:pert-Smat-exp}, analogously to what happens in higher dimensions. Here too, the sub-leading corrections to these S matrices have an expansion in $h$ which can also be compared with perturbative worldsheet scattering computations~\cite{Sundin:2013ypa,Roiban:2014cia,Sundin:2015uva}. In such computations, it remains to be fully understood how to regularise certain massless divergences~\cite{Sundin:2016gqe}.} The left-massless/right-massless S matrix has a conventional perturbative expansion~\eqref{eq:pert-Smat-exp}, which becomes trivial in the strict BMN limit. On the other hand, left-massless/left-massless and  right-massless/right-massless S matrices remain {\em non-trivial} and {\em non-diagonal} at leading order
\begin{equation}
S=S^{(0)}+h^{-2}S^{(1)}+h^{-4}S^{(2)}+\dots\,.
\label{eq:pert-Smat-exp2}
\end{equation}
The leading-order S matrices above are integrable and relativistic, and we will denote them by $S_{LL}$ and $S_{RR}$.~\footnote{Since in this limit the theory is relativistic, $S_{LL}$ and $S_{RR}$ depend only on the difference in rapidities of the two excitations.} Direct comparison of these S matrices with worldsheet perturbative calculations is not possible: after all, massless particles of the same chirality cannot scatter with one another since both move at the speed of light. Nevertheless, viewed as an algebraic object, the S matrices are well defined.

This is exactly the situation which is described in \cite{ZamoMasslessTBA} and corresponds to how Zamolodchikov proposed to interpret massless scattering in relativistic integrable $1+1$-dimensional systems. The right-right and left-left amplitudes turn out to be completely non-perturbative and the expectations based on the Feynman diagrammatic expansion fail. Nevertheless, such amplitudes are essential to obtain the description of critical points of the massless trajectories. As reviewed in \cite{Fendley:1993xa,Fendley:1993jh,Borsato:2016xns}, such S matrices carry an inherent scale invariance, due to the same-sign shift in the rapidities of the two scattering particles (in the process of obtaining the massless scattering from a massive relativistic one). Such S matrices are therefore exclusively characterised
by the properties of the infrared fixed point of the theory, and one can think of them as encoding the non-perturbative dynamical information of the critical theory. While, for instance, in the case of the flow from the tricritical to critical Ising model \cite{Zamolodchikov:1991vx}, the right-right and left-left S matrices are indeed trivial and the mixed ones drive the genuine flow, in the opposite situation of the $su(2)_k$ theory with $k=1$ \cite{ZamoMasslessTBA} the right-right and left-left amplitudes are non-trivial and the mixed ones instead trivialise: the TBA describes in this case a theory at its CFT point for all values of the cylinder radius, as the left and right modes entirely decouple.

To recapitulate, on a decompactified worldsheet the $\AdS_3$ closed string spectrum is gapless and its small-momentum excitations are massless relativistic left- and right-movers equipped with difference-form S matrices $S_{LL}$ and $S_{RR}$, with $S_{LR}$ trivial. This closely resembles the integrable description of certain $\CFT_2$'s that arise as infra-red (IR) fixed-points of renormalization-group flows~\cite{Zamolodchikov:1989cf}. In a similar line of reasoning, we therefore conclude that the small-momentum excitations are described by a two-dimensional conformal field theory, which we will denote by $\CFT_2^{(0)}$. 

The energy spectrum of $\CFT_2^{(0)}$ is determined through the BEs that follow from $S_{LL}$ and $S_{RR}$, up to wrapping corrections. When the worldsheet is compactified, L\"uscher-type corrections involving exchanges of virtual particles that wrap the compact worldsheet spatial direction need to be accounted for. In integrable theories this can be done through the Thermodynamic Bethe Ansatz (TBA)~\cite{Zamolodchikov:1989cf}, which in the context of integrable holographic models was found in~\cite{Bombardelli:2009ns,Gromov:2009bc, Arutyunov:2009ur, Bombardelli:2009xz,Gromov:2009at}. These latter TBAs have been used to construct the Quantum Spectral Curve (QSC)~\cite{QSC,Gromov:2014caa,Cavaglia:2014exa,Bombardelli:2017vhk}, a powerful method for determining the exact spectrum including wrapping corrections (see for example \cite{Marboe:2014gma,Marboe:2014sya,Gromov:2015wca,Anselmetti:2015mda,Hegedus:2016eop,Marboe:2016igj,Lee:2017mhh,Bombardelli:2018bqz,Lee:2018jvn} and the review \cite{Gromov:2017blm}). Such methods are at present unavailable for the $\AdS_3$ integrable models, also due to the presence of gapless excitations~\cite{Abbott:2015pps}.

In this paper we will investigate wrapping effects on the low-momentum $\CFT_2^{(0)}$ states. Since $S_{LL}$ and $S_{RR}$ are relativistic, we will be able to adapt conventional methods to write down a TBA and use it to calculate the central charge of $\CFT_2^{(0)}$. We expect that once a complete non-relativistic TBA for the $\AdS_3$ models is found, it should reduce at small momenta to the relativistic TBA for $\CFT_2^{(0)}$ that we find here. As a result, the relativistic TBA we construct here should provide guidance on the way in which massless modes should be incorporated into the complete $\AdS_3$ TBA .

The integrable description of $\CFT_2^{(0)}$ that we present in this paper, has a number of striking similarities to the massless ${\cal N}=2$ super-sine-Gordon model~\cite{Kobayashi:1991st,Kobayashi:1991rh,Kobayashi:1991jv} at $\beta^2_{{\cal N}=2}=16 \pi$. Recall that at this point, the S matrix of the massless ${\cal N}=2$ super-sine-Gordon model is a tensor product of two massless (${\cal N}=0$) sine-Gordon S matrices at $\beta_{{\cal N}=0}=\beta_*$, where we define
\beq
\beta_*^2\equiv 16\pi/3\,.
\label{beta*}
\eeq
As is well known, at this point the massless sine-Gordon model in fact describes a free compact boson at $r^2=\frac{3}{4}$.~\footnote{This is the value of the radius for which the free boson theory has ${\cal N}=2$ supersymmetry, which should not be confused with the ${\cal N}=2$ supersymmetry of the super-sine-Gordon model itself.} We show that the matrix part of the $\CFT_2^{(0)}$ S matrix is almost identical to (two copies of) the massless ${\cal N}=2$ super-sine-Gordon model at $\beta^2_{N=2}=16 \pi$; the only differences come from certain constant phases related to the statistics of the excitations. Furthermore, we find that the dressing factor of $\CFT_2^{(0)}$ is the square of the corresponding bosonic sine-Gordon factor - the square being due to the doubling of nodes in the Dynkin diagram. What is more, the TBA equations for the ground state of $\CFT_2^{(0)}$ and its central charge turn out to be identical to (two copies of) those of the super-sine-Gordon model at $\beta^2_{N=2}=16 \pi$.

On general grounds we expect the spectrum of $\CFT_2^{(0)}$ to be that of four free bosons with zero winding and momentum and their superpartners. Therefore, finding a TBA that comes from an S matrix for two copies of the massless sine-Gordon theory at the free boson point, together with the fact that the energies of certain excited states are integer multiples of $2\pi/R$, provides a strong consistency check on the validity of our approach. Additionally, we would like to emphasize that, although the theory is expected to be free, the free excitations emerging from the TBA are by no means the scattering excitations used to construct the S matrix. The same phenomenon occurs in the $su(2)_{k=1}$ model~\cite{ZamoMasslessTBA}. Based on these insights, a further analysis of the degeneracies of the spectrum, as well as the inclusion of winding and momentum modes deserves to be undertaken. We intend to return to these issues in the future.

This paper is organised as follows. In section~\ref{summary} we derive explicit expressions for the matrix parts of $S_{LL}$ and $S_{RR}$ in the relativistic limit. In sections~\ref{sec:Crossing} and~\ref{sec:Relativistic3} we show how in the BMN limit, the massless dressing factor~\cite{Borsato:2016xns} reduces to the well-known dressing factor found by Zamolodchikov and Zamolodchikov~\cite{Zamolodchikov:1978xm}. In section~\ref{sec:Thermodynamic} we formulate the TBA and use it to compute the central charge of $\CFT_2^{(0)}$, as well as the energies of the first excited states. We conclude in section~\ref{sec:Conclusions} and present some of our technical findings in appendices.

\section{Massless R matrix}
\label{summary}

Worldsheet excitations on $\AdS_3\times\Sphere^3\times\Torus^4$ with RR flux have mass $m^2=1$ or  $m^2=0$. Both types of excitations transform in short representations of the centrally-extended $\alg{su}(1|1)^4_{\mbox{\scriptsize c.e.}}$ algebra of symmetries that commute with the Hamiltonian~\cite{Borsato:2014exa,Borsato:2014hja}. The structure of the central extensions is such that $\alg{su}(1|1)^4_{\mbox{\scriptsize c.e.}}\cong \left(\alg{su}(1|1)^2_{\mbox{\scriptsize c.e.}}\right)^2$. As a result, short representations  can be written as tensor products of two short representations of $\alg{su}(1|1)^2_{\mbox{\scriptsize c.e.}}$, and for the most part we will focus on this smaller algebra.

In this section we begin by reviewing the $\alg{su}(1|1)^2_{\mbox{\scriptsize c.e.}}$ algebra, its massless short representations, as well as the S matrix for scattering two such excitations.~\footnote{Since all $m^2=0$ short representations are isomorphic to one another, we will write all the expressions using only the so-called ${\rho}_L(m=0)$ representations~\cite{Borsato:2014hja}. In order not to clutter the notation, we will drop the subscript $L$ from most expressions. Note that the labels $L$ and $R$ are {\em not} related to worldsheet chirality.} We then review the relativistic limit of the massless S matrix and finally we summarize how the above structure can be understood in terms of the quantum super-Poincar\'e algebra introduced in~\cite{Stromwall:2016dyw}.

\subsection{The exact massless R matrix}
\label{sec:Theqs}

The centrally extended $\alg{su}(1|1)_{L}\times \alg{su}(1|1)_{R}$ algebra has  non-zero commutators
\begin{eqnarray}
\label{alge}
&&\{\mathfrak{Q}_{L}, \mathfrak{S}_{L}\} = \mathfrak{H}_{L}\,, \quad \{\mathfrak{Q}_{R}, \mathfrak{S}_{R}\} = \mathfrak{H}_{R}\,, \qquad \{\mathfrak{Q}_{L}, \mathfrak{Q}_{R}\} = \mathfrak{P}\,, \qquad  \{\mathfrak{S}_{L}, \mathfrak{S}_{R}\} = \mathfrak{K}\,,
\end{eqnarray}
where on the right-hand sides we have the four central elements.~\footnote{This algebra is in fact the conventional ${\cal{N}}=2$ supersymmetry algebra in 1+1 dimensions upon identifying
\begin{equation}
\alg{Q}_L \to Q_+, \qquad \alg{G}_L \to Q_-, \qquad \alg{Q}_R \to \bar{Q}_+, \qquad \alg{G}_R \to \bar{Q}_-\,,\nonumber
\end{equation}
and has appeared in relation to integrability before, for example in~\cite{Fendley:1990zj}. Our central extensions $\alg{P}$ and $\alg{K}$ correspond to $2 \Delta W$ and $2 \Delta W^*$ - see for instance equation (2.1) of~\cite{Fendley:1990zj}), where the algebra is specialised to a massive relativistic dispersion relation). We would like to thank Paul Fendley and Matthias Gaberdiel for discussions related to this point.}

A representation of~\eqref{alge} on a boson-fermion doublet $\{|\phi\rangle, |\psi\rangle\}$ takes the form
\begin{eqnarray}
\label{leftrep}
&&\mathfrak{Q}_{L} =-\mathfrak{S}_{R} =  \sqrt{h \sin \tfrac{p}{2}}\begin{pmatrix}0&0\\1&0\end{pmatrix}\,,\quad
\mathfrak{S}_{L}=-\mathfrak{Q}_{R} =  \sqrt{h \sin \tfrac{p}{2}}\begin{pmatrix}0&1\\0&0\end{pmatrix}\,, \nonumber \\
&&\qquad \qquad \mathfrak{H}_{L} = \mathfrak{H}_{R} = - \mathfrak{P} = - \mathfrak{K} = h \, \sin \frac{p}{2}\,.
\end{eqnarray}
Above, $p$ is the {\it momentum}, which takes values in $\left[0\,,\,2\pi\right]$, while $\mathfrak{H} = \mathfrak{H}_L + \mathfrak{H}_R$ is the {\it energy}. The shortening condition implies that the dispersion relation for a massless excitation is
\begin{equation}
\label{eq:full-massless-disp}
\mathfrak{H}=2\left|h \sin\tfrac{p}{2}\right|\,.
\end{equation}
Up to an overall dressing factor, the R matrix $R$ is given by
\begin{equation}\label{eq:RLL}
  \begin{aligned}
    &R |\phi\rangle \otimes |\phi\rangle\ = {} \, |\phi\rangle \otimes |\phi\rangle, \\
    &R |\phi\rangle \otimes |\psi\rangle\ = - {} \, A_{p_1,p_2}
|\phi\rangle \otimes |\psi\rangle + \, 
B_{p_1,p_2} |\psi\rangle \otimes |\phi\rangle, \\
    &R |\psi\rangle \otimes |\phi\rangle\ = {} \, A_{p_1,p_2}
|\psi\rangle \otimes |\phi\rangle +  \, 
B_{p_1,p_2}|\phi\rangle \otimes |\psi\rangle, \\
    &R |\psi\rangle \otimes |\psi\rangle\ = - {} \, |\psi\rangle \otimes |\psi\rangle\,,
\end{aligned}
\end{equation}
where
\begin{eqnarray}
\label{fu}
&&A_{p_1,p_2} = \csc \frac{p_1 + p_2}{4} \, \sin \frac{p_1 - p_2}{4},\qquad B_{p_1,p_2} = \csc \frac{p_1 + p_2}{4} \, \sqrt{\sin \frac{p_1}{2} \sin \frac{p_2}{2}}\,.
\end{eqnarray}
This form of the R matrix is fixed by compatibility with the centrally extended $\alg{su}(1|1)^2$ symmetry
\begin{equation}\label{definiLL}
  \Delta_N^{\text{op}} (\mathfrak{a})\,  R\ =\ R\, \Delta_N (\mathfrak{a})\,,\qquad \qquad \forall \, \, \mathfrak{a} \in 
\mathfrak{su}(1|1)^2_{\mbox{\scriptsize c.e.}}\,.
\end{equation}
Above $\Delta_N^{op} = \Pi (\Delta_N)$, with $\Pi$ the graded permutation on the tensor-product algebra $\Pi (\mathfrak{a} \otimes \mathfrak{b}) = (-)^{|\mathfrak{a}| |\mathfrak{b}|} \mathfrak{b} \otimes \mathfrak{a}$. The coproducts are specified as follows: 
 \begin{align}
\begin{split}
\label{coprod}
     &\Delta_N(\mathfrak{P})= \, \mathfrak{P} \otimes e^{i \frac{p}{2}} + e^{-i \frac{p}{2}} \otimes \mathfrak{P}, \qquad\qquad\Delta_N(\mathfrak{K})= \, \mathfrak{K} \otimes e^{i \frac{p}{2}} + e^{-i \frac{p}{2}} \otimes \mathfrak{K},  \\
      &\Delta_N(\mathfrak{H}_{R})= \, \mathfrak{H}_{R} \otimes {e^{i \frac{p}{2}}} + {e^{-i \frac{p}{2}}} \otimes \mathfrak{H}_{R}, \qquad \Delta_N(\mathfrak{H}_{L})= \, \mathfrak{H}_{L} \otimes {e^{i \frac{p}{2}}} + {e^{-i \frac{p}{2}}} \otimes \mathfrak{H}_{L},\\
        &\Delta_N(\mathfrak{Q}_{L}) = \, \mathfrak{Q}_{L} \otimes e^{i \frac{p}{4}} + e^{-i \frac{p}{4}} \otimes \mathfrak{Q}_{L}, \qquad \Delta_N(\mathfrak{S}_{L}) = \, \mathfrak{S}_{L} \otimes e^{i \frac{p}{4}} + e^{-i \frac{p}{4}} \otimes \mathfrak{S}_{L}\,,\\
     &\Delta_N(\mathfrak{Q}_{R}) = \, \mathfrak{Q}_{R} \otimes {e^{i \frac{p}{4}}} + {e^{-i \frac{p}{4}}} \otimes \mathfrak{Q}_{R}, \qquad \Delta_N(\mathfrak{S}_{R})= \, \mathfrak{S}_{R} \otimes {e^{i \frac{p}{4}}} + {e^{-i \frac{p}{4}}} \otimes \mathfrak{S}_{R}\,.
\end{split}
  \end{align}
Since $p$ appears on the rhs above, we will also require
\begin{equation}
 \Delta_N(p) = \, p \otimes \mathfrak{1} + \mathfrak{1} \otimes p\,.
\end{equation}
The above coproducts provide a prescription for how the symmetry algebra acts on two-particle states, in such a way that it is a representation of (\ref{alge}). $R$ satisfies the Yang-Baxter equation and {\it braiding unitarity}: $\Pi(R)(p_2,p_1) \, R(p_1,p_2) = \mathfrak{1} \otimes \mathfrak{1}$. The R matrix also satisfies $\Pi(R)(p_2,p_1) = R(p_1,p_2)$. To describe the scattering of massless $AdS_3$ modes, $R$ needs to be multiplied by a suitable dressing factor, which we will denote by $\Phi$, whose form is determined by a crossing equation~\cite{Borsato:2016xns} up to CDD factors. Dressed in this way and evaluated in the {\it physical region} of momenta, $R$ represents (up to a permutation of the outgoing particles) the physical S matrix, scattering particles $1$ and $2$ - with momenta $p_1$ and $p_2$, respectively.

\subsection{The relativistic limit of the massless R matrix}
\label{sec:Relativistic}

In investigating worldsheet S matrices it is useful to consider the relativistic, or near-BMN regime 
\begin{equation}
\label{unde}
p \to \epsilon \, q, \quad h \to \frac{c}{\epsilon}, \qquad \mbox{with}\qquad 
\epsilon \to 0\,.
\end{equation}
In this limit it is well known that S matrices describing the scattering of {\em massive} excitations become proportional to the identity, and sub-leading terms can be matched to perturbative worldsheet scattering processes ($\alpha'$ corrections)~\cite{Sundin:2013ypa}. Similarly, the S matrices for mixed massive/massless scattering trivialise in this limit.~\footnote{Because of complications related to regularising massless particles in loops, matching to perturbative computations remains an outstanding challenge~\cite{Sundin:2016gqe}.}  The relativistic limit of massless/massless scattering  is more subtle~\cite{Borsato:2016xns} because it depends on the relative sign of the momenta of the two excitations. When $p_1>0$ and $p_2<0$, or vice versa, to leading order in $\epsilon$ the S matrix is proportional to identity with sub-leading perturbative corrections, much as in the massive case. On the other hand when $p_1,p_2>0$ or $p_1,p_2<0$ the S matrix remains non-trivial as $\epsilon\rightarrow 0$. It is this novel behaviour of the massless worldsheet S matrix in the relativistic limit that is the main focus of this paper.

In the relativistic limit~\eqref{unde}, with $p>0$,  the $\alg{su}(1|1)^2_{\mbox{\scriptsize c.e.}}$ generators are
\begin{eqnarray}
\label{leftreprel}
&&
\mathfrak{Q}_{L} =-\mathfrak{S}_{R} =  \sqrt{\frac{c q}{2}}\begin{pmatrix}0&0\\1&0\end{pmatrix},\qquad
\mathfrak{S}_{L} =-\mathfrak{Q}_{R} =  \sqrt{\frac{c q}{2}}\begin{pmatrix}0&1\\0&0\end{pmatrix}, 
\nonumber \\
&& \qquad \qquad \qquad \mathfrak{H}_{L} = \mathfrak{H}_{R} = - \mathfrak{P} = - \mathfrak{K} \equiv e_0 = \frac{c q}{2}\,,
\end{eqnarray}
and the dispersion relation~\eqref{eq:full-massless-disp} becomes that of a conventional massless left-moving (on the worldsheet) relativistic excitation
\begin{eqnarray}
H= c q\,.
\end{eqnarray} 
With $p_1,p_2>0$, the R matrix (ignoring for the moment the scalar factor) reduces to~\footnote{Similar expressions can be found when $p_1,p_2<0$.}
\begin{equation}\label{eq:RLLlim}
  \begin{aligned}
    &R |\phi\rangle \otimes |\phi\rangle\ = {} \, |\phi\rangle \otimes |\phi\rangle, \\
    &R |\phi\rangle \otimes |\psi\rangle\ = - \frac{q_1 - q_2}{q_1 + q_2} |\phi\rangle \otimes |\psi\rangle +  \, \frac{2 \sqrt{q_1 q_2}}{q_1 + q_2} |\psi\rangle \otimes |\phi\rangle, \\
    &R|\psi\rangle \otimes |\phi\rangle\ = {} \, \frac{2 \sqrt{q_1 q_2}}{q_1 + q_2} |\phi\rangle \otimes |\psi\rangle + \frac{q_1 - q_2}{q_1 + q_2}  |\psi\rangle \otimes |\phi\rangle, \\
    &R |\psi\rangle \otimes |\psi\rangle\ = - {} \, |\psi\rangle \otimes |\psi\rangle.
\end{aligned}
\end{equation}
Introducing the relativistic rapidity
\begin{equation}
\label{rap}
q = e^\theta,
\end{equation}
the R matrix takes the difference form
\begin{equation}\label{eq:RLLlimtheta}
  \begin{aligned}
    &R |\phi\rangle \otimes |\phi\rangle\ = {} \, |\phi\rangle \otimes |\phi\rangle, \\
    &R |\phi\rangle \otimes |\psi\rangle\ = -\tanh\tfrac{\vartheta}{2} |\phi\rangle \otimes |\psi\rangle + \sech\tfrac{\vartheta}{2}|\psi\rangle \otimes |\phi\rangle, \\
    &R |\psi\rangle \otimes |\phi\rangle\ = {} \, \sech\tfrac{\vartheta}{2} |\phi\rangle \otimes |\psi\rangle + \tanh\tfrac{\vartheta}{2} |\psi\rangle \otimes |\phi\rangle, \\
    &R |\psi\rangle \otimes |\psi\rangle\ = - {} \, |\psi\rangle \otimes |\psi\rangle\,,
\end{aligned}
\end{equation}
where 
\begin{equation}
\label{eq:diff-form}
\vartheta \equiv \theta_1 - \theta_2\,.
\end{equation}
We will denote by $R$ both the non-relativistic R matrix, and its relativistic limit, since it should be clear from the context which R matrix we mean.

\subsection{The $q$-super-Poincar\'e algebra and boosts}
\label{sec:qsp-rev}

In \cite{Stromwall:2016dyw}, an algebraic reformulation of the results summarised in section~\ref{sec:Theqs} was given in terms of two copies of a 1+1 dimensional {\it $q$-deformed super-Poincar\'e algebra}. Each copy satisfies the following relations:
\begin{eqnarray}
\label{algeq}
&&\{\mathfrak{Q}_{R}, \mathfrak{S}_{R}\} \ = \ \mathfrak{H}_{R}, \quad \{\mathfrak{Q}_{L}, \mathfrak{S}_{L}\} \ = \ \mathfrak{H}_{L}, \quad [\mathfrak{J}_{R}, p] \ = \ i \mathfrak{H}_{R}, \nonumber \\
&& [\mathfrak{J}_{L}, p] \ = \ i \mathfrak{H}_{L},\nonumber \quad [\mathfrak{J}_A, \mathfrak{H}_B] \ = \frac{e^{i p} - e^{-i p}}{2 \mu},\nonumber \\
&& [\mathfrak{J}_A, \mathfrak{Q}_B] \ = \frac{i}{2 \sqrt{\mu}} \frac{e^{i \frac{p}{2}} + e^{- i \frac{p}{2}}}{2} \mathfrak{Q}_B, \qquad [\mathfrak{J}_A, \mathfrak{S}_B] = \frac{i}{2 \sqrt{\mu}} \frac{e^{i \frac{p}{2}} + e^{- i \frac{p}{2}}}{2} \, \mathfrak{S}_B\,,\nonumber \\
\label{algeqc}
&&\{\mathfrak{Q}_{L}, \mathfrak{Q}_{R}\} \ = \ \mathfrak{P}~, \qquad  \{\mathfrak{S}_{L}, \mathfrak{S}_{R}\} \ = \ \mathfrak{K},\nonumber \\ 
&&[\mathfrak{J}_{L}, \mathfrak{P}] \ = [\mathfrak{J}_{R}, \mathfrak{P}] \ = \ [\mathfrak{J}_{L}, \mathfrak{K}] \ = [\mathfrak{J}_{R}, \mathfrak{K}]= \frac{e^{- i p} - e^{i p}}{2 \mu}\,,
\end{eqnarray}
where $\mu \equiv \frac{4}{h^2}$, 
$(A,B) = (L,L), (R,R)$
and the boost operators act as
\begin{equation}
\mathfrak{J}_R = i \mathfrak{H}_R \, \partial_p, \qquad \mathfrak{J}_L = i \mathfrak{H}_L \, \partial_p\,.
\end{equation}
The (suitably normalised) quadratic Casimir is given by  
\begin{eqnarray}
\mathfrak{C}_2 \equiv \mathfrak{H}^2  - 4 h^2 \sin \frac{p}{2}\,. \nonumber
\end{eqnarray}
The massless representation is characterised by the vanishing of the Casimir eigenvalue (massless {\it dispersion relation}). The coproduct for the boost operator, say, $\mathfrak{J}_L$ reads (cf. \cite{Young:2007wd})
\begin{eqnarray}
\label{deltaJ}
&&\Delta_N (\mathfrak{J}_L) = \mathfrak{J}_L \otimes e^{i \frac{p}{2}} +  e^{-i \frac{p}{2}} \otimes \mathfrak{J}_L + \, \frac{1}{2} \, \mathfrak{Q}_L \, e^{- i \frac{p}{4}} \otimes \mathfrak{S}_L \, e^{i \frac{p}{4}} +  \frac{1}{2} \, \mathfrak{S}_L \, e^{- i \frac{p}{4}} \otimes \mathfrak{Q}_L \, e^{i \frac{p}{4}}.
\end{eqnarray}

The result of \cite{Fontanella:2016opq} were used to introduce a geometric picture in the scattering problem. The equations
\begin{equation}
\label{boost}
\Delta_N(\mathfrak{J}_L) (R) = 0 = \Delta_N^{op}(\mathfrak{J}_L) (R),
\end{equation}
for $R$ normalised as in (\ref{eq:RLL}), can be re-written in the form of a covariant derivative on a 2-dimensional manifold $\mathcal{B}$ with coordinates $(p_1,p_2)$~\footnote{The two equations (\ref{7}) are related since $\Pi(R)(p_2,p_1) = R(p_1,p_2)$. Alternatively, starting from  equations~\eqref{7}, one can derive $\Pi(R)(p_2,p_1) = R(p_1,p_2)$. Braiding unitarity is then a constraint equation.} 
\begin{equation}
\label{7}
D_M R \equiv \bigg[\frac{\partial}{\partial p_M} + \Gamma_M\bigg]R = 0, \qquad M =1\,,\,2\,.
\end{equation}
Above
\begin{equation}
\Gamma_M = g_{M} \big[ E_+ \otimes E_- + E_- \otimes E_+\big],
\end{equation}
with
\begin{equation}
g_1 = -\frac{1}{4} \sqrt{\frac{\sin \frac{p_2}{2}}{\sin \frac{p_1}{2}}} \, \csc \frac{p_1 + p_2}{4}\,, \qquad g_2 = - g_1(p_2,p_1)\,,
\end{equation}
and
\begin{equation}
\label{gl110}
E_+ \equiv E_{12} = \begin{pmatrix}0&1\\0&0\end{pmatrix}, \qquad E_- \equiv E_{21} = \begin{pmatrix}0&0\\1&0\end{pmatrix}\,.\nonumber 
\end{equation}
From~(\ref{7}) we can write an integral formula for the R matrix, in terms of the line-integral over any given (suitably differentiable) curve $\gamma(\lambda) : [0, 1] \rightarrow \mathcal{B}$:
\begin{equation}
\label{int_R}
R \big[\gamma (\lambda)\big] = \Pi_s\,  {\cal P}\exp\left( \int_{\gamma(0)}^{\gamma(\lambda)} dp^M \Gamma_M \right), 
\end{equation}
where $\Pi_s$ is the graded permutation operator acting on two-particle states as $\Pi (|v\rangle \otimes |w\rangle) = (-)^{|v| |w|} |w\rangle \otimes |v\rangle$, and $\cal{P}\exp$ denotes the path-ordering of the exponential.~\footnote{The sign in the exponent of (\ref{int_R}) is justified since we extracted $\Pi_s$ in front for convenience, and one has
\begin{equation}
\{\Pi_s, \Gamma_M\} = 0,\qquad \Big[\frac{\partial}{\partial p_M} - \Gamma_M\Big]\Pi_s \circ R = 0, \qquad\Pi_s^2=\mathfrak{1}\otimes \mathfrak{1}\,.
\end{equation}}  The starting point of integration is chosen to reproduce the property $R(p,p)=\Pi_s$. 
The putative connection $\Gamma_M$ is locally flat ({\it pure gauge}), since its curvature $F_{MN}$ is vanishing:
\begin{equation}
\label{flatness}
F_{12} = \partial_1 \Gamma_2 - \partial_2 \Gamma_1 + [\Gamma_1,\Gamma_2] = 0.
\end{equation}

Including a dressing factor $\Phi$ in the R matrix modifies~\eqref{7} in a straightforward way
\begin{equation}
\label{fase}
\bigg[\frac{\partial}{\partial p_M} + \Gamma_M - \frac{\partial}{\partial p_M} \log \Phi\bigg]\tilde{R} = 0, \qquad \tilde {R} \equiv \Phi R\,.
\end{equation}
The R matrix undergoes crossing when one of the momenta leaves the physical region and was not discussed previously. We analyse this effect on the above differential equation and connection in the next section.

\subsection{Relativistic limit of the $q$-super-Poincar\'e algebra}

We conclude this summary by considering the effect of the relativistic limit on the $q$-super-Poincar\'e algebra. The boost operators become equal to one another and we denote them by $\mathfrak{b}$
\begin{equation}
\label{eq:rel-lim-boost}
\mathfrak{J}_A \to \frac{i \mathfrak{b}}{2 \epsilon}\,,\qquad\mbox{where}\qquad
\mathfrak{b}=c q \frac{\partial}{\partial q}\,.
\end{equation}
The coproduct reduces to
\begin{equation}
\Delta_N(\mathfrak{J}_A) \to \frac{i}{2 \epsilon} \, (\mathfrak{b} \otimes \mathfrak{1} + \mathfrak{1} \otimes \mathfrak{b}) = \frac{i c}{2 \epsilon} \, \big(q_1 \partial_{q_1} + q_2 \partial_{q_2}\big) \to \frac{i c}{2 \epsilon} \, \big(\partial_{\theta_1} + \partial_{\theta_2}\big),
\end{equation}
{\it i.e.} the R matrix, which satisfies 
$\Delta_N(\mathfrak{J})R=0$ \cite{Fontanella:2016opq},
has to become of difference form in the strict relativistic limit. This is indeed the case, as we saw explicitly in equation~\eqref{eq:RLLlimtheta}. Notice also that 
$\Delta(\mathfrak{J})$ and $\Delta^{op}(\mathfrak{J})$ become coincident in the relativistic limit.

In the relativistic limit, the covariant derivatives in equation~\eqref{7} reduce to
\begin{equation}
D_M \to \frac{\delta_M}{\epsilon}, \qquad \delta_M = \partial_{q_M} + \gamma_M,
\end{equation}
with
\begin{eqnarray}
&&\gamma_1 = - \sqrt{\frac{q_2}{q_1}} \, \frac{\big(E_+ \otimes E_- + E_- \otimes E_+\big)}{q_1 + q_2}, \qquad \gamma_2 = \sqrt{\frac{q_1}{q_2}} \, \frac{\big(E_+ \otimes E_- + E_- \otimes E_+\big)}{q_1 + q_2}.
\end{eqnarray}
One can verify that
\begin{equation}
\label{remark}
\delta_M R =0\,, \qquad M=1,2\,.
\end{equation}
Equivalently, in terms of rapidities $\theta_M$, we have
\begin{equation}
\label{remarko}
d_M R =0\,, \qquad d_M = \partial_{\theta_M} + A_M,
\end{equation}
with
\begin{eqnarray}
&&A_1 = -\tfrac{1}{2}\sech\tfrac{\theta_1 - \theta_2}{2}
\big(E_+ \otimes E_- + E_- \otimes E_+\big)= - A_2.
\end{eqnarray}
Just as the R matrix~\eqref{eq:RLLlimtheta}, the connection $A_M$ is also of difference form.

Let us remark that equation (\ref{remark}) would be rather hard to detect starting from the strict relativistic case, but it emerges quite naturally when deriving it from the $q$-Poincar\'e algebra. As a matter of fact, because of the difference form imposed by $\Delta(\mathfrak{J})R=0$, both conditions (\ref{remark}) coincide with the single ordinary differential equation
\begin{equation}
\label{read}
\bigg[\frac{\partial}{\partial \vartheta} 
-\tfrac{1}{2}\sech\tfrac{\theta}{2}
\big(E_+ \otimes E_- + E_- \otimes E_+\big)\bigg]R(\vartheta)=0, \qquad  \vartheta \equiv \theta_1 - \theta_2,
\end{equation}
which can be immediately integrated to
\begin{equation}
\label{gude}
R(\vartheta) = \Pi_s \, e^{- \big(E_+ \otimes E_- + E_- \otimes E_+\big) \, \mbox{gd}\big(\frac{\vartheta}{2}\big)}\,,
\end{equation}
where
\begin{equation}
\mbox{gd}(x) \equiv \int_0^x \frac{dy}{\cosh y} = 2 \arctan \tanh \frac{x}{2}
\end{equation}
is the {\it Gudermannian} function. By explicitly working out (\ref{gude}), we obtain 
\begin{equation}
R = \begin{pmatrix}1&0&0&0\\0&\sin \sigma &\cos \sigma&0\\0&\cos \sigma&-\sin \sigma&0\\0&0&0&-1\end{pmatrix}, \qquad \sigma \equiv -\mbox{gd}\bigg(\frac{\vartheta}{2}\bigg),\nonumber
\end{equation}
which can be seen to coincide with (\ref{eq:RLLlimtheta}).

\section{Dressing factor and Crossing}
\label{sec:Crossing}

In this section we discuss the crossing symmetry that is used to determine the R matrix dressing factor. We begin by explaining how crossing is implemented in the geometric formulation of the R matrix reviewed in section~\ref{sec:qsp-rev}. We then show that the massless dressing factor found in~\cite{Borsato:2016xns} reduces to the famous sine-Gordon scalar factor that enters the S matrix for solitons and anti-solitons~\cite{Zamolodchikov:1978xm}.

\subsection{Crossing and the $q$-super-Poincar\'e algebra}

With $p\in [0,2\pi]$, the supercharges in the {\it crossed region} are defined as
\begin{eqnarray}
\label{accor}
&&\mathfrak{Q}_{\bar{L},-p}^{str} = - C \mathfrak{Q}_{L,p} C^{-1} = - i \sqrt{h \sin\frac{p}{2}} E_-\,,\qquad 
\mathfrak{S}_{\bar{L},-p}^{str} = - C \mathfrak{Q}_{L,p} C^{-1}= i \sqrt{h \sin\frac{p}{2}} E_+, \nonumber\\
&&\mathfrak{Q}_{\bar{R},-p}^{str} = - C \mathfrak{Q}_{R,p} C^{-1}= - i \sqrt{h \sin\frac{p}{2}} E_+\,,\qquad  
\mathfrak{S}_{\bar{R},-p}^{str} = - C \mathfrak{Q}_{R,p} C^{-1}= i \sqrt{h \sin\frac{p}{2}} E_-, 
\end{eqnarray}
where the supertranspose of a matrix $M$ is defined as
\begin{equation}
M^{str}_{ij} = (-)^{ij+i} \, M_{ji}\,,
\end{equation}
and the charge conjugation matrix as 
\begin{equation}
C = \begin{pmatrix}1&0\\0&i\end{pmatrix}\,.
\end{equation}

Up to a dressing factor, the R matrix for the scattering of a crossed particle with an uncrossed one is given by
\begin{equation}\label{eq:RRL}
  \begin{aligned}
    &R_c |\phi\rangle \otimes |\phi\rangle\ = - {} \, A_{p_1,p_2} |\phi\rangle \otimes |\phi\rangle - \,B_{p_1,p_2} |\psi\rangle \otimes |\psi\rangle, \\
    &R_c |\phi\rangle \otimes |\psi\rangle\ =  |\phi\rangle \otimes |\psi\rangle, \\
    &R_c |\psi\rangle \otimes |\phi\rangle\ = -  |\psi\rangle \otimes |\phi\rangle, \\
    &R_c |\psi\rangle \otimes |\psi\rangle\ = - {} \, B_{p_1,p_2} |\phi\rangle \otimes |\phi\rangle + \, A_{p_1,p_2} |\psi\rangle \otimes |\psi\rangle,
\end{aligned}
\end{equation}
where $p_1$ is in the crossed region and $p_2$ in the physical region. $R_c$ satisfies
\begin{equation}\label{definiRL}
   \Delta_N^{\text{op}} (\mathfrak{a})\,  R_c \ =\ R_c\,  \Delta_N (\mathfrak{a})\,,
\qquad \forall \, \, \mathfrak{a} \in 
\mathfrak{su}(1|1)^2_{\mbox{\scriptsize c.e.}}\,.
\end{equation}
The crossing equation reads
\begin{equation}
\label{cro}
R_{LL} \, \big[C^{-1}\otimes \mathfrak{1}\big] \, R_{\bar{L}L,(-p_1,p_2)}^{str_1} \big[C \otimes \mathfrak{1}\big] = \frac{\sin \frac{p_2 + p_1}{4}}{\sin \frac{p_2 - p_1}{4}} \, \mathfrak{1} \otimes \mathfrak{1}\,.
\end{equation}
Similarly to $R$, the crossed R matrix can be shown to satisfy 
\begin{eqnarray}
\label{boostRL}
&&\Delta_N (\mathfrak{J}_L) (R_{c}) = \Delta_N^{\text{op}} (\mathfrak{J}_L) (R_{c}) = 0\,,
\end{eqnarray}
with an analogous expression for $\mathfrak{J}_R$. As in the previous section, this condition can be re-written in a more geometrical form
\begin{equation}
\label{77}
\bigg[\frac{\partial}{\partial p_M} - \Gamma_M\bigg]R_{c} = 0, \qquad M =1,2\,.
\end{equation}
We perform the continuation to the crossed region according to $\sqrt{\sin p_1/2} = i \sqrt{ | \sin p_1/2 | }$.
Integrating along a contour $\gamma$ gives the following expression for the $R_c$-matrix:
\begin{equation}
\label{int_Rrl}
R_c \big[\gamma (\lambda)\big] = \Theta \, {\cal P} \exp\left(\int_{\gamma(0)}^{\gamma(\lambda)} dp^M \Omega_M \right)\,, 
\end{equation}
where the path starts at $(p,p)$ and ends at $(p_1,p_2)$, and the matrix $\Theta $ is defined as
\begin{equation}
\Theta = E_+ \otimes E_+ - E_- \otimes E_- + E_{11} \otimes E_{22} - E_{22} \otimes E_{11},  \nonumber
\end{equation}
with~\footnote{We have again used the fact that $\{\Theta, \Omega_M\}=0$ to extract the matrix $\Theta$ in front and adjust the sign of the exponent in (\ref{int_Rrl}).}
\begin{equation} 
\label{gl111}
E_{11} \equiv \begin{pmatrix}1&0\\0&0\end{pmatrix}, \qquad E_{22} \equiv \begin{pmatrix}0&0\\0&1\end{pmatrix}.
\end{equation} 

Including a dressing factor, which we call $\Psi$ to distinguish it~\footnote{The simple relationship between $\Phi$ and $\Psi$ will be fixed in section \ref{sec:Comparison}.} from $\Phi$ -- the difference being the arbitrarily chosen normalisation of (\ref{eq:RRL}) w.r.t. (\ref{eq:RLL}) -- we write
\begin{eqnarray}
\label{77f}
&&\bigg[\frac{\partial}{\partial p_M} + \Omega_M - \frac{\partial}{\partial p_M} \log \Psi \bigg]\tilde{R}_{c} = 0, \qquad \tilde{R}_{c} = \Psi R_{c}\,.
\end{eqnarray}
In fact, in order for (\ref{cro}) to be compatible with crossing symmetry, one needs to impose
\begin{equation}
\label{crosf}
\Phi_{(p_1,p_2)} \, \Psi_{(-p_1,p_2)} = \frac{\sin \frac{p_2 - p_1}{4}}{\sin \frac{p_2 + p_1}{4}} \equiv f^{-1}_{p_1,p_2}\,,
\end{equation}
where the continuation to negative momenta was described in detail in \cite{Borsato:2016xns}.

\subsection{Relativistic limit and crossing}
\label{sec:Relativistic2}

In the relativistic limit, crossing symmetry on superalgebra generators~\eqref{accor} takes the form
\begin{eqnarray}
\label{in}
&&\mathfrak{Q}_{\bar{L},-q}^{str} = - C \mathfrak{Q}_{L,q} C^{-1} = - i \sqrt{\frac{c q}{2}} E_-\,,\qquad 
\mathfrak{S}_{\bar{L},-q}^{str} = - C \mathfrak{Q}_{L,q} C^{-1}= i \sqrt{\frac{c q}{2}} E_+, \nonumber\\
&&\mathfrak{Q}_{\bar{R},-q}^{str} = - C \mathfrak{Q}_{R,q} C^{-1}= - i \sqrt{\frac{c q}{2}} E_+\,,\qquad 
\mathfrak{S}_{\bar{R},-q}^{str} = - C \mathfrak{Q}_{R,q} C^{-1}= i \sqrt{\frac{c q}{2}} E_-, 
\end{eqnarray}
where the crossing map reduces to the familiar relativistic one
\begin{equation}
q \to - q, \qquad \theta \to i \pi + \theta\,.
\end{equation}
Ignoring the dressing factor, the relativistic limit of the crossed R matrix $R_c$~\eqref{eq:RRL} is 
\begin{equation}\label{eq:RRLr}
  \begin{aligned}
    &R_c |\phi\rangle \otimes |\phi\rangle\ = - {} \, \tanh \frac{\vartheta}{2} |\phi\rangle \otimes |\phi\rangle  - {} \, \mbox{sech} \frac{\vartheta}{2} |\psi\rangle \otimes |\psi\rangle, \\
    &R_c |\phi\rangle \otimes |\psi\rangle\ =  |\phi\rangle \otimes |\psi\rangle, \\
    &R_c |\psi\rangle \otimes |\phi\rangle\ = -  |\psi\rangle \otimes |\phi\rangle, \\
    &R_c |\psi\rangle \otimes |\psi\rangle\ = - {} \, \mbox{sech} \frac{\vartheta}{2} |\phi\rangle \otimes |\phi\rangle  + {} \, \tanh \frac{\vartheta}{2} |\psi\rangle \otimes |\psi\rangle\,,
\end{aligned}
\end{equation}
and it satisfies the differential equation
\begin{equation}
\label{reado}
\bigg[\frac{\partial}{\partial \vartheta} + \tfrac{1}{2}\sech\tfrac{\vartheta}{2}\big(E_+ \otimes E_+ + E_- \otimes E_-\big)\bigg]R_c(\vartheta)=0\,,
\end{equation}
which can be solved analogously to equation~\eqref{gude}.

Expanding on an idea put forward in \cite{Fontanella:2016opq}, we consider the two expression for the R matrix, namely (\ref{eq:RLLlimtheta}) and (\ref{eq:RRLr}), as pertaining to two separate patches of a fiber bundle, \footnote{We thank Jock McOrist for discussions on this point.} with the R matrix being a covariantly-constant section, and the connection being simply read-off from (\ref{read}) and (\ref{reado}), respectively. Going from one patch to the other amounts to a non-trivial transformation on the matrices. One can also implement such transformation by the constant transition function
\begin{equation}
R_{c}(\vartheta) = P R(\vartheta) P^{-1},
\end{equation}
with
\begin{equation}
P=\begin{pmatrix}
0&1&0&0\\1&0&0&0\\0&0&0&1\\0&0&-1&0
\end{pmatrix}.
\end{equation}

Turning to the dressing factors, the full crossing equation reads
\begin{equation}
\label{cror}
R (\vartheta) \, \big[C^{-1}\otimes \mathfrak{1}\big] \, R_c^{str_1}(i \pi + \vartheta) \big[C \otimes \mathfrak{1}\big] = - \coth \frac{\vartheta}{2} \, \mathfrak{1} \otimes \mathfrak{1},
\end{equation}
hence the dressing factors need to satisfy
\begin{equation}
\label{needs}
\Phi(\vartheta) \, \Psi(\vartheta + i \pi) = - \tanh \frac{\vartheta}{2}\,.
\end{equation}

\section{Relativistic limit of the massless dressing phase}
\label{sec:Relativistic3}

In this section, we derive the relativistic limit of the phase for massless-massless scattering constructed in~\cite{Borsato:2016xns}. In a large-$h$ expansion, the two leading terms in the dressing phase~\cite{Borsato:2016xns} are referred to as Arutyunov-Frolov-Staudacher (AFS)~\cite{Arutyunov:2004vx} and Hern\'andez-L\'opez (HL)~\cite{Hernandez:2006tk} phases, and they correspond to the ${\cal{O}}(h)$ and ${\cal{O}}(1)$ orders, respectively. Since the AFS term tends to $1$ in the relativistic limit, we shall focus on the HL term in what follows. The higher order terms become trivial in the relativistic limit.

In order to solve the crossing equation, a specific path was 
chosen~\cite{Borsato:2016xns} along which to perform the analytic continuation of the phase from the physical region $\mbox{Re}(p) \in (0, 2\pi)$ into the crossed region $\mbox{Re} (p) \in (-2\pi,0)$. Such a path in the $p$-plane was singled out as going from a real $p \in (0, 2\pi)$ to $-p$, intersecting the imaginary axis  for $\mbox{Im} (p )<0$.

In the relativistic limit (\ref{unde}) the physical region in the $q$-plane is the entire half-plane $\mbox{Re}( q )>0$, and the path used for crossing goes from a real $q>0$ to $-q$, intercepting the imaginary axis for $\mbox{Im}( q) <0$. In terms of the rapidity variable $\theta$ defined in equation~\eqref{rap}, the physical region is mapped into the strip $\mbox{Im}( \theta) \in (-\frac{\pi}{2}, \frac{\pi}{2})$ in the $\theta$-plane, and the path used for crossing goes from a real $\theta \in (-\infty,\infty)$ to $\theta - i \pi$, intercepting the lower branch cut $\mbox{Im}( \theta) = - \frac{\pi}{2}$.  

Below we obtain the relativistic limit of the massless dressing phase and show that it reduces to the famous  scalar factor of the sine-Gordon model obtained by the Zamolodchikovs~\cite{Zamolodchikov:1978xm}.

\subsection{\label{sec:Integral}Integral representation}

The massless HL phase has the following integral representation~\cite{Borsato:2016xns}
\begin{equation}\label{eq:hl-generic}
\begin{aligned}
\theta^{\HL} (x^\pm,y^\pm) &=\int\limits_{-1+i\epsilon}^{1+i\epsilon} \frac{dz}{4\pi}G_-(z,y^+)\bigl( g(z,x^+)-g(z,x^-) \bigr)
-\int\limits_{-1-i\epsilon}^{1-i\epsilon}  \frac{dz}{4\pi}G_+(z,y^-) \bigl( g(z,x^+)-g(z,x^-) \bigr) 
\\
& \qquad -\frac{i}{2} \left( G_-(\tfrac{1}{x^-},y^+)-G_+(\tfrac{1}{x^+},y^-) \right)\,,
\end{aligned}
\end{equation}
where $x^+ x^-=1=y^+y^-$ and~\footnote{The function $g$ does \emph{not} depend on the choice of sign $\pm$
that enters $G_\pm$.}
\begin{equation}
G_\pm(z,y)\equiv\log\left(\pm i(y-z)\right)-\log\left(\pm i(y-\tfrac{1}{z})\right)\,,\quad g(z,x)\equiv\frac{\partial}{\partial z}G_{\pm}(z,x)=\frac{1}{z-x}-\frac{1}{z-\tfrac{1}{x}}+\frac{1}{z}\,.
\end{equation}
In the relativistic limit we define
\begin{equation}
x^+=e^{\frac{ip_1}{2h}}\,,\qquad
y^-=e^{\frac{ip_2}{2h}}\,,
\end{equation}
and take the limit $h\rightarrow\infty$, while keeping the real part of the momenta $p_1$ and $p_2$ positive. Relegating the details to Appendix~\ref{app:rel-lim-int-hl}, we find 
\begin{equation}
\label{eq:hl-rel-limit-45}
\theta^{\HL}_{ \mbox{rel}} (p_1,p_2) \equiv \lim_{h\rightarrow\infty}\theta^{\HL}(x^\pm\,,y^\pm)= \frac{2}{\pi}\int\limits_{0}^{i \infty} 
dr\frac{p_1\log\frac{p_2-r}{p_2+r}}{p_2^2-r^2}-\frac{\pi}{2}
\,.
\end{equation}
Introducing massless rapidity variables
\begin{equation}
p_1=e^{\theta_1}\,,\qquad
p_2=e^{\theta_2}\,,\qquad
r=ie^{\phi}\,,
\end{equation}
we may write
\begin{equation}
\theta^{\HL}_{\mbox{rel}} (\theta_1,\theta_2)=\frac{2i}{\pi}\int\limits_{-\infty}^{\infty} 
d\phi
\frac{ e^{\theta_1+\phi } }{e^{2 \theta_1}+e^{2 \phi }}\log \left(\frac{e^{\theta_2}-i e^{\phi }}{e^{\theta_2}+i e^{\phi }}\right)
-\frac{\pi}{2}\,.
\end{equation}
Redefining the integration variable $\phi\rightarrow \phi+\theta_2$ we have
\begin{equation}
\label{eq:our-int-rep}
\theta^{\HL}_{\mbox{rel}} (\theta_1,\theta_2)
\equiv\theta^{\HL}_{\mbox{rel}} (\vartheta)
=\frac{2i}{\pi}\int\limits_{-\infty}^{\infty} 
d\phi
\frac{ e^{\vartheta+\phi } }{e^{2 \vartheta } +e^{2 \phi }}
\log \left(\frac{1-ie^{\phi }}{1+i e^{\phi }}\right)
-\frac{\pi}{2}\,,
\end{equation}
with $\vartheta=\theta_1-\theta_2$ showing that in the relativistic limit the dressing phase is of difference form, as expected for a  relativistic theory. The corresponding relativistic dressing factor is defined for rapidities in the physical strip as
\begin{equation}
\sigma^{\HL}_{\mbox{rel}}(p_1,p_2)\equiv \sigma^{\HL}_{\mbox{rel}}(\vartheta)=
\exp\left(\tfrac{i}{2}\theta^{\HL}_{\mbox{rel}} (\vartheta)
\right)\,.
\label{4.8}
\end{equation}
The dressing phase~\eqref{eq:our-int-rep} takes the form of a conventional Riemann-Hilbert type integral~\eqref{eq:conv-rh-int}, with a cut along the imaginary momentum axis. We may use the Sochocki-Plemelj theorem~\cite{Sochocki,Plemelj} to determine the value of the dressing factor after analytically continuing through the cut at $\mbox{Im}(\vartheta) = \tfrac{\pi}{2}$
\begin{equation}
\label{eq:an-cont-bel}
\sigma^{\HL}_{\mbox{rel}}\left(\vartheta+i\left(\tfrac{\pi}{2}-\epsilon\right)\right)
=-\coth\left(\frac{\vartheta+\tfrac{i\pi}{2}}{2}\right)
\sigma^{\HL}_{\mbox{rel}}\left(\vartheta+i\left(\tfrac{\pi}{2}+\epsilon\right)\right)\,.
\end{equation}
Similarly,  continuing through the cut at $\mbox{Im}(\vartheta) = -\tfrac{\pi}{2}$ we have
\begin{equation}
\label{eq:an-cont-abv}
\sigma^{\HL}_{\mbox{rel}}\left(\vartheta-i\left(\tfrac{\pi}{2}-\epsilon\right)\right)
=-\tanh\left(\frac{\vartheta-\tfrac{i\pi}{2}}{2}\right)
\sigma^{\HL}_{\mbox{rel}}\left(\vartheta-i\left(\tfrac{\pi}{2}+\epsilon\right)\right)\,.
\end{equation}
From these relations we can immediately deduce the crossing equations
\begin{equation}
\label{eq:rel-cross-hl-dress}
\sigma^{\HL}_{\mbox{rel}}(\vartheta)\,
\sigma^{\HL}_{\mbox{rel}}(\vartheta+i\pi)=i\tanh\tfrac{\vartheta}{2}\,,
\qquad\qquad
\sigma^{\HL}_{\mbox{rel}}(\vartheta)\,
\sigma^{\HL}_{\mbox{rel}}(\vartheta-i\pi)=i\coth\tfrac{\vartheta}{2}\,.
\end{equation}
Using equations~\eqref{eq:an-cont-bel} and~\eqref{eq:an-cont-abv}, and the fact that the integral~\eqref{eq:our-int-rep} can be computed for any value of $\vartheta$, the dressing factor on the whole rapidity plane is given by the value of the integral times the terms one picks up by crossing the cuts\footnote{The two equations given in (\ref{eq:rel-cross-hl-dress}) are equivalent to one another upon shifting the rapidity $\vartheta$ by $\pm i \pi$ as long as the dressing factor is explicitly meromorphic. We have checked that both relations are satisfied by our expression (\ref{4.8}) in order to ensure that the apparent cuts do not spoil this property.}
\begin{equation}\label{eq:int-exp-dress-fact}
\sigma^{\scriptsize\HL}_{\mbox{\scriptsize rel}}(\vartheta)=
e^{\frac{i}{2}\theta^{\scriptsize\HL}_{\mbox{\scriptsize rel}} (\vartheta)} \tanh ^{n(\vartheta)}
\left(-\tfrac{\vartheta }{2}\right)\,.
\end{equation}
Above,  $n(\vartheta)$ is defined in terms of the ceiling function~\footnote{The ceiling of a real number $x$ is defined as the smallest integer greater than or equal to $x$, and is denoted by $\lceil x \rceil$.}
\begin{equation}
n(\vartheta)=-\left\lceil\mbox{Im}\left(\tfrac{\vartheta}{\pi}\right)-\tfrac{1}{2}\right\rceil\,.
\end{equation}

\subsection{\label{sec:Comparison}Comparison with Zamolodchikov's phase factor}

We shall now compare the relativistic limit of the dressing factor, which we have obtained in the previous sections, with the famous scalar factor obtained by Zamolodchikov for the sine-Gordon model (sG), with the coupling set to $\beta_*$ given in equation~\eqref{beta*}, and find them to agree. At this value of the coupling the sine-Gordon scalar factor, which multiplies the scattering matrix between a sG soliton and a sG anti-soliton~\cite{Zamolodchikov:1978xm} (see~\cite{Bombardelli:2016scq} for a recent review) can be written as~\footnote{One obtains this formula by setting
\begin{equation}
\gamma = 16 \pi \qquad \iff \qquad \beta^2 = \beta_*^2 = \frac{16 \pi}{3}\nonumber
\end{equation}
in formula (4.11) of~\cite{Zamolodchikov:1978xm} and redefining the rapidity variable to include a minus sign.}
\begin{eqnarray}
\label{zamo}
S (\vartheta) = \prod_{\ell=1}^\infty \frac{\Gamma^2(\ell - \tau) \, \Gamma(\frac{1}{2} + \ell + \tau) \,\Gamma(- \frac{1}{2} + \ell + \tau)}{\Gamma^2(\ell + \tau) \, \Gamma(\frac{1}{2} + \ell - \tau) \,\Gamma(- \frac{1}{2} + \ell - \tau)},
\end{eqnarray}
where  
\begin{eqnarray}
\tau \equiv \frac{\vartheta}{2 \pi i}.
\end{eqnarray}
Expression (\ref{zamo}) solves the crossing equation
\begin{eqnarray}
\label{crosszamo}
S(\vartheta) \, S(\vartheta + i \pi) = i \tanh \frac{\vartheta}{2}\,,
\end{eqnarray}
which is the same as what the relativistic limit of the HL phase satisfies~\eqref{eq:rel-cross-hl-dress}. Therefore, the two dressing factors can differ by at most CDD factors.
We have in fact verified numerically that the formula (\ref{zamo}) exactly reproduces the relativistic limit of the massless phase we derived in the previous sections. More precisely, 
\begin{eqnarray}
\sigma^{\scriptsize\HL}_{\mbox{\scriptsize rel}} (\vartheta) = S(\vartheta)\,,
&\qquad\qquad  &   \mbox{Im}(\vartheta) \in (-\pi/2 , \pi/2)\,,\\
\Phi(\vartheta)=S(\vartheta)\,,&\qquad\qquad&\Psi(\vartheta)=iS(\vartheta)\,.
\end{eqnarray} 
As far as we are aware, the integral expression~\eqref{eq:our-int-rep} for the Zamolodchikov dressing factor has not previously appeared in the literature and is different from other known integral formul\ae{} such as those given in~\cite{Fendley:1991ve} or~\cite{Karowski:1977tv,Karowski:1977fu}. 

\subsection{\label{sec:Comparison2}Comparison with the literature on 2D ${\cal{N}}=2$ theories}

Our S matrix is closely related to the S matrix of the massless ${\cal N}=2$ super-sine-Gordon model~\cite{Kobayashi:1991st,Kobayashi:1991rh,Kobayashi:1991jv} at a special value of its coupling.~\footnote{Similar S matrices have appeared in other contexts. This is to be expected since the super-algebras used to fix the S matrices are the same (see section~\ref{sec:Theqs}). For example, the matrix $R_{LL}$ in the relativistic limit (\ref{eq:RLLlimtheta}) coincides with a subsector of the R matrix obtained in~\cite{Fendley:1990zj} for the scattering of solitons in integrable deformations of ${\cal N}=2$ minimal models, though the theory considered there is massive. The massless ${\cal N}=2$ super-sine-Gordon S matrix appears, for example, in the study of integrable flows of ${\cal N}=2$ Landau Ginzburg theories~\cite{Fendley:1993pi}.} The difference between our R matrix and the massless ${\cal N}=2$ super-sine-Gordon R matrix at coupling $\beta=\beta_*$, where $\beta_*$ has been introduced in equation~(\ref{beta*}), is located  in the entries 
\begin{equation}
|\phi\rangle \otimes |\psi\rangle \to |\phi\rangle \otimes |\psi\rangle
\,,
\qquad\mbox{and}\qquad
|\psi\rangle \otimes |\phi\rangle \to |\psi\rangle \otimes |\phi\rangle\,.
\end{equation}
In our case these R matrix entries can be read off from equation~\eqref{eq:RLLlimtheta}
\begin{equation}
\mp \Phi(\vartheta) \tanh \frac{\vartheta}{2}\,.
\end{equation}
On the other hand, the corresponding entries of the massless ${\cal N}=2$ super-sine-Gordon model S matrix~\cite{Kobayashi:1991st,Kobayashi:1991rh,Kobayashi:1991jv,Fendley:1991ve} at $\beta=\beta_*$, are both equal to 
\begin{equation}
-i \Phi(\vartheta) \tanh \frac{\vartheta}{2}\,.
\end{equation}
This difference comes from different statistics of the scattering particles, and results in different braidings of the coproducts.~\footnote{We thank Paul Fendley for communication about this point.} This is in fact the only difference between our S matrix and the super-sine-Gordon one, because, as we found in section~\ref{sec:Comparison}, our dressing factor matches the corresponding sine-Gordon one.

Recall that the S matrix of the ${\cal N}=2$ super-sine-Gordon model at any coupling $\beta_{\mathcal{N}=2}$ factorises into two sine-Gordon S matrices
\begin{equation}
S_{ssG} (\beta_{\mathcal{N}=2}) = S_{sG} (\beta_{\mathcal{N}=0}) \otimes S_{sG}(\beta_{\mathcal{N}=0}=\beta_*)\,,
\label{eq:SssG}
\end{equation}
where one of the sine-Gordon factors is at the particular value of the coupling constant $\beta_*$ given in equation~(\ref{beta*}), while the second factor's coupling constant $\beta_{\mathcal{N}=0}$ is related to the ${\cal N}=2$ coupling constant by~\cite{Kobayashi:1991st,Kobayashi:1991rh,Kobayashi:1991jv,Fendley:1992dm}
\begin{equation}
 \beta^2_{\mathcal{N}=2}=\frac{\beta^2_{\mathcal{N}=0}}{1-\frac{\beta^2_{\mathcal{N}=0}}{8\pi}}\,.
\end{equation}
This type of factorisation is familiar from other supersymmetric integrable models; see for example~\cite{Shankar:1977cm,Schoutens:1990vb,Ahn:1990uq,Ahn:1990gn}. Notice that at $\beta^2_{\mathcal{N}=2}=16\pi$, or equivalently at $\beta^2_{\mathcal{N}=0}=16\pi/3$, $S_{ssG}$ is a tensor product of two sine-Gordon S matrices at the special point $\beta_*$. It is well known~\cite{Dunning} that at this value of the coupling the massless sine-Gordon theory corresponds to a free boson, with the S matrix reducing to the non-perturbative S matrices of the type introduced by Zamolodchikov~\cite{ZamoMasslessTBA}. Since the massless ${\cal N}=2$ sine-Gordon S matrix at $\beta^2_{\mathcal{N}=2}=16\pi$ is just a tensor product of two such "free" factors, we expect it will also give an integrable description of a free CFT. As we discussed above, the relativistic S matrix for $\CFT^{(0)}$ is very similar to the one of the massless ${\cal N}=2$ super-sine-Gordon theory at $\beta^2_{\mathcal{N}=2}=16\pi$. We take this as evidence that the $\CFT^{(0)}$ will analogously be a free theory, with the natural candidate the zero-momentum, zero-winding subsector of the supersymmetric $\Torus^4$ theory.

Furthermore, as we shall show in the next section, the similarity with  $\mathcal{N}$ = 2 super-sine-Gordon will extend also to the Thermodynamic Bethe Ansatz (TBA)  equations: in particular, for the ground state we shall get equations identical to those of the $\mathcal{N}$ = 2 super-sine-Gordon model \cite{Fendley:1993pi} in the massless limit, though we expect that the excited states will be different, due to the differences at the level of S matrix and Bethe equations.

\section{\label{sec:Thermodynamic}Thermodynamic Bethe ansatz}

In this section, we provide the TBA  equations \cite{Zamolodchikov:1989cf} (see \cite{vanTongeren:2016hhc} for a recent review) restricted to the massless sector. Having established a relationship with a standard relativistic field-theory construction related to ${\cal{N}}=2$ theories, we would like to exploit this to move the first steps into the finite-size program for this sector. It will eventually be necessary to extend this framework to the whole theory in order to completely solve the model, as it was done for higher-dimensional cases \cite{Bombardelli:2009ns,Gromov:2009bc,Arutyunov:2009ur,Bombardelli:2009xz,Gromov:2009at} (see also the review \cite{Bajnok:2010ke}).

Let us get inspiration from the treatment of \cite{Fendley:1991ve,Fendley:1993pi}, where the TBA was used to obtain the Casimir energy of the 2D theory compactified on a spatial circle of length $R$. According to Zamolodchikov's idea  \cite{Zamolodchikov:1989cf}, one can use the asymptotic data of the scattering problem to derive integral equations for the finite-size spectrum, utilising the principle of the double Wick-rotation. This amounts to exchanging space with time, turning a problem which is periodic with period $R$ in space and infinite time $L \to \infty$, into one which is decompactified in space and with periodic time, {\it i.e.} at finite temperature $\frac{1}{R}$. Thanks to relativistic invariance, we are guaranteed to be able to use the same principle of double Wick-rotation in our relativistic-limit situation.

Based on this reasoning, the ground-state energy of the original model (which is the leading contribution to the partition function at large time) can be read-off from the minimum free energy $F_{min}$ at large $L$ of the doubly Wick-rotated model:
\begin{equation}
E_0(R) = \lim_{L \to \infty} \frac{R \, F_{min}}{L}.
\label{freeenergy}
\end{equation}

For ${\cal{N}}=2$ theories, for instance, this procedure reproduces the correct central charge for the massless flows which \cite{Fendley:1993pi} were concerned about. The massless scattering theory describes a renormalisation group flow between a UV and an IR fixed point, and the TBA computes the ground state energy at arbitrary intermediate points along the flow. This ground state (Casimir) energy then is shown to correctly approach the UV and IR CFT central charges at the two respective extrema of the flow.

The first fundamental ingredient to perform a similar analysis in our case is the formulation of a set of Bethe equations describing the large volume spectrum of the massless sector in the relativistic limit, that is the subject of the next section.

\subsection{Relativistic Bethe equations \label{sec:Transfer}}

The Bethe equations can be constructed employing the tool of the transfer matrix, which is built as the trace of a string of S matrices for an ordered sequence of interacting particles. Let us briefly outline the calculation in our case.

If one considers $N$ particles, taken to be all bosonic for the moment, on a circle of length $L$, interacting one with each other via an integrable scattering matrix, one is brought to impose the following quantisation conditions on the momenta:
\begin{eqnarray}
\label{eigen}
e^{i p_k L} \, T(p_k | p_1,...,p_N) |\psi\rangle = |\psi \rangle, \qquad k = 1,...,N,
\end{eqnarray} 
where $p_i$ is the momentum of the $i$-th particle on the circle
\begin{eqnarray}
p_i = e^{\theta_i},
\end{eqnarray} 
while
\begin{eqnarray}
\label{tracea}
T(p_0|p_1,...,p_N) = \mbox{tr}_0 M(p_0|p_1,...,p_N)
\end{eqnarray} 
is the transfer matrix, namely the trace over the auxiliary $0$-th space of the monodromy matrix 
\begin{eqnarray}
\label{traccia}
\Big[M_a^b(p_0|p_1,...,p_N)\Big]_{c_1 ... c_N}^{d_1 ... d_N} = \sum_{\{k\}} S_{a c_1}^{d_1 k_1} (\theta_0 - \theta_1) \,  S_{k_1 c_2}^{d_2 k_2} (\theta_0 - \theta_2)... S_{k_{N-1} c_N}^{d_N b} (\theta_0 - \theta_N),
\end{eqnarray}
and $S$ is the two-body S matrix. Equations (\ref{eigen}), (\ref{tracea}) and (\ref{traccia})  are saying that revolving each particle around the circle of length $L$, while scattering all the other ones in sequence, amounts to the identity acting on an eigenstate $|\psi \rangle$ of the transfer matrix. Normally one would exclude the same particle $k$ in the scattering sequence, however we can include it since $S_{a c}^{d b}(0) = \delta_a^d \delta_c^b$, which acts by effectively permuting the two scattering particles and has the result of cutting the product (\ref{traccia}) precisely in correspondence with particle $k$, as it is needed. 

\begin{figure}
\centerline{\includegraphics[width=10cm]{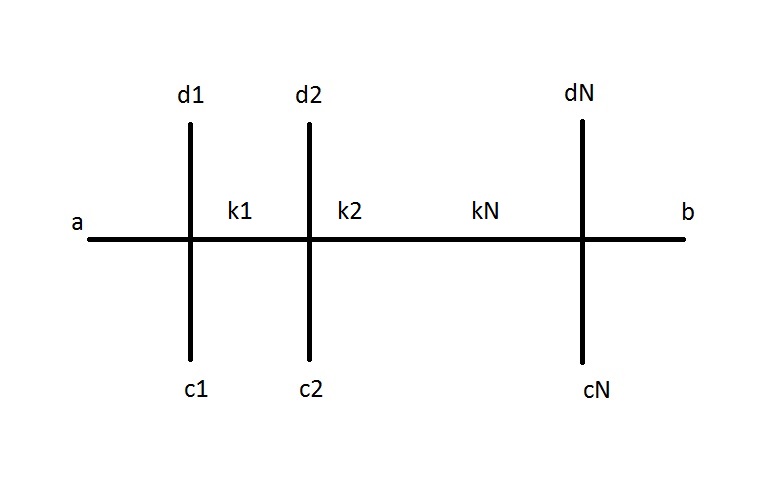}}
\caption{The transfer matrix is obtained by identifying $a=b$ in the monodromy matrix,  and summing over $\sum_a$. The indices $a$ and $b$ are in the auxiliary $0$-th space, while the indices $c_i$ and $d_i$ pertain to the chain of frame particles (often referred to as the {\it quantum space}).}
\label{figura}
\end{figure}

From this treatment it is clear that the next task is to find the eigenstates $|\psi \rangle$ of the transfer matrix. For non-diagonal scattering, when the S matrix is not just a scalar but, as in our case, it does transform non-trivially the particles' internal degrees of freedom, diagonalisation is best achieved via the so-called Algebraic Bethe Ansatz (ABA) technique. One can prove that, if one constructs the tensor
\begin{eqnarray}
\Sigma_{ij} = S_{a c}^{d b} (\theta_i - \theta_j) E_{ba} \otimes E_{dc},
\end{eqnarray} 
where $E_{x y}$ are the standard matrix unities, then 
\begin{equation}
\Sigma_{ij} = R_{ij}
\end{equation}
and
\begin{eqnarray}
\Sigma_{0N} ... \Sigma_{01} = \Big[M_a^b(p_0|p_1,...,p_N)\Big]_{c_1 ... c_N}^{d_1 ... d_N} E_{ba} \otimes E_{d_1 c_1} \otimes ... \otimes E_{d_N c_N}.
\end{eqnarray}
In the supersymmetric case, we therefore now take
\begin{eqnarray}
\label{maint}
R_{0N} ... R_{01}
\end{eqnarray}
 as the appropriate definition of the monodromy matrix to be used, and switch to the supertrace. We perform the full algebraic Bethe Ansatz (ABA) for the transfer matrix resulting from such definition in Appendix A. The result is as follows:
\begin{eqnarray}
&&{\cal{T}}(p_0|p_1,...,p_{K_0}) |q_1,...,q_M \rangle = \Lambda |q_1,...,q_M \rangle
\nonumber\\
&&\Lambda = \Lambda(q_1,...,q_M; p_0|p_1,...,p_{K_0}) = \bigg( 1-\prod_{i=1}^{K_0} \tanh \frac{\theta_0 - \theta_i}{2}\bigg) \prod_{i=1}^{K_0} \Phi(\theta_0 - \theta_i) \, \prod_{i=1}^M \mbox{coth} \frac{\beta_i - \theta_0}{2},
\end{eqnarray}
where $M \equiv K_1+K_3=0,1,2,...$ is the total number of level-1 magnons with momenta $q_i = e^{\beta_i}$. These are magnon excitations created by  the upper-triangular entry of the monodromy matrix ${\cal{M}}$, which we call $B(q_i|p_1,...,p_{K_0})$, over the pseudo-vacuum $ |0\rangle = |\phi \rangle \otimes ... \otimes |\phi \rangle$ formed out of $K_0$ bosons with momenta $p_k$:
\beq
|q_1,...,q_M \rangle = B(q_1|p_1,...,p_N) ... B(q_M|p_1,...,p_N) |0\rangle\,.
\eeq 
From the point of view of the nested Bethe ansatz, the $K_0$ particles work at the next level as an effective new chain, of length $K_0$ and with impurities $p_i$, where the level-1 magnons now move. That is why the $K_0$ particles are also called {\it frame particles}, when regarded under this light. The situation is conveniently captured by a diagram of the type in Figure \ref{figura}. 

There is also a quantisation condition for the level-1 magnon momenta (level-1 Bethe equations):
\begin{eqnarray}
\label{BAE2}
\prod_{i=1}^{K_0} \tanh \frac{\beta_k- \theta_i}{2} = 1, \qquad k=1,...,M,
\end{eqnarray}
 meaning that the level-1 magnons only interact with the $K_0$ frame particles (impurities on the level-1 chain), but not one with each other.

It is now clear that
\begin{eqnarray}
\Lambda(q_1,...,q_M; p_k|p_1,...,p_{K_0}) = \prod_{i=1}^{K_0} \Phi(\theta_k - \theta_i) \, \prod_{i=1}^M \mbox{coth} \frac{\beta_i - \theta_k}{2},
\end{eqnarray}
since one of the products is simply entirely annihilated by one of the factors being $0$, specifically $\tanh \frac{\theta_k - \theta_k}{2}$. We see therefore that (\ref{eigen}) can be replaced by the following system of equations:
\begin{eqnarray}
\label{syste}
e^{i L e^{\theta_k}} \, \prod_{i=1}^{K_0} \Phi(\theta_k - \theta_i) \, \prod_{i=1}^M \mbox{coth} \frac{\beta_i - \theta_k}{2} = 1, \qquad k=1,...,K_0,
\end{eqnarray}
quantising the momenta when the system is in the eigenstate characterised by $M$ level-1 magnons with rapidities $\beta_i$, subject to (\ref{BAE2}). 

\begin{figure}
\centerline{\includegraphics[width=5cm]{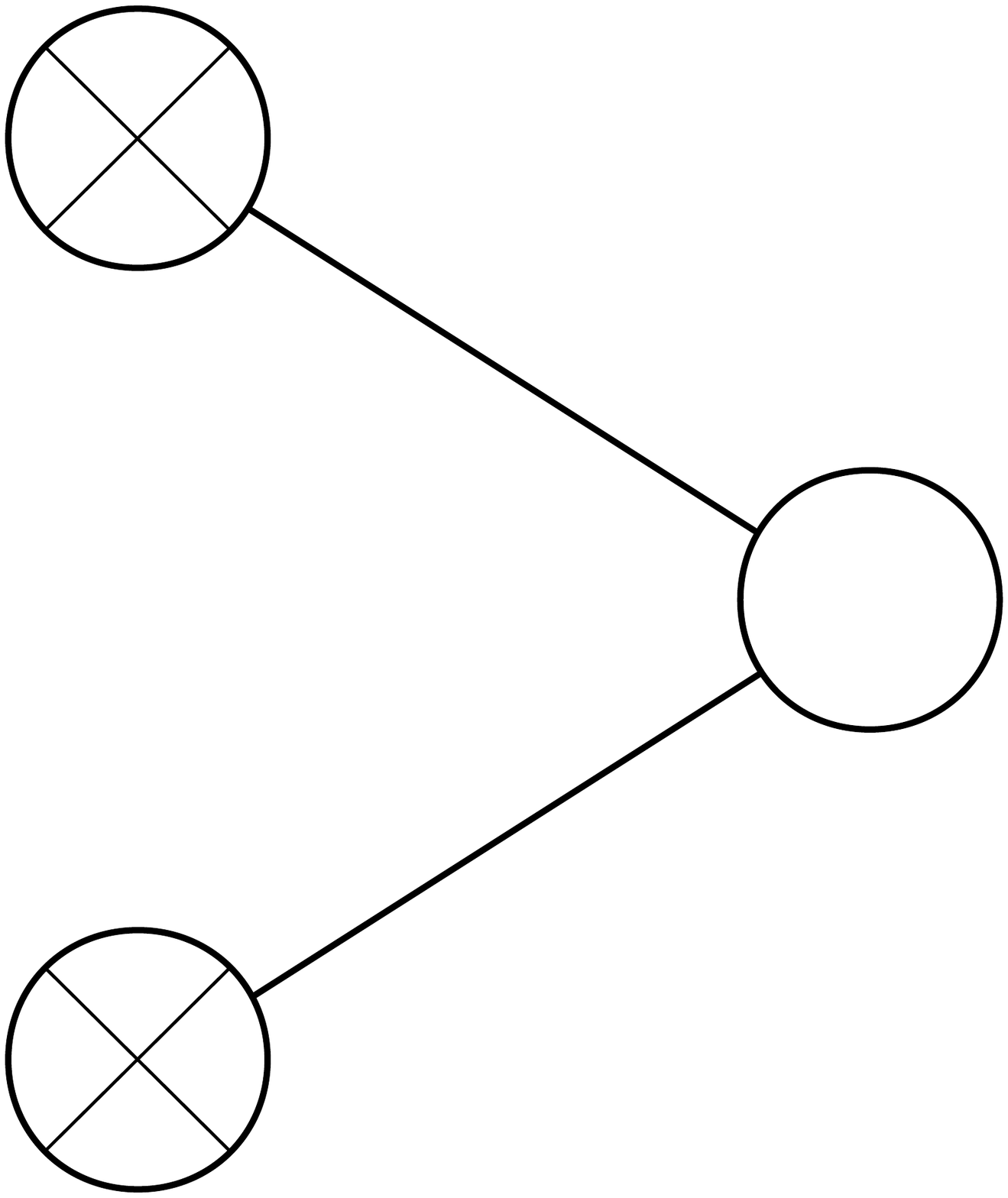}}
\caption{The Dynkin diagram associated to the Bethe equations (\ref{betherel1})-(\ref{betherel3}): the central node corresponds to the momentum carrying variable $\theta$, while the crossed nodes denote the fermionic nodes associated to the auxiliary variables $\beta_1$ and $\beta_3$.}
\label{fig:dynkin}
\end{figure}

The same set of Bethe equations can be obtained directly by applying the relativistic limit to the all-loop Bethe equations for the massless sector \cite{Borsato:2016kbm, Borsato:2016xns}: 
\begin{eqnarray}
&&1=\prod_{j=1}^{K_0}\frac{ y_{1,k} - z_{j}^+}{ y_{1,k} - z_{j}^-}\,,\label{bethe1}\\
&&e^{ip_kL}=\prod_{\stackrel{j=1}{j\neq k}}^{K_0}\sigma_{\circ\circ}^2(z_k^{\pm},z_j^{\pm})\frac{ z_{k}^+ - z_{j}^-}{z_{k}^- - z_{j}^+}\prod_{j=1}^{K_1}\frac{ z_{k}^- - y_{1,j}}{z_{k}^+ - y_{1,j}}
\prod_{j=1}^{K_3}\frac{ z_{k}^- - y_{3,j}}{z_{k}^+ - y_{3,j}}
\,,~~~~\label{bethe3}\\
&&1=\prod_{j=1}^{K_0}\frac{ y_{3,k} - z_{j}^+}{ y_{3,k} - z_{j}^-}\,.
\label{bethe5}
\end{eqnarray}
The relativistic limit corresponds to taking the following small momentum limit on the dynamical variables
\begin{equation}
\label{parame1}
z_{k}^{\pm} = e^{\pm i \epsilon p_k}\,,\quad y_{i,k} = e^{i \epsilon v_{i,k}}\,,\quad \mbox{with}\ \epsilon\rightarrow 0\,,\quad p_k=\pm e^{\theta_k}\,,\quad v_{i,k}=\pm e^{\beta_{i,k}}\,.
\end{equation}
Applying this limit to the Bethe eqs. (\ref{bethe1})-(\ref{bethe5}), we get  scattering phases depending on differences of rapidities and in particular, for the left-movers ($p_k=-e^{\theta_k}$, $v_{i,k}=-e^{\beta_{i,k}}$)
\begin{eqnarray}
1&=&\prod_{j=1}^{K_0}\tanh\left(\frac{\beta_{1,k}-\theta_j}{2}\right)\,,\label{betherel1}\\
e^{-iLe^{\theta_k}}&=&(-1)^{K_0-1}\prod_{\stackrel{j=1}{j\neq k}}^{K_0}S^2(\theta_k-\theta_j)\prod_{j=1}^{K_1}\coth\left(\frac{\beta_{1,j}-\theta_{k}}{2}\right)
\prod_{j=1}^{K_3}\coth\left(\frac{\beta_{3,j}-\theta_{k}}{2}\right)
\,,~~~~~~~~\label{betherel2}\\
1&=&\prod_{j=1}^{K_0}\tanh\left(\frac{\beta_{3,k}-\theta_j}{2}\right)\,,
\label{betherel3}
\end{eqnarray}
where $S(\theta)$ is the Zamolodchikov's sine-Gordon scalar factor, as shown in Section \ref{sec:Comparison}.
It is easy to check that we get exactly the same Bethe equations as (\ref{syste}) and two copies of (\ref{BAE2}), corresponding to the Dynkin diagram represented in Figure \ref{fig:dynkin}. It is this form of the quantisation condition which we shall submit to the TBA analysis of section \ref{Thermo}, following \cite{ZamoMasslessTBA}.

Similar Bethe equations were studied in \cite{Zamolodchikov:1991vh} and \cite{Fendley:1991ve}, for example.
As in those cases, the auxiliary fermionic Bethe roots $\beta_{1,k}$ ($\beta_{3,k}$) organize on two lines at $z_{1,k}+i\pi/2$ ($z_{3,k}+i\pi/2$) and $z_{1,k}-i\pi/2$ ($z_{3,k}-i\pi/2$), with $z_{1,k}$ ($z_{3,k}$) real. 
Then we shall use the notation $\pm1$ ($\pm3$) to denote the Bethe roots of type 1 (3)  placed at $\pm i\pi/2$ respectively. 

If we had started from the dual Bethe equations, also derived in \cite{Borsato:2016kbm, Borsato:2016xns}, for the so-called ''fermionic grading",  in the relativistic limit we would have obtained slightly different equations for the momentum carrying node: 
\begin{eqnarray}
e^{-iLe^{\theta_k}}&=&\prod_{\stackrel{j=1}{j\neq k}}^{K_0}S^2(\theta_k-\theta_j)\prod_{j=1}^{\tilde K_1}\tanh\left(\frac{\beta_{1,j}-\theta_{k}}{2}\right)
\prod_{j=1}^{K_3}\coth\left(\frac{\beta_{3,j}-\theta_{k}}{2}\right)
\,.~~~~~~~~
\label{betherelfermionic}
\end{eqnarray}
However, the differences with respect to (\ref{betherel2}) will not imply any change in the procedure exposed in the next section, also since we are allowed to simply relabel the auxiliary variables: in the dual frame the roots previously labelled by $+1$ can be mapped to roots of type $-1$ and vice versa.

It is also possible to derive the dual Bethe equations directly in the relativistic limit, following section 3.2 of \cite{Borsato:2016xns}. It is sufficient in fact to adopt the same duality transformation employed there, taken in the relativistic limit and switching off the massive roots. If one sets $N_2 = N_{\bar{2}}=0$ in $P(\zeta)$, formula (3.20) in \cite{Borsato:2016xns}, one obtains
\begin{eqnarray}
\label{simu}
P(\zeta) = \prod_{j=1}^{K_0} (\zeta - z_j^+) \nu_j^{-\frac{1}{2}} - \prod_{j=1}^{K_0} (\zeta - z_j^-). 
\end{eqnarray} 
This polynomial clearly vanishes when evaluated on the auxiliary Bethe roots $x_{1,k}$, by virtue of the auxiliary Bethe equations (\ref{bethe1}), and also when evaluated at $\zeta=0$, because of the level matching condition:
\begin{equation}
P(x_{1,k}) = 0, \qquad P(0)=0.
\end{equation}  
Since $P$ is a polynomial of degree $K_0-1$ (the highest power cancels out), it must be
\begin{equation}
\label{ltan}
P(\zeta) = \zeta \, \prod_{k=1}^{K_1} (\zeta - x_{1,k}) \prod_{k=1}^{K_0-K_1-1} (\zeta - \tilde{x}_{1,k}).
\end{equation}
This means that it must simultaneously happen, from (\ref{simu}) and (\ref{ltan}), that
\begin{equation}
\frac{P(z_i^+)}{P(z_i^-)} = \frac{z_i^+ \, \prod_{k=1}^{K_1} (z_i^+ - x_{1,k}) \prod_{k=1}^{K_0-K_1-1} (z_i^+ - \tilde{x}_{1,k})}{z_i^- \, \prod_{k=1}^{K_1} (z_i^- - x_{1,k}) \prod_{k=1}^{K_0-K_1-1} (z_i^- - \tilde{x}_{1,k})} = \frac{- \prod_{j=1}^{K_0} (z_i^+ - z_j^-)}{\prod_{j=1}^{K_0} (z_i^- - z_j^+) \nu_j^{-\frac{1}{2}}}.
\end{equation}
The second equality can be used to convert the factor $\prod_{k=1}^{K_1} \frac{z_i^+ - x_{1,k}}{z_i^- - x_{1,k}}$ in the momentum-carrying equation (\ref{bethe3}) in terms of the dual roots $\tilde{x}$.

In the relativistic limit, it is enough to parametrise the roots in the same way as in (\ref{parame1}): set 
\begin{equation}
\zeta = e^{i \epsilon \, e^u},
\end{equation}
and let $\epsilon \to 0$. The factors of $\nu$ in (\ref{simu}) tend to $1$ being exponential of momenta, then one is left with 
\begin{eqnarray}
&&P(\zeta) \to (i \epsilon)^{K_0} \bigg(\prod_{j=1}^{K_0} (e^u - e^{\theta_j}) - \prod_{j=1}^{K_0} (e^u + e^{\theta_j})\bigg) = \\
&&(i \epsilon)^{K_0} \, \prod_{k=1}^{K_1} 2 \, e^\frac{u - \beta_{1,k}}{2} \sinh\bigg(\frac{u - \beta_{1,k}}{2}\bigg)\prod_{k=1}^{K_0-K_1-1} 2 \, e^\frac{u - \tilde{\beta}_{1,k}}{2} \sinh\bigg(\frac{u - \tilde{\beta}_{1,k}}{2}\bigg) \equiv Q(u)\,,\nonumber
\end{eqnarray}
where we have used the same argument based on the auxiliary Bethe equations (\ref{betherel1}). One does not have a polynomial in the limit, therefore we have used the $\sinh$ function, which has the appropriate zeroes and periodicity. We can now use the fact that the limit of $\frac{P(z_i^+)}{P(z_i^-)}$ is $(-1)^{K_0+1}$ to obtain
\begin{equation}
(-1)^{K_0+1} = \frac{Q(\theta_i)}{Q(\theta_i + i \pi)} = \frac{\prod_{k=1}^{K_1} \sinh\bigg(\frac{\beta_{1,k}- \theta_i}{2}\bigg)\prod_{k=1}^{K_0-K_1-1} \sinh\bigg(\frac{\tilde{\beta}_{1,k}- \theta_i}{2}\bigg)}{\prod_{k=1}^{K_1} \cosh\bigg(\frac{\beta_{1,k}- \theta_i}{2}\bigg)\prod_{k=1}^{K_0-K_1-1} \cosh\bigg(\frac{\tilde{\beta}_{1,k}- \theta_i}{2}\bigg)},
\end{equation}
which can now be used to dualise the momentum-carrying equation (\ref{betherel2}).

\subsection{Thermodynamics \label{Thermo}}

Now, let us consider the thermodynamic limit of (\ref{betherel1})-(\ref{betherel3}), whereby 
\begin{equation}
L \to \infty, \qquad K_0 \to \infty, \qquad K_1, K_3 \to \infty.
\end{equation}
In this limit, the system (\ref{betherel1})-(\ref{betherel3}) is replaced by a set of integral equations. 
Taking the logarithm of (\ref{betherel1})-(\ref{betherel3}), and then applying the thermodynamic limit, amounts to introducing the density $\rho_0(\theta) = \frac{\Delta n}{\Delta \theta}$  of allowed frame particle states per unit rapidity, and analogously  $\rho_{\pm1}$, $\rho_{\pm3}$ for level-1 magnons, corresponding respectively to pairs of solutions $\beta_{\pm n,i}=z_i\pm\frac{i\pi}{2}\,,n=1,3$. Actually, $\rho_0$ and $\rho_{\pm1}$, $\rho_{\pm3}$ include the densities of both particles and holes
\begin{eqnarray}
&& \rho_0(\theta)=\rho_0^r(\theta)+\rho_0^h(\theta)=\frac{\Delta n}{\Delta\theta}\,,\label{rrhodef1} \\
&& \rho_{\pm1}(z)=\rho_{\pm1}^r(z)+\rho_{\pm1}^h(z)=\frac{\Delta m_{\pm1}}{\Delta z}\,, \\
&& \rho_{\pm3}(z)=\rho_{\pm3}^r(z)+\rho_{\pm3}^h(z)=\frac{\Delta m_{\pm3}}{\Delta z}\label{rrhodef3}\,,
\end{eqnarray}
and satisfy the following integral equations (see Appendix \ref{app:TBA} for a derivation)
\begin{eqnarray}
\label{rho1}
&&\rho_0 = \frac{e^\theta}{2\pi} + 2\int_{-\infty}^\infty d\theta' \, \phi_0 (\theta - \theta') \, \rho_0^r (\theta') +\sum_{\pm, n=1,3}\int_{-\infty}^\infty d z \, \phi_{\pm} (\theta - z) \, \rho_{\pm n}^r (z), \\
&&\rho_{\pm n}^r(z)+\rho_{\pm n}^h(z) =\mp \int_{-\infty}^\infty d\theta \phi_{\pm}(z-\theta)\rho_{0}^r(\theta)\,;\ n=1,3\label{rho2}\,,
\end{eqnarray}
where the kernels are given by
\begin{eqnarray}
\label{inte}
\phi_0 (\theta) \equiv \frac{1}{2\pi i} \frac{d}{d \theta}\log S(\theta)= \frac{\theta}{2\pi^2 \sinh \theta}, \qquad \phi_{\pm} (\theta) \equiv \frac{1}{2\pi i} \frac{d }{d \theta} \log \tanh \frac{\theta\pm i\frac{\pi}{2}}{2}= \mp\frac{1}{2\pi \cosh \theta}. 
\end{eqnarray}
Due to (\ref{BAE2}), the level-1 magnons are effectively free (apart from their interaction with the frame particles), then they cannot form bound states and there are only densities of fundamental particles $\rho_{\pm n}$ appearing in (\ref{inte}), and not infinite towers of bound states densities as in \cite{ZamoMasslessTBA}, for instance. 
If we define a unique kernel $\phi\equiv \phi_-=\frac{1}{2\pi\cosh(\theta)}$ for the interactions with and among auxiliary densities $\rho_{\pm n}$, then the densities equations assume a simpler form:
\begin{eqnarray}
 &&\rho^r_0(\theta)+\rho^h_0(\theta) = \frac{e^{\theta}}{2\pi} + 2\phi_0*\rho_0^r + \sum_{n=1,3}\phi*(\rho_{-n}^r-\rho_{+n}^r)\label{rho0simple}\\
 &&\rho_{\pm n}^r(\beta)+\rho_{\pm n}^h(\beta) =\phi*\rho_{0}^r\,;\ n=1,3\,,
\label{rho2simple}
\end{eqnarray}
where we introduced the symbol $*$ to denote the standard convolution.
Now we can use the property $\phi_0 = \phi*\phi$ and (\ref{rho2simple}) to simplify further the equation for $\rho_0$ in the following way: \footnote{This is one of the main differences with the case studied in \cite{Zamolodchikov:1991vh}, where one would have $\rho_0 = \frac{e^{\theta}}{2\pi} + \sum_{n=1,3}\phi*(\rho_{-n}^r)$ instead.}
\begin{equation}
 \rho^r_0(\theta)+\rho^h_0(\theta) = \frac{e^{\theta}}{2\pi} + \sum_{n=1,3}\phi*(\rho_{-n}^r+\rho_{+n}^h)\,, 
 \label{rho1simple}
\end{equation}
where we basically managed to get rid of the self-interacting convolution of $\rho_0$.

The procedure continues by minimising the free energy, which, as advertised at the beginning of the section, returns the ground state energy of the original model before the double Wick rotation. The free energy $F$ in the thermodynamic limit gets two contributions: \footnote{We denote the energy of the Wick-rotated theory in the thermodynamic limit as $\widetilde E$, to avoid confusion with the energy $E$ of the physical theory.}
\begin{equation}
Z = \mbox{tr} \, e^{- R \widetilde E} \to \int {\cal{N}} e^{- R \widetilde E} = \int e^{- R \widetilde E + \log {\cal{N}}}\,,
\end{equation}
hence 
\begin{equation}
\label{free}
-F = - R \, \widetilde E + \log {\cal{N}}.
\end{equation} 
The measure term ${\cal{N}}$ gives an {\it entropy factor}, accounting for the combinatorics of all the possible ways the allowed states $\Delta n = \rho_0 (\theta) \Delta \theta$ -- respectively, $\Delta m_{\pm n} = \rho_{\pm n} (z) \Delta z$ -- are filled by the available frame particles $\Delta \ell_0 = \rho_0^r(\theta) \Delta \theta$ -- respectively, level-1 magnons with densities $\Delta \ell_i = \rho_i^r(\theta) \Delta \theta$ -- namely
\begin{equation}
{\Delta n_i \choose \Delta \ell_i} = \frac{\big(\Delta n_i\big)!}{\Delta \ell_i! \, \big(\Delta n_i-\Delta \ell_i\big)!}
\end{equation}
for each species $i=0,\pm n$. By applying Stirling's approximation of the factorial due to the large occupation numbers, one gets
\begin{equation}
\log \mathcal{N} = \mathcal{S} \sim \sum_{i} \int \big[ \rho_i \log \rho_i - \rho_i^r \log \rho_i^r  - ( \rho_i^h) \log (\rho_i^h)\big].  
\end{equation}
However, the energy turns out to receive contributions only from the frame particles, and not from the level-1 magnons, which are only contributing to the entropy:
\begin{equation}
\widetilde E = M\int d\theta \, e^\theta \, \rho_0^r(\theta).
\end{equation}
Minimising the free energy (\ref{free}) in the thermodynamic limit, subject to the constraints (\ref{rho1simple}) and (\ref{rho2simple}), gives a system of 10 variations, ${\it i.e.}$ with respect to $\rho_0^r$, $\rho_0^h$, $\rho_{\pm n}^r$ and $\rho_{\pm n}^h$, with $n=1,3$. The resulting TBA equations read (see Appendix \ref{app:TBA} for a derivation)
\begin{equation}
 \varepsilon_0=\nu_0(\theta)-\sum_{n=1,3}\phi*(L_{+n}+L_{-n})\,;\quad \varepsilon_{\pm n}=-\phi*L_0,\ n=1,3\,,\label{tba}
\end{equation}
where we have defined
\beq
\nu_0 (\theta) \equiv M R \, e^\theta\,, \qquad
\varepsilon_A\equiv\log\frac{\rho_A^h}{\rho_A^r}\,,\qquad
L_A\equiv\log(1+e^{-\varepsilon_A})\,,
\eeq
 with the multi-index $A=(0,\pm n)$. In a case with generic chemical potentials $\gamma_A$, the definition of $L_A$ would simply generalise to $L_A^{\gamma}\equiv\log(1+e^{i\gamma_A-\varepsilon_A})$.

In terms of the solutions of (\ref{tba}), the exact ground-state energy for left-movers is given by
\begin{equation}
 E_{0,\mathrm{left}}(R)=-\frac{M}{2\pi}\int d\theta e^{\theta} \log(1+e^{-\varepsilon_0(\theta)})\,, 
 \label{energy0}
\end{equation}
while the total ground-state energy reads
\begin{equation}
E_{0}(R) = E_{0,\mathrm{left}}(R)+ E_{0,\mathrm{right}}(R)=
-\frac{M}{\pi}\int d\theta e^{\theta} \log(1+e^{-\varepsilon_0(\theta)})\,.
 \label{energy}
\end{equation}
Actually, taking into account that our theory contains two massless momentum-carrying roots, the total ground state energy is $2E_0(R)$, with $E_0(R)$ given by (\ref{energy}). 

\begin{figure}
\centerline{\includegraphics[width=5cm]{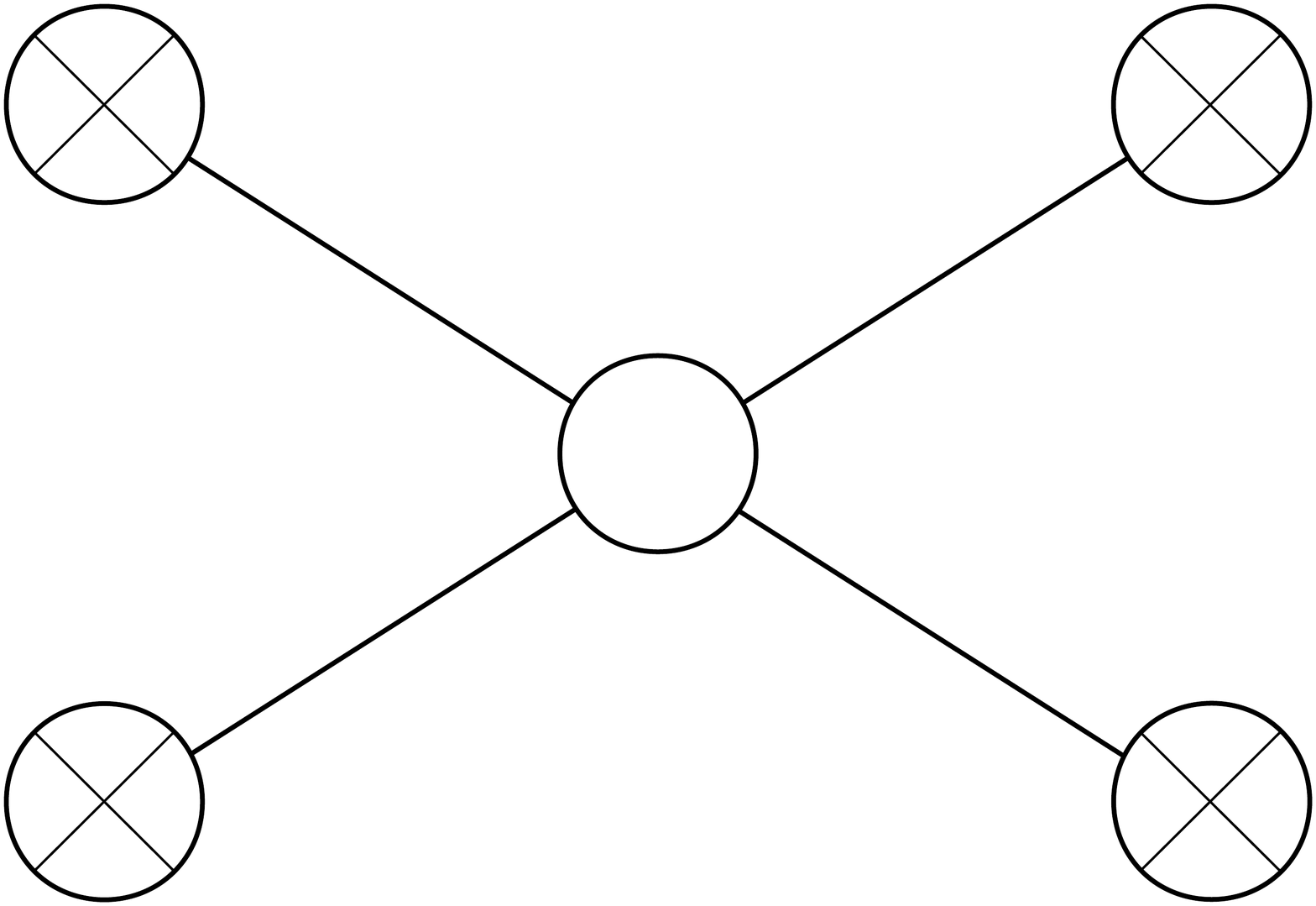}}
\caption{The diagram associated to the TBA equations (\ref{tba}): the central node correspond to $\varepsilon_0$, the crossed nodes correspond to the fermionic pseudoenergies $\varepsilon_{\pm n}$ with $n=1,3$ and the lines represent the equations linking the pseudoenergies via the kernel $\phi$.}
\label{fig:tbadiagram}
\end{figure}

\subsection{Central charge from the TBA}
\label{sec:central}

Equations (\ref{tba}) can be represented by the $\hat{D}_4$-type diagram in Figure \ref{fig:tbadiagram}, associating the pseudoenergies to the nodes and the equations to the lines of the diagram.

The same TBA-diagram describes the ground state TBA equations for the UV limit of $\mathcal{N}=2$ super-sine-Gordon (ssG)~\cite{Kobayashi:1991st,Kobayashi:1991rh,Kobayashi:1991jv} with $\beta^2_{N=2}=16\pi$ ($\beta^2_{N=0}=\beta_*^2=16\pi/3$) \cite{Fendley:1992dm,Fendley:1993pi}. Therefore, we expect that our TBA will give as a result the same central charge $c=3$, at least in the case with trivial chemical potentials. 

In order to calculate the central charge from the TBA equations (\ref{tba}), we use the well known "dilogarithm trick" (see for example \cite{Fendley:1993jh} for an explanation), for which it is necessary to fix the values of the pseudoenergies at $\theta=\pm\infty$.
Obviously, $\varepsilon_0$ is constant for fixed values of $\theta$, then the equations for $\varepsilon_{\pm n}$ reduce to
\begin{equation}
 \varepsilon_{\pm n}=-\frac{1}{2}\log(1+e^{-\varepsilon_0})\label{constban}\,.
\end{equation}
At $\theta=+\infty$, in particular, the driving term $\nu_0(\theta)$ in the first of (\ref{tba}) diverges and then we have $\varepsilon_0(\infty)=\infty$, and $\varepsilon_{\pm n}(\infty)=0$ from (\ref{constban}).
At $\theta=-\infty$, instead, the driving term of the central node equation vanishes, leaving the following equation for $\varepsilon_0(-\infty)$ at constant $\varepsilon_{\pm n}(-\infty)$:
\begin{equation}
 \varepsilon_0(-\infty)=-2\log(1+e^{-\varepsilon_{\pm n}(-\infty)})\,.\label{constba0}
\end{equation}
A real solution of (\ref{constban}) and (\ref{constba0}) is $\varepsilon_{0,min}\equiv\varepsilon_0(-\infty)\sim 2\,\varepsilon_{\pm n,min}=-\infty$.
Now, taking the derivative of the first of (\ref{tba}), we can replace $e^\theta$ in (\ref{energy}) by
\begin{equation}
 e^{\theta}=\frac{\varepsilon'_0+4\phi*L'_1}{MR}\,,
\end{equation}
where, for simplicity, we called $L_1\equiv L_{\pm n}$, since the TBA equations (\ref{tba}) tell us that all the $\varepsilon_{\pm n}$ are equal, and we denote them by $\varepsilon_1\equiv\varepsilon_{\pm n}$.
In this way we get
\begin{equation}
 E_0(R)=-\frac{1}{\pi R}\int d\theta (\varepsilon'_0L_0 - 4\varepsilon_1*L'_1)\,,
\end{equation}
where we replaced $\phi*L_0$ by $-\varepsilon_1$, as given by the second of (\ref{tba}).
This integral can be written as
\begin{equation}
 E_0(R)=-\frac{1}{\pi R}\int_{\varepsilon_{0,min}}^{+\infty} d\varepsilon_0L_0 - \frac{4}{\pi R}\int_{\frac{\varepsilon_{0,min}}{2}}^{0} d\varepsilon_1 \frac{e^{-\varepsilon_1}\varepsilon_1}{1+e^{-\varepsilon_1}}\,.
\label{E0}
\end{equation}
The first integral above gives $-\frac{\pi}{6 R}-\frac{\varepsilon_{0,min}^2}{2\pi R}$ as a result, while the second one yields $-\frac{\pi}{3 R}+\frac{\varepsilon_{0,min}^2}{3\pi R}$, so the divergences exactly compensate each other. 
Given the definition of central charge in terms of the partition function $Z(R,L)=\mbox{Tr}e^{-RH_L}$ of a $(1+1)$-dimensional theory on a torus with infinite spatial dimension $L$ and euclidean time $R=1/T$, where $T$ is the temperature, and its relation to the ground state energy      
\beq
c\equiv\lim_{L\rightarrow\infty}\frac{6R}{\pi L}\log\mbox{Tr}e^{-RH_L}=-\frac{6R}{\pi}E_0(R)\,,
\label{c}
\eeq 
we obtain $c=3$, as expected.
Finally, since in our case the total ground state energy is given by $2E_0(R)$, the total central charge is actually doubled to $c=6$. 

In fact, in this way we are calculating the ground state energy of the sector with antiperiodic boundary conditions on the fermions \cite{Fendley:1997ys}, see \cite{Arutyunov:2009ur, Frolov:2009in} for a discussion in the $AdS_5$ case.
The ground state energy of the sector with periodic fermions, instead, is calculated by Witten's index \cite{Witten:1982df}, rather than the usual free energy as in (\ref{freeenergy}).
Witten's index is obtained by adding non-trivial chemical potentials to the auxiliary fermionic pseudoenergies , so that, in our case,  $\varepsilon_{\pm n}\rightarrow \varepsilon_{\pm n} \pm i \pi$.

A consistent solution of the ground-state TBA equations (\ref{tba}) with these chemical potentials is given by the constants $\varepsilon_0=+\infty, \varepsilon_1=0$. These yield $E_0(R)=0$ exactly, for any value of $R$, as expected for the vacuum energy in a supersymmetric theory with periodic fermions. More details about this solution and more general chemical potentials will be discussed in the next section.

\subsection{Twisted theory and excited states}
\label{sec:exc}

In the case of generic chemical potentials $i\gamma_{\pm n}$ added to the fermionic pseudoenergies, our ground state TBA equations for the left-movers (\ref{tba}) become
\begin{equation}
 \varepsilon_0=\nu_0(\theta)-\sum_{n=1,3}\phi*(\log(1+e^{i\gamma_{+n}} e^{-\varepsilon_+n})+\log(1+e^{i\gamma_{-n}} e^{-\varepsilon_-n}))\,;\quad \varepsilon_{\pm n}=-\phi*L_0,\ n=1,3\,,
\label{twistTBA}
\end{equation}
where $\gamma_{\pm n}$ are real constants. In what follows these will be called twists, for shortness sake, and we shall consider the left-movers' sector only.

The main motivation for us to consider such twisted version of the ground state TBA equations (\ref{tba}) is to calculate the energies of the excited states.
Equations (\ref{twistTBA}), indeed, turn out to be very similar to those studied in \cite{Fendley:1997ys} to determine the excited states' energies of the sine-Gordon model at its ${\cal N}=2$ supersymmetric point, {\it i.e.} at $\beta = \beta_*$ given in equation~(\ref{beta*}), in the massless limit: the only difference is that we have two more equations for the additional auxiliary fermionic variables of type 3.
Thanks to this similarity, in this section  we shall follow the analysis performed in section 4 of \cite{Fendley:1997ys} (see also \cite{Dunning}) for the massless limit.

We shall then consider the case \footnote{This case corresponds also to the UV limit of a twisted version of the $\mathcal{N}=2$ super-sine-Gordon with $\beta=\beta_*$, mentioned in sections \ref{sec:Comparison2} and \ref{sec:central}, with twists $\alpha_F=(k+2)\alpha_T=\gamma$ and $k=0$ in the notations of \cite{Fendley:1993pi}. This means also that more general twists are possible, and then other sectors of excited states may remain to be explored.} $\gamma_{+1}=\gamma_{+3}=-\gamma_{-1}=-\gamma_{-3}=\gamma$ and allow the Y-functions $Y_0\equiv e^{-\varepsilon_0}$ and $Y_{\pm n}\equiv e^{-\varepsilon_{\pm n}}$ to develop zeros as $\gamma$ increases.
It is then useful to derive from the TBA equations (\ref{twistTBA}) a set of functional relations connecting the Y-functions, the so-called Y-system, valid also when $Y_0$ and $Y_1$ have zeros:
\beqa
&&Y_0\left(\theta+\frac{i\pi}{2}\right)Y_0\left(\theta-\frac{i\pi}{2}\right)=\prod_{n=1,3}(1+e^{i\gamma_{+n}}Y_{+n}(\theta))(1+e^{i\gamma_{-n}}Y_{-n}(\theta))\,,\\
&&Y_{\pm n}\left(\theta+\frac{i\pi}{2}\right)Y_{\pm n}\left(\theta-\frac{i\pi}{2}\right)=1+Y_0(\theta)\,.
\label{Ysystem}
\eeqa
At the level of TBA equations, instead, if $Y_0$ and $Y_{\pm n}$ have zeros in the strip $|\mbox{Im}(\theta)|\le\pi/2$, then they satisfy a modified set of twisted TBA equations, given by 
\beqa
\hspace{-0.7cm}&&\log Y_0(\theta)=-\nu_0(\theta)+2\sum_{j=1}^J\log\tanh\left(\frac{\theta-x_j}{2}\right)-2\left(\phi*\log[(1+e^{i\gamma}Y_1)(1+e^{-i\gamma}Y_1)]\right)(\theta)\,,\label{ex1}\\
\hspace{-0.7cm}&&\log Y_1(\theta)= \sum_{k=1}^K \log\tanh\left(\frac{\theta-y_k}{2}\right)+(\phi * L_0)(\theta) \,,\label{ex2}
\eeqa
where, as in the previous section, since all the $Y_{\pm n}(\theta)$ are equal, we denote them all as $Y_1(\theta)$, and the positions of the zeros $\{x_j\}_{j=1}^J$ and $\{y_k\}_{k=1}^K$ are fixed by
\beq
Y_0\left(x_j+\frac{i\pi}{2}\right)=-e^{i\gamma}\ \mbox{or}\ -e^{-i\gamma}\,;\quad  Y_1\left(y_k+\frac{i\pi}{2}\right)=-1\,.
\eeq
These conditions follow from the Y-system (\ref{Ysystem}) evaluated at the locations of the zeros, and, using (\ref{ex1})-(\ref{ex2}), they can be written as integral equations
\beqa
\hspace{-0.7cm}&&e^{y_k}=-(2N_k+1)\pi-2i\sum_{j=1}^J\log\tanh\left(\frac{y_k-x_j}{2}+\frac{i\pi}{4}\right)-\int\frac{d\theta}{\pi}\frac{\ln\left[(1+e^{i\gamma}Y_1(\theta))(1+e^{-i\gamma}Y_1(\theta))\right]}{\sinh(y_k-\theta)}\,,~~~~~~\label{ex3}\\
\hspace{-0.7cm}&&\gamma-(2M_j+1)\pi=i\sum_{k=1}^K\log\tanh\left(\frac{x_j-y_k}{2}+\frac{i\pi}{4}\right)+\int\frac{d\theta}{2\pi}\frac{\ln\left(1+Y_0(\theta)\right)}{\sinh(x_k-\theta)}\,.
\label{ex4}
\eeqa
We shall adopt the same conjecture of \cite{Fendley:1997ys} about all the zeros coming from $\theta=-\infty$.
Therefore, in order to understand at which values of $\gamma$ they come into play, it is essential
 to solve the TBA system (\ref{ex1})-(\ref{ex2}) at $\theta=-\infty$:
\beq
Y_0^2(-\infty)=\left[1+2\cos(\gamma)Y_1(-\infty)+Y_1^2(-\infty)\right]^2\,;\quad Y_1^2(-\infty)=1+Y_0(-\infty)\,,
\label{minima}
\eeq
with $Y_0(-\infty)=e^{-\varepsilon_{0,min}}$ and $Y_1(-\infty)=e^{-\varepsilon_{1,min}}$.

\begin{figure}
\begin{minipage}{0.45 \linewidth}
\centering
\includegraphics[scale=0.5]{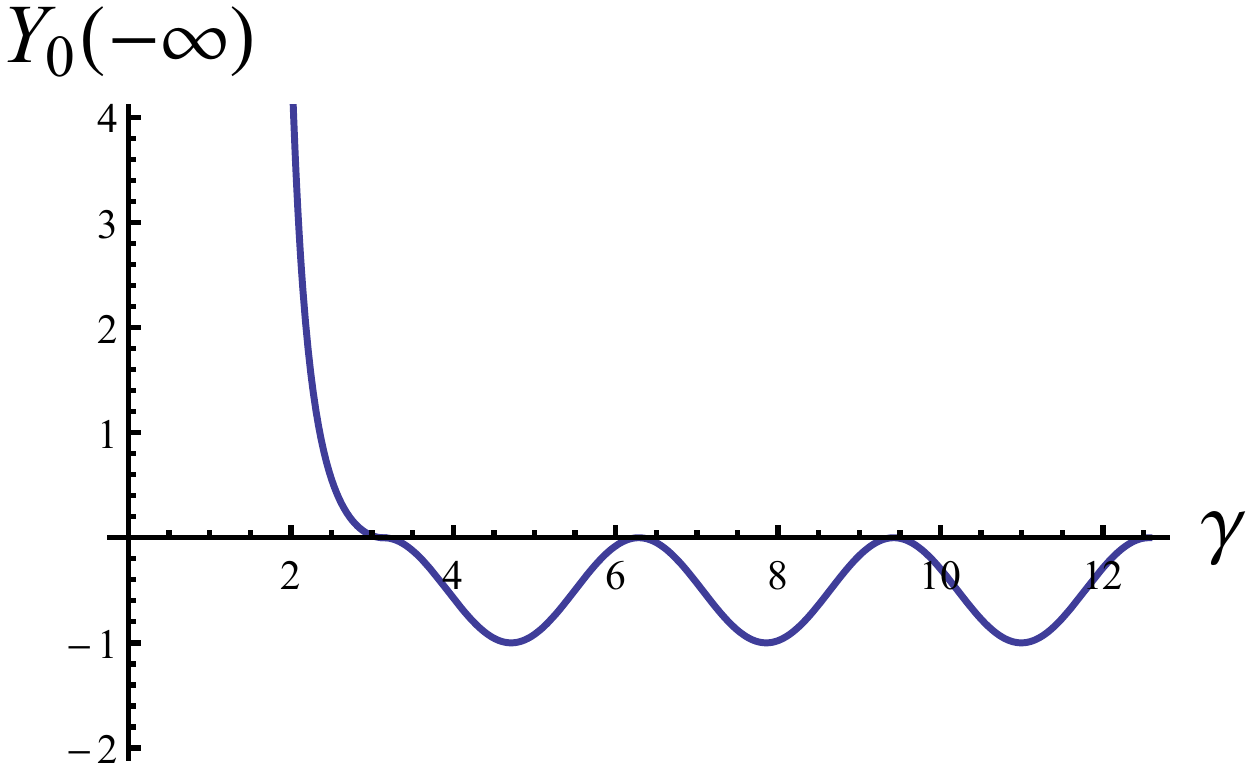}
\end{minipage}
\hspace{0.2cm}
\begin{minipage}{0.45 \linewidth}
\centering
\includegraphics[scale=0.5]{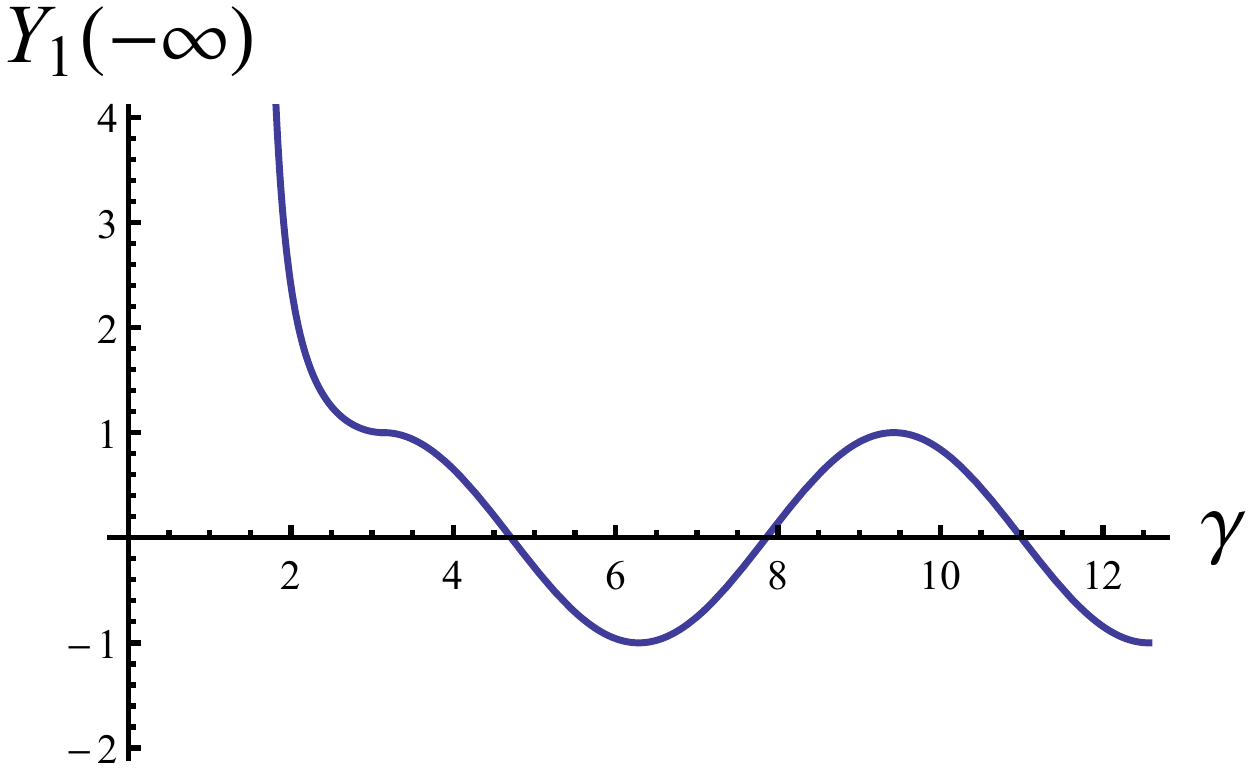}
\end{minipage}\\
\centering
\includegraphics[scale=0.5]{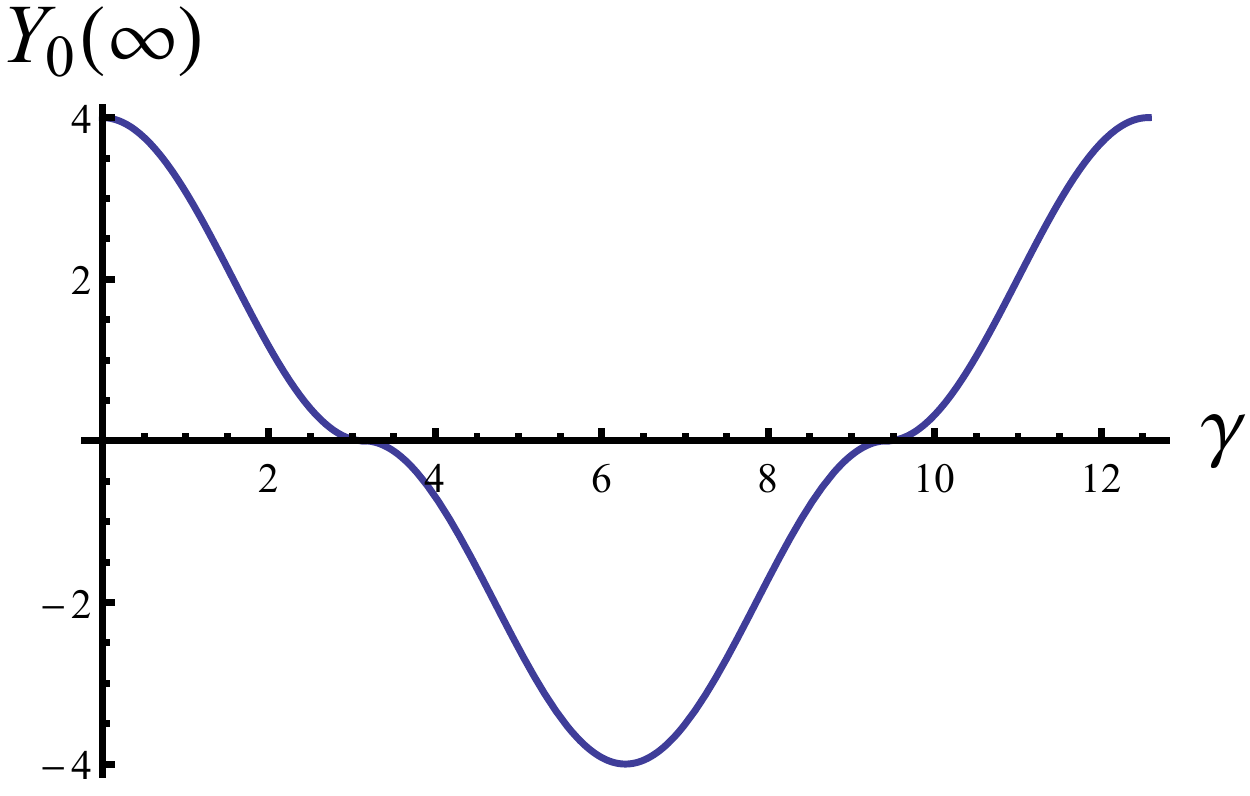}
\caption{Finite solutions chosen for $Y_0(-\infty)$ and $Y_1(-\infty)$ from equations (\ref{minima}), and the solution chosen for $Y_0(\infty)$ from (\ref{maxima}), in the interval $\gamma\in(0,4\pi)$. 
}
\label{fig:minmax}
\end{figure} 

In the opposite limit ($\theta\rightarrow\infty$), we get
\beq
Y_0(\infty)=\pm 4\cos^2\left(\frac{\gamma}{2}\right)e^{-e^\theta}\,;\ Y_1(\infty)=\pm 1\,.
\label{maxima}
\eeq
This tells us that the positions of $Y_0(\theta)$'s zeros can reach infinity at $\gamma=(2k+1)\pi$, $k=0,1,\dots$, while those of $Y_1(\theta)$ cannot go to infinity for any value of $\gamma$.

For $\gamma\in (0,\pi/2)$, equation (\ref{minima}) has no finite solutions and $\varepsilon_{0,min}\sim 2\,\varepsilon_{1,min}=-\infty$ as for the untwisted case studied in section \ref{sec:central}.
So, no zeros can come from $\theta=-\infty$ and the energy is given by
\beq
E_{\mathrm{left}}(R,\gamma)=-\frac{1}{2\pi R}\int_{\varepsilon_{0,min}}^{+\infty} d\varepsilon_0\,L_0 -
\frac{1}{\pi R}\int_{\varepsilon_{1,min}}^{0} d\varepsilon_1\left( \frac{e^{i\gamma-\varepsilon_1}\varepsilon_1}{1+e^{i\gamma-\varepsilon_1}}+\frac{e^{-i\gamma-\varepsilon_1}\varepsilon_1}{1+e^{-i\gamma-\varepsilon_1}}\right)\,.
\label{E}
\eeq
Using the singular solutions found for $\varepsilon_{0,min}$ and $\varepsilon_{1,min}$, then (\ref{E}) gives as a result
\beq
E_{\mathrm{left}}(R,\gamma)= -\frac{\pi}{4R} + \frac{\gamma^2}{4\pi R} = \frac{2\pi}{R}\left(\frac{\gamma^2}{4\pi^2}-\frac{1}{8}\right)\,.
\label{E0pi/2}
\eeq
For $\gamma>\pi/2$, instead, (\ref{minima}) have a pair of finite solutions:
\beqa
&&1)\quad(\varepsilon_{0,min},\varepsilon_{1,min})=(-2\log(\tan\gamma),-\log(-\sec\gamma))\,;
\label{minsol1}\\
&&2)\quad(\varepsilon_{0,min},\varepsilon_{1,min})=(-\log(-\sin^2\gamma),-\log(-\cos\gamma))\,.
\label{minsol2}
\eeqa
In particular, for $\gamma\in (\pi/2,\pi)$ we choose the first solution (\ref{minsol1}), since it is connected at $\gamma=\pi/2$ with the singular one found above for $\gamma\in (0,\pi/2)$. 
Plugging it into formula (\ref{E}), one gets
\beq
E_{\mathrm{left}}(R,\gamma)= -\frac{(\gamma-\pi)^2}{2\pi R} = -\frac{\pi}{2R}\left(\frac{\gamma}{\pi}-1\right)^2\,,
\label{E0pi}
\eeq
so that at $\gamma=\pi$ we have the expected Witten's index $E(R,\pi)=0$.

Equations (\ref{minima}) imply that the zeros of $Y_0(\theta)$ enter from $-\infty$ at $\gamma=(2k+1)\pi$, $k=0,1,\dots$, while zeros of $Y_1(\theta)$ cannot enter from $-\infty$ for $\gamma<\pi$. 

\begin{figure}
\centering
\includegraphics[scale=0.3]{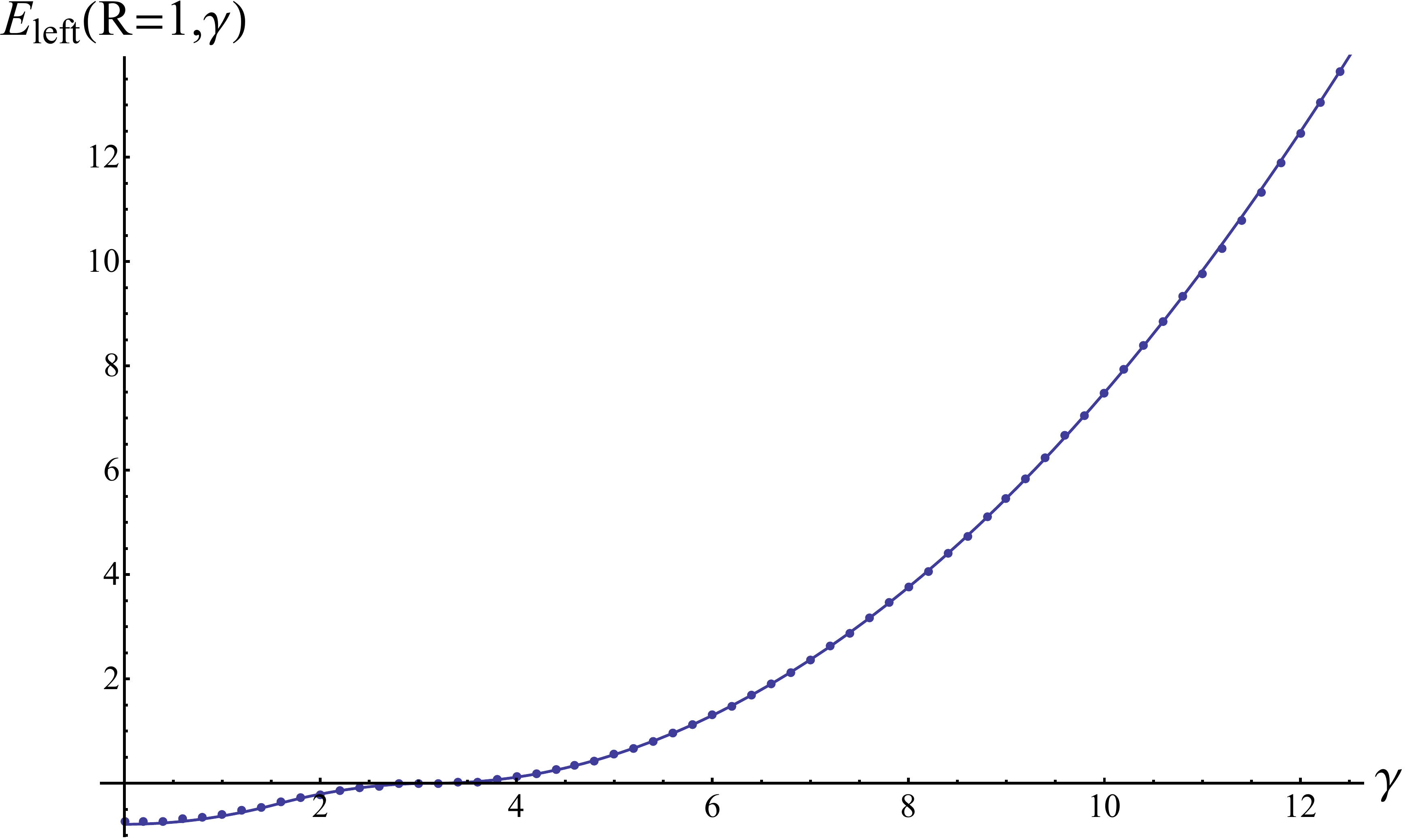}
\caption{Comparison of our numerical results with analytic formulas (\ref{E0pi/2}), (\ref{E0pi}) and (\ref{par}), for $R=1$. 
}
\label{fig:Er}
\end{figure} 

At $\gamma=\pi$, $Y_0(-\infty)$ is zero, but actually $Y_0(\theta)$ is zero at any $\theta$.
As in \cite{Fendley:1997ys}, this means that the first zero $x_1$ of $Y_0(\theta)$ at $\gamma=\pi$ enters at $-\infty$ and goes straight to $+\infty$, ensuring that $Y_0(\theta)=0$ for any $\theta$, and its effect is just to change the sign of  $Y_0(\theta)$, or equivalently to add a $-i\pi$ in the r.h.s. of the TBA equation for $\varepsilon_0$:
\begin{equation}
\varepsilon_0=\nu_0(\theta)-i\pi-2\phi*\ln\left[(1+e^{i\gamma}Y_1(\theta))(1+e^{-i\gamma}Y_1(\theta))\right]\,.
\label{eps0ex}
\end{equation}
Also the second equation of (\ref{minima}) has to be modified accordingly, by changing sign of $Y_0(-\infty)$ in the l.h.s.. This implies that we have to choose the solution (\ref{minsol2}) for the lower limits $(\varepsilon_{0,min},\varepsilon_{1,min})$, so that (\ref{E}) gives
\beq
E_{\mathrm{left}}(R,\gamma)=\frac{(\gamma-\pi)^2}{2\pi R}\ ;\ \ \mbox{for}\ \gamma\in (\pi,3\pi/2)\,.
\label{E13/2}
\eeq
At $\gamma=3\pi/2$, the first zero $y_1$ of $Y_1(\theta)$, corresponding to the solution of $Y_0(y_1+ i\pi/2)=-1$, enters from $\theta=-\infty$, then the TBA equations for $\varepsilon_{\pm n}$ in (\ref{twistTBA}) have to be modified as 
in (\ref{ex2}) with $K=1$, 
and the (left-movers) energy formula changes as
\beq
RE_{\mathrm{left}}(R,\gamma)=\sum_{k=1}^K e^{y_k}-\frac{1}{2\pi}\int d\theta\, e^{\theta}\, L_0(\theta)\,,
\label{Emod}
\eeq
where the positions of the zeros $y_k$ are determined by solving (\ref{ex3}).
In particular, since $y_1$ enter from $-\infty$, equation (\ref{ex3}) for $K=1$ simplifies to $2N_1+1=1$, then $N_1=0$. 
In general, $x_j$ and $y_k$ will enter from $\theta=-\infty$ at $\gamma=j\pi$ and $\gamma=k\pi/2$ respectively, and at these values equations (\ref{ex3})-(\ref{ex4}) simplify respectively to 
\beq
(2N_k+1)=J\,;\quad \gamma-(2M_j+1)\pi=-\pi K\,,
\eeq
that are solved by $N_k=k-1$ and $M_j=j-1$.
We verified this structure of the zeros and computed the energy (\ref{Emod}) also numerically for $R=1$ (see Appendix \ref{app:num} and Figure \ref{fig:Er}), by solving iteratively (\ref{ex1})-(\ref{ex4}). However, it is possible to derive analytically a relatively closed formula for the energy at any value of $\gamma$ (see Appendix \ref{app:excited} for the derivation): the result reads
\beqa
&&\hspace{-0.5cm}RE_{\mathrm{left}}(R,\gamma)=-\frac{1}{2\pi}\int_{\varepsilon_{0,min}}^{\infty} d \varepsilon_0\,L_0 - \frac{1}{\pi R}\int_{\varepsilon_{1,min}}^{0} d\varepsilon_1\left( \frac{e^{i\gamma-\varepsilon_1}\varepsilon_1}{1+e^{i\gamma-\varepsilon_1}}+\frac{e^{-i\gamma-\varepsilon_1}\varepsilon_1}{1+e^{-i\gamma-\varepsilon_1}}\right)\nonumber\\
&&\hspace{-0.5cm}-\frac{1}{2\pi}\log((-1)^K)\log(Y_0^2(-\infty))+2(J-1)K\pi+2\sum_{j=J_{\infty}+1}^J(\gamma-(2M_j+1)\pi)-\sum_{k=1}^K2N_k\pi\,,
\label{Efin}
\eeqa
where $J_{\infty}$ is the number of zeros $x_j$ at $+\infty$, the lower limits of integration $\varepsilon_{0,min}=2\,\varepsilon_{1,min}=\infty$ for $\gamma\in(0,\pi/2)$, while they have to be chosen as in (\ref{minsol1}) for $\gamma\in(\pi/2,\pi)$ and as in (\ref{minsol2}) for $\gamma\ge\pi$ (see Figure \ref{fig:minmax}).
These choices for the lower limits have been verified also numerically.
Moreover, the two integrals in (\ref{Efin}) can be easily evaluated in terms of dilogarithms:
\beqa
&&\hspace{-0.5cm}RE_{\mathrm{left}}(R,\gamma)=\frac{1}{2\pi}\left\{\mbox{Li}_2[-Y_0(-\infty)]+2\mbox{Li}_2[-e^{i\gamma}Y_1(-\infty)]+2\mbox{Li}_2[-e^{-i\gamma}Y_1(-\infty)
\right\}\nonumber\\
&&\hspace{-0.5cm}-\frac{1}{2\pi}\left\{2\mbox{Li}_2(-e^{i\gamma})+2\mbox{Li}_2(-e^{-i\gamma})-\log[Y_0^2(-\infty)]\log[(-1)^KY_1(-\infty)]\right\}\nonumber\\
&&\hspace{-0.5cm}+2(J-1)K\pi+2\sum_{j=J_{\infty}+1}^J(\gamma-(2M_j+1)\pi)-\sum_{k=1}^K2N_k\pi\,.
\label{Efin1}
\eeqa
Now, let us explain how formula (\ref{Efin1}) matches the results (\ref{E0pi/2}) for $\gamma\in(0,\pi/2)$, (\ref{E0pi}) for $\gamma\in(\pi/2,\pi)$, and gives 
\beq
E_{\mathrm{left}}(R,\gamma)=\frac{(\gamma-\pi)^2}{2\pi R}
\label{par}
\eeq
for $\gamma\in(\pi, 4\pi)$, by taking into account the behaviour of the Y-functions' zeros, that has been also verified numerically (see Appendix \ref{app:num}).

\begin{itemize}
\item In fact, for $\gamma<3\pi/2$ there are no zeros, except for the first atypical zero $x_1$ at $+\infty$, then $J=J_{\infty}=1$, $K=0$ and the energy is given by the first two lines of (\ref{Efin1}), using $Y_0(-\infty)=Y_1^2(-\infty)=\infty$ for $\gamma\in(0,\pi/2)$ and $Y_0(-\infty)=\tan^2(\gamma)$, $Y_1(-\infty)=-\sec(\gamma)$ for $\gamma\in(\pi/2,\pi)$.

\item For $\gamma\in(3\pi/2,2\pi)$, instead, $K=1$, but the discrete contributions still vanish since $J=J_{\infty}=1$ and $N_1=0$, while we need to take into account the last term of the second line in (\ref{Efin1}) with $K=1$.

\item For $\gamma\in(2\pi,5\pi/2)$, we have the appearance of the first finite zero $x_2$, with $M_2=1$, then $J=J_{\infty}+1=2$ and we have to add a total contribution $2\pi+2\gamma-6\pi$. 
 
\item For $\gamma\in(5\pi/2,3\pi)$, a second zero $y_2$ with $N_2=1$ enters, then the contribution from the last term of the second line vanishes, while we have to add $4\pi+2\gamma-6\pi-2\pi$.  
 
\item For $\gamma\in(3\pi,7\pi/2)$, the zero $x_2$ has gone to $\infty$, 
but a third zero $x_3$ appears with $M_3=2$, then $J=J_{\infty}+1=3$. Therefore, we have to add $8\pi+2\gamma-10\pi-2\pi$. 
 
\item Finally,  for $\gamma\in(7\pi/2,4\pi)$, a third zero $y_3$ enters, then we have to take again into account the last term of the second line in (\ref{Efin1}), while the discrete terms give $12\pi+2\gamma-10\pi-6\pi$.
\end{itemize}
Let us notice that, from $\gamma=2\pi$ to $\gamma=4\pi$, to get the result (\ref{par}) we have to add simply  $2\gamma-4\pi$ to the first two lines of (\ref{Efin1}).

These results have been tested, for $R=1$, by the numerical analysis discussed in Appendix \ref{app:num}, see Figure \ref{fig:Er}, and let us guess that formula (\ref{par}) is valid for any value of $\gamma\ge \pi$.
This let us also conjecture that the energies of some excited states belonging to the sector with periodic (antiperiodic) fermions can be calculated by $E(\gamma,R)=4\,E_{\mathrm{left}}(\gamma,R)=2\,\frac{(\pi-\gamma)^2}{\pi R}$ at odd (even) integer values of $\gamma/\pi\ge 2$:
\beq
E_n(R)=E(n\pi,R)=\frac{2\pi}{R}(1-n)^2\,,\qquad n=2,3,\dots\,.
\eeq

\section{\label{sec:Conclusions}Conclusions}

In this paper we have investigated $\AdS_3/\CFT_2$ states whose energies are closest to the BMN vacuum. On a decompactified world-sheet these correspond to the gapless (massless) excitations that distinguish $\AdS_3/\CFT_2$ from higher-dimensional holographic duals. At low energies, they behave as massless relativistic left- or right-movers on the world-sheet. Remarkably, in this limit the exact worldsheet S matrix remains non-trivial. More precisely, while massless left/right, massive and mixed-mass scattering does trivialise, the scattering of massless excitations of the same worldsheet chirality is described by a non-trivial integrable relativistic S matrix. 

This S matrix is essentially {\em non-perturbative} in its form: after all, relativistic excitations moving at the speed of light in the same direction cannot scatter with one another! Instead, as first proposed by Zamolodchikov~\cite{ZamoMasslessTBA}, the S matrix should be thought of as an auxiliary algebraic tool which can be used to determine the spectrum of the gapless excitations using Bethe Ansatz methods. Since the S matrix for left/right massless scattering is trivial in the low-energy limit, we further conclude, following Zamolodchikovs' approach, that the low-energy spectrum should be that of a two-dimensional conformal field theory, which we have denoted as $\CFT_2^{(0)}$.

In order to understand the $\CFT_2^{(0)}$ further, we have analysed how finite-size, or wrapping, corrections enter the spectral problem in the low-energy limit. Since the theory is relativistic in this limit, we were able to construct the TBA and corresponding Y-system both for the ground-state and for the excited states building on some of the original constructions in the integrable literature~\cite{Dunning,Fendley:1997ys}. We find that the central charge of $\CFT_2^{(0)}$ is $c=6$ from calculating the vacuum energy for antiperiodic fermions - with the vacuum energy being zero for periodic fermions in agreement with a supersymmetric theory - and that some excited states energies are given by integer multiples of $2\pi/R$. These findings, together with the target-space supersymmetry of the spectrum in the relativistic limit, point towards $\CFT_2^{(0)}$ being a free $\CFT_2$, perhaps just a space-time supersymmetric $T^4$ theory. We hope to return to a more detailed analysis of the exact identification of $\CFT_2^{(0)}$ in the future.

An outstanding problem in integrable $\AdS_3/\CFT_2$ holography has been the challenge of incorporating finite-size effects, with  perturbative calculations proving difficult due to the presence of gapless/massless excitations~\cite{Abbott:2014rca}. Our analysis shows that to overcome these obstacles one needs to adopt an essentially non-perturbative approach. We were able to do this at low energies and have shown that wrapping effects do not spoil integrability. It would be very interesting to extend our findings beyond the relativistic limit, to a complete TBA and QSC for the theory, as for the higher dimensional $\AdS/\CFT$ spectral problems \cite{Bombardelli:2009ns,Gromov:2009bc, Arutyunov:2009ur, Bombardelli:2009xz,Gromov:2009at,QSC,Gromov:2014caa,Cavaglia:2014exa,Bombardelli:2017vhk}.

Throughout this paper we have focussed on the $\AdS_3\times \Sphere^3\times\Torus^4$ theory supported by RR flux. It would be particularly interesting to generalise our analysis to backgrounds supported by NSNS flux. In the presence of non-zero RR moduli, the exact worldsheet S matrix of this theory is known~\cite{OhlssonSax:2018hgc}. In the low-energy limit the S matrix of massless modes remains non-trivial and we are currently investigating the resulting $\CFT_2^{(0)}$~\cite{BOSSTtoappear}. This should also lead to a better understanding of the pure NSNS theory in the limit of zero RR modulus. Here, the {\em non-perturbative} massless S matrices $S_{LL}$ and $S_{RR}$ remain non-trivial and non-diagonal. A careful analysis of this limit should help to provide an integrable description of the WZW theory, as well as determine the status of a recent proposal  based on an almost trivial  S matrix~\cite{Baggio:2018gct}. Given the non-perturbative nature of our findings, one may additionally hope to shed light on the $k=1$ theory and its relation to the symmetric orbifold $\CFT_2$, as recently investigated in~\cite{Giribet:2018ada,Gaberdiel:2018rqv}. It would also be interesting to identify the role these gapless excitations play in the Higgs branch spin-chain~\cite{Sax:2014mea}. Finally, generalising our construction to the $\AdS_3\times \Sphere^3\times\Sphere^3\times\Sphere^1$ background supported by RR flux should be straightforward and one may also consider extending the analysis to mixed-flux backgrounds~\cite{Hoare:2013pma,Hoare:2013ida,Hoare:2013lja,Lloyd:2014bsa,Borsato:2015mma}.

\section*{Acknowledgements}

We would like to thank Michael Abbott, Ines Aniceto, Niklas Beisert, Patrick Dorey, Clare Dunning, Paul Fendley, Andrea Fontanella, Ben Hoare, Tim Hollowood, Romuald Janik, Florian Loebbert, Jock McOrist, Olof Ohlsson Sax, Peter Orland, Andrea Prinsloo, Francesco Ravanini, Ryo Suzuki, Arkady Tseytlin, Kostya Zarembo for illuminating discussions. D. B. is particularly indebted to Roberto Tateo for many enlightening discussions, hints and comments, and to Riccardo Conti for sharing a {\tt Mathematica} code. We thank the Galileo Galilei Institute for Theoretical Physics (GGI) for the hospitality and INFN for partial support where parts of this work, within the program "New Developments in AdS3/CFT2 Holography" were undertaken. We also thank the anonymous referee for a careful reading of the manuscript and several questions that helped us to clarify and improve our findings and presentation.

D. B. is partially supported by the INFN project SFT. B. S. acknowledges funding support from an STFC Consolidated Grant ``Theoretical Physics at City University" ST/P000797/1. A. T. thanks the STFC for support under the Consolidated Grant project nr. ST/L000490/1. We acknowledge useful conversations with the participants of the Nordita program ``Holography and Dualities 2016: New Advances in String and Gauge Theory".

No data beyond those presented and cited in this work are needed to validate this study.

\appendix

\section{Relativistic Dressing Phase}

In this appendix we collect the computational details used to determine the relativistic limit of the HL dressing factor.

\subsection{Relativistic limit of integral expression for HL phase}
\label{app:rel-lim-int-hl}

The integral representation of the HL phase~\eqref{eq:hl-generic} takes the form of a Riemann-Hilbert integral, schematically given by
\begin{equation}
\label{eq:conv-rh-int}
\phi(w)=\int_C \frac{\varphi(\zeta)}{\zeta-w}\,,
\end{equation}
with $\varphi$ an analytic function. Crossing is closely related to the Sochocki-Plemelj theorem \cite{Sochocki, Plemelj} with the value of $\phi$ jumping as $w$ goes from one side of the contour $C$ to the other. It is helpful to split the integration interval in equation~\eqref{eq:hl-generic} into two 
\begin{equation}
z\in\left[-1,1\right]=\left[-1,e^{-1}\right]\cup\left[e^{-1},1\right]\,.
\end{equation}
In the relativistic limit the momenta $p$ and $q$ are small (compared to $h$) and so crossing can only take place for $z\in[e^{-1},1]$ since we can always increase the value of $h$ to ensure this. The integrals over $z\in[-1,e^{-1}]$ then involve analytic functions only and can be performed by expanding the integrands at large $h$. For  $z\in[-1,0]$ integrals we find
\begin{equation}
\begin{aligned}
&\int\limits_{-1+i\epsilon}^{0+i\epsilon} \frac{dz}{4\pi}G_-(z,y^+)\bigl( g(z,x^+)-g(z,x^-) \bigr)
-\int\limits_{-1-i\epsilon}^{0-i\epsilon}  \frac{dz}{4\pi}G_+(z,y^-) \bigl( g(z,x^+)-g(z,x^-) \bigr) 
\\
&=\int\limits_{-1+i\epsilon}^{0+i\epsilon} \frac{dz}{4\pi}
\left(\frac{2i p_1}{h (z -1)^2}+{\cal O}(\tfrac{1}{h^3})\right) \left(-\frac{i p_2(z+1)}{2 h (z -1)}+\log (i (z -1))-\log \left(i (\tfrac{1}{z}-1 )\right)+{\cal O}(\tfrac{1}{h^3})\right) 
\\
&\,\,\,
-\int\limits_{-1-i\epsilon}^{0-i\epsilon}  \frac{dz}{4\pi}
\left(-\frac{2i p_1}{h (z -1)^2}+{\cal O}\left(\tfrac{1}{h^3}\right)\right) \left(-\frac{i p_2(z+1)}{2 h (z -1)}+\log (i (z -1))-\log \left(i (\tfrac{1}{z}-1 )\right)+{\cal O}\left(\tfrac{1}{h^3}\right)\right) 
\\
&=-\frac{p_1 p_2}{2 h^2}-\frac{p_1 p_2 \left(p_2^2-p_1^2\right)}{192 h^4}+\dots
\,.
\end{aligned}
\end{equation}
Similarly for  $z\in[0,e^{-1}]$ integrals we have
\begin{equation}
\begin{aligned}
&\int\limits_{0+i\epsilon}^{e^{-1}+i\epsilon} \frac{dz}{4\pi}G_-(z,y^+)\bigl( g(z,x^+)-g(z,x^-) \bigr)
-\int\limits_{0-i\epsilon}^{e^{-1}-i\epsilon}  \frac{dz}{4\pi}G_+(z,y^-) \bigl( g(z,x^+)-g(z,x^-) \bigr) 
\\
&=\int\limits_{-1+i\epsilon}^{0+i\epsilon} \frac{dz}{4\pi}
\left(\frac{2i p_1}{h (z -1)^2}+{\cal O}(\tfrac{1}{h^3})\right) \left(-\frac{i p_2(z+1)}{2 h (z -1)}+\log (i (z -1))-\log \left(i (\tfrac{1}{z}-1 )\right)+{\cal O}(\tfrac{1}{h^3})\right) 
\\
&\,\,\,
-\int\limits_{-1-i\epsilon}^{0-i\epsilon}  \frac{dz}{4\pi}
\left(-\frac{2i p_1}{h (z -1)^2}+{\cal O}\left(\tfrac{1}{h^3}\right)\right) \left(-\frac{i p_2(z+1)}{2 h (z -1)}+\log (i (z -1))-\log \left(i (\tfrac{1}{z}-1 )\right)+{\cal O}\left(\tfrac{1}{h^3}\right)\right) 
\\
&=\frac{4 \pi  p_1}{(e-1) h}-\frac{2 e p_1 p_2}{(e-1)^2 h^2}
-\frac{e (1+e) \pi  p_1^3}{6 (e-1)^3 h^3}
+\frac{e p_1 p_2 \left((1+e+e^2) p_1^2+e p_2^2\right)}{12 (e-1)^4 h^4}+\dots
\,.
\end{aligned}
\end{equation}
Since the integrals over $z\in[-1,e^{-1}]$ are trivial under crossing in the large-$h$ limit,  they give rise to a (sub-leading) CDD-factor,
\begin{equation}\label{eq:hl-cdd}
\begin{aligned}
\theta^{CDD} (p_1,p_2) &=
\frac{4 \pi  p_1}{(e-1) h}-\frac{(1+e)^2 p_1 p_2}{2 (e-1)^2 h^2}+{\cal O}\left(\frac{1}{h^{3}}\right)
\,.
\end{aligned}
\end{equation}

On the other hand, the integral over $z\in[e^{-1},1]$ does contribute to crossing in the relativistic limit. We define a new integration variable
\begin{equation}
z=e^{\frac{ir}{2h}}\,.
\end{equation}
Further, for $z\in[e^{-1},1]$ we can take $\epsilon\rightarrow 0$ in the integrals, as this was introduced to regularize the singularity at $z=0$. 
Then expanding the integrand gives
\begin{equation}
\begin{aligned}
&\int\limits_{e^{-1}}^{1} \frac{dz}{4\pi}\left(G_-(z,y^+)-G_+(z,y^-)\right)
\bigl( g(z,x^+)-g(z,x^-) \bigr)
\\
&=\int\limits_{2i h}^{0} 
\frac{dr}{\pi} 
\frac{2 p_1 (\log (p_2+r)-\log (p_2-r))}{p_1^2-r^2}
+\frac{1}{24h^2}\frac{p_1 ((p_1^2-r^2) (\log (p_2-r)-\log (p_2+r))-2 p_2 r)}{p_1^2-r^2}
+{\cal O}\left(\frac{1}{h^4}\right)\,.
\end{aligned}
\end{equation}
The leading large-$h$ term, when combined with the non-integral part of the HL phase to ensure anti-symmetry, then gives equation~\eqref{eq:hl-rel-limit-45}.

\subsection{The dilogarithm form of the HL phase}

We can obtain another useful expression for the relativistic phase in terms of dilogarithms, starting from the expression for the HL phase given  in~\cite{Arutyunov:2006iu}
\begin{equation}
\chi_1(x,y) \equiv \frac{1}{\pi} \bigg[\log \frac{y-1}{y+1} \, \log \frac{x-\frac{1}{y}}{x-y} + \mbox{Li}_2 \frac{\sqrt{y} - \sqrt{\frac{1}{y}}}{\sqrt{y} - \sqrt{x}}
-\mbox{Li}_2 \frac{\sqrt{y} + \sqrt{\frac{1}{y}}}{\sqrt{y} - \sqrt{x}}+\mbox{Li}_2 \frac{\sqrt{y} - \sqrt{\frac{1}{y}}}{\sqrt{y} + \sqrt{x}}-\mbox{Li}_2 \frac{\sqrt{y} + \sqrt{\frac{1}{y}}}{\sqrt{y} + \sqrt{x}}\bigg]\,,\nonumber
\end{equation}
valid for 
\begin{equation}
\label{valid}
|xy|>1, \qquad \mbox{Re}(\sqrt{x}\sqrt{y}) > 1\,.
\end{equation} 
The dressing phase can be then expressed in terms of the function
\begin{eqnarray}
\chi(x,y) = \frac{1}{2} \big[\chi_1(x,y) - \chi_1(y,x)\big]
\end{eqnarray}
as
\begin{equation}
\Phi = e^{2 i \chi}.
\label{Phichi}
\end{equation}
We set 
\begin{eqnarray}
x = e^{i \frac{p_1}{2}}, \qquad y=e^{i \frac{p_2}{2}}, 
\end{eqnarray}
and take the simultaneous relativistic limit
\begin{equation}
p_i \sim \epsilon \, q_i, \qquad h \sim \frac{c}{\epsilon}.
\end{equation}
In doing this, however, we make a specific choice. By relying on the fact that the final answer will have to display difference-form in the variable $\theta$ because of ordinary relativistic invariance, we use the freedom of setting
\begin{equation}
q_i = e^{\theta_i}, \qquad 
\theta_1 = \log 2 -  \frac{i\pi}{2} + i0, \qquad \theta_2 = - \frac{i\pi}{2} + \theta, \nonumber
\end{equation}
where we restrict to 
\begin{eqnarray}
\mbox{Im}( \theta) \in \left(0,\tfrac{\pi}{2}\right)\,.
\end{eqnarray}
This means that we are sending $x$ to $1$ from real values greater than $1$, and $y$ to $1$ from values of the real part greater than $1$, which also means that both $x$ and $y$ are consistently approaching the boundary of the region (\ref{valid}). As they do, the variable $v_1 \equiv u_1 -2$ approaches the branch cut on the positive real axis from above. Specifically,
\begin{eqnarray}
x \sim 1 + \epsilon + i0, \qquad y \sim 1 + \kappa \, \epsilon, \qquad \kappa \equiv \frac{e^\theta}{2}, 
\label{xykappa}
\end{eqnarray}
with 
\begin{eqnarray}
\mbox{Re}(\kappa) >0, \qquad \mbox{Im} (\kappa) >0.
\end{eqnarray}
Relativistic invariance will imply that the dependence we obtain in the sole variable $\theta$ will account for the whole dependence on $\theta_1 - \theta_2$.

The leading order in the $\epsilon$-expansion reads
\begin{eqnarray}
\label{reduce}
&&\chi(x,y) \to \frac{1}{2\pi} \bigg[\log \frac{\kappa \, \epsilon}{2} \, \log \frac{1 + \kappa}{1 - \kappa} + \mbox{Li}_2 \frac{2 \kappa}{\kappa - 1}-\mbox{Li}_2 \frac{4}{\epsilon (\kappa - 1)}+\mbox{Li}_2 \frac{\kappa \, \epsilon}{2}+\mbox{Li}_2 \frac{4}{\epsilon (1 - \kappa)}-\mbox{Li}_2 \frac{\epsilon}{2}\nonumber\\
&&\qquad \qquad \qquad -\log \frac{\epsilon}{2} \, \log \frac{\kappa + 1}{\kappa - 1} - \mbox{Li}_2 \frac{2}{1- \kappa}\bigg].
\end{eqnarray}
We now can use the fact the we are in the region $\mbox{Re} (\kappa) >0 \cup \mbox{Im} (\kappa) >0$, and that 
\begin{eqnarray}
&&\mbox{Li}_2 (z) \sim z, \qquad z \to 0,\nonumber \\
&&\mbox{Li}_2 (z) \sim \frac{\pi^2}{6} + (1-z)\log (1-z), \qquad z \to 1,\nonumber \\
&&\mbox{Li}_2 (z) \sim - \frac{\pi^2}{6} - \frac{1}{2} \log^2 (-z), \qquad |z| \to \infty.\nonumber 
\end{eqnarray} 
Moreover, we set the branch cut of the logarithm on the negative real axis, with argument approaching $+ i \pi$ from above and $- i \pi$ from below. Taking all of this into account, a careful analysis allows to reduce the expression (\ref{reduce}) to
\begin{eqnarray}
\label{formula}
&&\chi_0(\kappa) = \frac{1}{2\pi} \bigg[\log \kappa \, \log \frac{1 + \kappa}{1 - \kappa} + \mbox{Li}_2 \frac{2 \kappa}{\kappa - 1}+ i \pi \log \frac{1}{1 - \kappa} + \frac{\pi^2}{2} + i \pi \log 2 - \mbox{Li}_2 \frac{2}{1- \kappa}\bigg],
\end{eqnarray}
which is manifestly a finite limit. Recalling that 
\begin{equation}
\kappa = \frac{e^\theta}{2},
\end{equation}
and by the arguments laid out earlier, formula (\ref{formula}) is the relativistic limit of the massless phase as a function of $\theta = \log 2 - \theta_1 + \theta_2 = \log 2 - \vartheta$. 

Let us investigate the discontinuities of $\chi_0$ and relate it to the relativistic crossing equation. It is convenient to continue working with the variable $\kappa$, which we now consider approaching the positive real axis from above, where it meets a branch cut of the phase (\ref{formula}). The idea is that we can continue the expression for the phase and see what values it approaches when we reach the branch cut from below. For this, we not only need the branch cut discontinuity of the logarithm, but also that the function $\mbox{Li}_2 (z)$ has a branch cut along $z \in (1,\infty)$ with 
\begin{eqnarray}
\label{85}
&&\mbox{Li}_2 (z-i0) \to \mbox{Li}_2 (z), \qquad z>1,\nonumber\\
&&\mbox{Li}_2 (z+i0) \to \mbox{Li}_2 (z) + 2 i \pi \log z, \qquad z>1.
\end{eqnarray}   
When this information is all put together, there is still a difference in the contribution to the jump-discontinuity of the phase, depending on whether $\mbox{Re} (\kappa) \in (0,1)$ or $\mbox{Re} (\kappa) > 1$ as $\mbox{Im} (\kappa)$ approaches $0$. We shall focus for convenience on the region 
\begin{equation}
\mbox{Re} (\kappa) \in (0,1), \qquad \mbox{Re} (\vartheta) >0. 
\end{equation} 
In this region, the contributions to the discontinuity come only from the term $- \frac{1}{2 \pi} \mbox{Li}_2 \frac{2}{1- \kappa}$, and the difference between the limit from above and the limit from below the cut is
\begin{equation}
\label{disco}
\chi_{0}|_{\theta + i0} - \chi_{0}|_{\theta - i0} = - i \log \frac{2}{1 - \kappa}, \qquad \theta <0.
\end{equation}

Now, following the argument spelled out in \cite{Borsato:2016xns}, we continue the r.h.s. of (\ref{disco}) to the crossed value of $\kappa$:
\begin{equation}
\label{non}
- i \log \frac{2}{1 - (-\kappa)} = i \log \frac{1 +\kappa}{2}\,.
\end{equation}
Moreover, we get the following functional identity
\begin{equation}
\label{same}
\chi_0(\kappa)+\chi_0(-\kappa) = \frac{i}{2}\log \frac{1}{1-\kappa} + \frac{i}{2}\log \frac{1}{1+\kappa} + \frac{\pi}{4} + i \log 2\,,
\end{equation}
by applying repeatedly the dilogarithm-identity
\begin{equation}
\mbox{Li}_2(z) + \mbox{Li}_2\Big(\frac{z}{z-1}\Big) = -\frac{i}{2} \, \log^2(1-z), \qquad z \in \mathbb{C}\setminus(0,\infty)\,.
\end{equation}
Therefore, adding the r.h.s. of (\ref{non}) and (\ref{same}), we obtain
\begin{equation}
\label{given}
\mbox{(\ref{non})}+\mbox{(\ref{same})}=\frac{i}{2} \log \frac{1+\kappa}{1-\kappa} + \frac{\pi}{4} = - \frac{i}{2} \log \tanh \frac{\vartheta}{2} + \frac{\pi}{4}.
\end{equation}
On the other hand, from (\ref{xykappa}), (\ref{Phichi}) and (\ref{needs}) with $\Psi(\vartheta) = i \Phi(\vartheta)$, we obtain
\begin{equation}
\chi(\vartheta) + \chi (\vartheta + i \pi) = - \frac{i}{2} \log \tanh \frac{\vartheta}{2} + \frac{\pi}{4}\,,
\end{equation}
which is precisely what is given by (\ref{given}), hence providing a solution to the crossing equation.

\section{Algebraic Bethe Ansatz}

In this appendix we summarise the Algebraic Bethe Ansatz procedure needed in section \ref{sec:Transfer}. We shall follow \cite{Faddeev:1996iy,Levkovich-Maslyuk:2016kfv}.

\subsection{General formulation}

To begin with, we define two functions
\begin{equation}
a(\vartheta) = \mbox{sech} \frac{\vartheta}{2}, \qquad  b(\vartheta) = \mbox{tanh} \frac{\vartheta}{2},
\end{equation}
such that the R matrix (\ref{eq:RLLlimtheta}) can be written as
\begin{equation}
\label{matrR}
R(\vartheta) = E_{11} \otimes E_{11} - E_{22} \otimes E_{22} - b(\vartheta) \, \big( E_{11} \otimes E_{22} - E_{22} \otimes E_{11}\big) - a(\vartheta) \, \big( E_{12} \otimes E_{21} - E_{21} \otimes E_{12}\big).
\end{equation}

The $N$-fold transfer matrix ${\cal{T}}$ (trace of the monodromy matrix ${\cal{M}}$), is given by
\begin{equation}
{\cal{T}}\big(\theta_0|\vec{\theta} \, \big) = \mbox{str}_0 \, {\cal{M}}\big(\theta_0|\vec{\theta} \, \big), \qquad {\cal{M}} \big(\theta_0|\vec{\theta} \, \big) = \prod_{i=1}^N R_{0i}(\theta_0 - \theta_i)
\end{equation}
and it is associated with the propagation of an auxiliary $0$-th particle with rapidity $\theta_0$ through an array of $i=1,...,N$ particles with rapidities $\theta_i$ - collectively grouped into a vector $\vec{\theta}$. The trace is taken in the auxiliary $0$ space.

Notice that the transfer matrix we define in this appendix differs from formula (\ref{maint}) in the main text by the ordering of the quantum spaces. Nevertheless, one can show that the two definitions are related by a similarity transformation, followed by a permutation of the rapidities $\theta_i$ associated to the quantum spaces. We will show in what follows that the eigenvalues of ${\cal{T}}$ and the Bethe-equation constraints are all of product form, hence they are invariant under permutations of the variables in the quantum spaces. This implies that the set of eigenvalues, which is all that it is needed for the thermodynamic Bethe ansatz (see section \ref{sec:Transfer}), will be the same for (\ref{maint}) as for the transfer matrix we will diagonalise here below. \footnote{An alternative way of seeing this occurrence is as follows. Both definitions of the monodromy matrix satisfy the fundamental {\it RTT} relations (\ref{RTT}) with the same R matrix. Although their respective entries are different, hence their eigenvectors will be different, nevertheless their eigenvalues and the corresponding Bethe equations are going to be derived purely relying on the {\it RTT} relations, hence they will be the same in both cases.}

One then chooses a pseudo-vacuum, namely a highest-weight eigenvector of the transfer matrix which may serve as a starting point. A natural choice in this case is the simple $N$-fold tensor-product state
\begin{equation}
|0\rangle = |\phi\rangle \otimes ... \otimes |\phi\rangle.
\end{equation} 
It is not difficult to see that such a state is an eigenstate of the transfer matrix, with eigenvalue
\begin{equation}
{\cal{T}}\big(\theta_0|\vec{\theta} \, \big) |0\rangle = \Lambda\big(\theta_0|\vec{\theta} \, \big) |0\rangle, \qquad \Lambda\big(\theta_0|\vec{\theta} \, \big) = 1 -\prod_{i=1}^N b(\theta_0 - \theta_i). 
\end{equation}

At this stage, one writes the monodromy matrix in the form
\begin{equation}
\label{entri}
{\cal{M}} \big(\theta_0|\vec{\theta} \, \big) = E_{11} \otimes A\big(\theta_0|\vec{\theta} \, \big) + E_{12} \otimes B\big(\theta_0|\vec{\theta} \, \big) + E_{21} \otimes C\big(\theta_0|\vec{\theta} \, \big) + E_{22} \otimes D\big(\theta_0|\vec{\theta} \, \big),
\end{equation}
where one has separated the $0$-th space upfront, with $A$, $B$, $C$ and $D$ being now operators acting exclusively on the {\it physical spaces} $1,...,N$. One postulates that the generic eigenvector of the transfer matrix is given by the $M$-{\it magnon} state
\begin{equation}
|\beta_1,...,\beta_M \rangle = \prod_{n=1}^M B\big(\beta_n|\vec{\theta} \, \big) |0\rangle.
\end{equation}
One can show that this state is an eigenvector of the transfer matrix for arbitrary $M$, by using the commutation relations of the operators $A$, $B$, $C$ and $D$. These commutation relations, in turn, follow from the fundamental relation
\begin{equation}
\label{RTT}
R_{00'} (\theta_0 - \theta_0') \, {\cal{M}} \big(\theta_0|\vec{\theta} \, \big) \, {\cal{M}} \big(\theta_0'|\vec{\theta} \, \big) \, = \, {\cal{M}} \big(\theta_0'|\vec{\theta} \, \big) \, {\cal{M}} \big(\theta_0|\vec{\theta} \, \big) \, R_{00'} (\theta_0 - \theta_0').
\end{equation}
written for two auxiliary spaces and $N$ physical ones. In terms of (\ref{entri}), and using the fact that 
\begin{equation}
a(\theta)^2 + b(\theta)^2 = 1,
\end{equation}
(\ref{RTT}) implies for instance 
\begin{eqnarray}
\label{AB}
A\big(\theta_0|\vec{\theta} \, \big) B\big(\theta_0'|\vec{\theta} \, \big) = \frac{a(\theta_0 - \theta_0')}{b(\theta_0 - \theta_0')} \, B\big(\theta_0|\vec{\theta} \, \big) A\big(\theta_0'|\vec{\theta} \, \big) \, - \, \frac{1}{b(\theta_0 - \theta_0')} \, B\big(\theta_0'|\vec{\theta} \, \big) A\big(\theta_0|\vec{\theta} \, \big)
\end{eqnarray}
and the same with $D$ replacing $A$. It is then possible to commute the transfer matrix
\begin{equation}
{\cal{T}} = A - D
\end{equation}
through all the $B$'s in the $M$-magnon state, and accumulate an eigenvalue
\begin{eqnarray}
\label{eige}
&&{\cal{T}}\big(\theta_0|\vec{\theta} \, \big) |\beta_1,...,\beta_M \rangle = \Lambda_M \big(\theta_0|\vec{\beta} \, |\vec{\theta} \, \big) |\beta_1,...,\beta_M \rangle + X, \nonumber\\
&&\Lambda\big(\theta_0|\vec{\beta} \, |\vec{\theta} \, \big) = \Big[1 - \prod_{i=1}^N b(\theta_0 - \theta_i)\Big] \prod_{n=1}^M \frac{1}{b(\beta_n - \theta_0)}, 
\end{eqnarray}
where we have used elementary properties of the functions $a(\theta)$ and $b(\theta)$ to cancel terms in the intermediate steps.
Of course, only for $X=0$ we can claim that $|\beta_1,...,\beta_M \rangle$ is an eigenstate. Since $X$ collects the contributions from the first term on the r.h.s. of (\ref{AB}), one realises that it is possible to set $X=0$ by requiring the {\it level-1} Bethe equations
\begin{equation}
\prod_{i=1}^N b(\beta_m - \theta_i) = 1, \qquad \forall \, \, n=1,...,M.
\end{equation}
One finally needs to multiply the eigenvalues we have found here by the product over the quantum spaces of the dressing factors, namely 
\begin{equation}
\prod_{i=1}^N \Phi(\theta_0 - \theta_i),
\end{equation}
reinstating the correct normalisation for the R matrix (\ref{matrR}).

\subsection{Lowest-level eigenstates}

In this subsection, we show how the Algebraic Bethe Ansatz we have performed in the previous subsection determines the transfer-matrix eigenstates in some specific example with low values of $N$. 

\subsubsection{Two physical sites}

Let us begin with $N=2$. It is easy to directly diagonalise the transfer matrix 
\begin{eqnarray}
{\cal{T}} = \mbox{str}_0 R_{01}(\theta_0 - \theta_1) R_{02}(\theta - \theta_2).
\end{eqnarray}
We find for the bosonic eigenstates
\begin{equation}
{\cal{T}} |\phi\rangle \otimes |\phi\rangle = (1 - b_{01} b_{02}) |\phi\rangle \otimes |\phi\rangle, \qquad {\cal{T}} |\psi\rangle \otimes |\psi\rangle = (-1 + b_{01} b_{02}) |\psi\rangle \otimes |\psi\rangle,
\end{equation}
having defined 
\begin{equation}
a_{ij} \equiv a(\theta_i - \theta_j), \qquad b_{ij} \equiv b(\theta_i - \theta_j).
\end{equation}
The (un-normalised) fermionic eigenstates are slightly more involved:
\begin{equation}
\label{use}
{\cal{T}} \Big(|\phi\rangle \otimes |\psi\rangle \pm    e^{\pm \frac{\theta_1 - \theta_2}{2}} |\psi\rangle \otimes |\phi\rangle\Big) = \big[\pm    e^{\pm \frac{\theta_1 - \theta_2}{2}} a_{01} a_{02} + b_{01} - b_{02}\big] \Big(|\phi\rangle \otimes |\psi\rangle \pm    e^{\pm \frac{\theta_1 - \theta_2}{2}} |\psi\rangle \otimes |\phi\rangle\Big).
\end{equation}
The fact that the eigenstates do not depend on $\theta_0$ is a hallmark of integrability: the transfer matrix commutes with itself at different values of the spectral parameter, generating therefore all the charges in involution. 

This is perfectly reproduced by the Algebraic Bethe ansatz. Clearly $|\phi\rangle \otimes |\phi \rangle = |0\rangle$ is the pseudo-vacuum (lowest-weight vector), whose eigenvalue is reproduced by formula (\ref{eige}) at $M=0$. Subsequently, we should look at the solutions of the auxiliary Bethe equations
\begin{equation}
b(\beta - \theta_1) b(\beta - \theta_2) = 1, \qquad \mbox{\it i.e.} \, \, \beta = \pm \infty,
\end{equation}
and use such solutions to construct the 1-particle eigenstates as
\begin{eqnarray}
B(\beta|\theta_1, \theta_2)|0\rangle. 
\end{eqnarray}
Explicit evaluation of the $B$ operator from the monodromy matrix gives
\begin{equation}
B(\beta|\theta_1, \theta_2)|0\rangle =- a(\beta - \theta_2) \Big(|\phi\rangle \otimes |\psi\rangle + \frac{a(\beta - \theta_1) b(\beta - \theta_2)}{a(\beta - \theta_2)} |\psi\rangle \otimes |\phi\rangle \Big).
\end{equation}
Plugging in $\beta = \pm \infty$ produces
\begin{equation}
B(\pm \infty|\theta_1, \theta_2)|0\rangle \propto \Big(|\phi\rangle \otimes |\psi\rangle \pm    e^{\pm \frac{\theta_1 - \theta_2}{2}} |\psi\rangle \otimes |\phi\rangle \Big).
\end{equation}
Finally, acting with $B(\infty|\theta_1, \theta_2) B(-\infty|\theta_1, \theta_2)$ produces a state proportional to $|\psi\rangle \otimes |\psi\rangle$: we have reached the highest-weight vector, and the spectrum is complete.
Given that
\begin{equation}
b(\pm \infty) = \pm 1,
\end{equation}
we also recover exactly all the eigenvalues from formula (\ref{eige}), as it can be verified by explicit calculation using (\ref{use}) and the hyperbolic-function identities.

\subsubsection{Three physical sites}

For $N=2$ only solutions at infinity are found of the auxiliary Bethe equations, while for $N=3$ one finds that
\begin{equation}
b(\beta - \theta_1) b(\beta - \theta_2) b(\beta - \theta_3)= 1
\end{equation}
is solved by
\begin{equation}
\label{plus}
\beta = \infty, \qquad e^\frac{\beta}{2} = - e^{- i \frac{\pi}{2} \pm i \frac{\pi}{4}} \, \sqrt{\frac{z_1 z_2 z_3}{|\vec{z}|}},  
\end{equation}
where we have defined
\begin{equation}
\vec{z} = (z_1, z_2, z_3), \qquad z_i = e^\frac{\theta_i}{2}.
\end{equation}
Let us define as $y$ the solution with the $+$ sign in the second formula of (\ref{plus}): the solution with the minus sign will therefore be equal to $-i y$. Correspondingly, the associated values of $\beta$ differ by $i \frac{\pi}{2}$.

The eigenvalues (\ref{eige}) can be expressed, using the auxiliary Bethe equations which appear in the formula as a multiplier, in terms of the location of their zeroes, which are precisely the auxiliary roots. The remainder of the formula, bearing the $M$-dependence, simply extracts one zero and adds another one at a different location. Let us show how it works. One can verify that, if one defines
\begin{equation}
\mu = e^{\frac{\theta_0}{2}}, 
\end{equation}
then
\begin{equation}
\Lambda(\theta_0|\vec{\theta}) = \frac{- 2 |\vec{z}|}{\sum_{i=1}^3 (\mu^2 - z_i^2)} \, (\mu^2 - y^2) (\mu^2 + y^2) \prod_{n=1}^M \frac{1}{b(\beta_n - \theta_0)} \equiv \Delta(\mu) (\mu^2 - y^2) (\mu^2 + y^2) \prod_{n=1}^M \frac{1}{b(\beta_n - \theta_0)}.
\end{equation}
There are $N$ zeroes ($3$ in this case) in the variable $\mu^2$, including the one at infinity. One also has, correspondingly,
\begin{equation}
b(\beta - \theta_0) = 1, \qquad b(\beta - \theta_0) = \frac{y^2 - \mu^2}{y^2 + \mu^2}, \qquad \qquad b(\beta - \theta_0) = \frac{y^2 + \mu^2}{y^2 - \mu^2}
\end{equation}
respectively for the $3$ solutions in (\ref{plus}). If we now list the eigenvalues associated to the eigenstates built adding auxiliary roots we obtain the following:
\begin{eqnarray}
&&M=0: \qquad \mbox{aux. roots} \, \, \mbox{none} \qquad \mbox{eigenv.} \, \, \Delta(\mu) (\mu^2 - y^2) (\mu^2 + y^2),\nonumber \\
&&M=1: \qquad \mbox{aux. roots} \, \, \infty \qquad \mbox{eigenv.} \, \, \Delta(\mu)(\mu^2 - y^2) (\mu^2 + y^2), \nonumber \\
&&M=1: \qquad \mbox{aux. roots} \, \, y \qquad \mbox{eigenv.} \, \, \Delta(\mu)(\mu^2 + y^2)^2, \nonumber \\
&&M=1: \qquad \mbox{aux. roots} \, \, - i y \qquad \mbox{eigenv.} \, \, \Delta(\mu)(\mu^2 - y^2)^2, \nonumber \\
&&M=2: \qquad \mbox{aux. roots} \, \, (\infty, y) \qquad \mbox{eigenv.} \, \, \Delta(\mu)(\mu^2 + y^2)^2, \nonumber \\
&&M=2: \qquad \mbox{aux. roots} \, \, (\infty, - i y) \qquad \mbox{eigenv.} \, \, \Delta(\mu)(\mu^2 - y^2)^2, \nonumber \\
&&M=2: \qquad \mbox{aux. roots} \, \, (y, -i y) \qquad \mbox{eigenv.} \, \, \Delta(\mu)(\mu^2 - y^2) (\mu^2 + y^2), \nonumber \\
&&M=3: \qquad \mbox{aux. roots} \, \, (\infty, y, -i y) \qquad \mbox{eigenv.} \, \, \Delta(\mu)(\mu^2 - y^2) (\mu^2 + y^2),
\end{eqnarray}
where for instance $(y, -i y)$ means that $y$ {\it and} $-i y$ are both chosen.
The explicit form of the eigenstates is quite complicated, and we shall not report it here. However it is easy to see that the Algebraic Bethe Ansatz reproduces the complete spectrum of $8$ states.

\subsubsection{Higher values of $N$}

It becomes very rapidly quite cumbersome to present the spectrum of the transfer matrix for higher values of $N$. Let us simply point out that $N=4$ is still characterised by auxiliary Bethe equations having solutions where only one pair of roots of the type $(y, -i y)$ is present (besides the roots at infinity). When one reaches $N=5$, two distinct pairs of solution appear, namely $(y_1, -i y_1)$ and $(y_2, -i y_2)$, characterised by two distinct {\it centres}. One also has a single root at $+ \infty$, making a total of $5$ possibilities to choose from for the auxiliary roots.   

The spectrum is then built accordingly, taking all possible combinations of $M$ out of these $5$ roots, with $M =0,...,5$. The total number of states is therefore $\sum_{M=0}^5 {5 \choose M} = 32 = 2^5.$ 

The eigenvalue can be written as
\begin{eqnarray}
&&\Lambda(\theta_0|\vec{\theta}) = \frac{- 2 |\vec{z}|}{\sum_{i=1}^5 (\mu^2 + z_i^2)} \, (\mu^2 - y_1^2) (\mu^2 + y_1^2) (\mu^2 - y_2^2) (\mu^2 + y_2^2)\prod_{n=1}^M \frac{1}{b(\beta_n - \theta_0)} \nonumber\\
&&\qquad \qquad \equiv \Delta'(\mu) (\mu^2 - y_1^2) (\mu^2 + y_1^2) (\mu^2 - y_2^2) (\mu^2 + y_2^2) \prod_{n=1}^M \frac{1}{b(\beta_n - \theta_0)},
\end{eqnarray}
and the same mechanism as in the $N=3$ case ensures that it is always a polynomial with $5$ zeroes in the variable $\mu^2$ (including the zero at infinity), and the various choice of $M$ auxiliary roots extract zeroes and add other zeroes. Let us list for example a few significant cases:

\begin{eqnarray}
&&M=0: \qquad \mbox{aux. roots} \, \, \mbox{none} \qquad \mbox{eigenv.} \, \, \Delta'(\mu) (\mu^2 - y_1^2) (\mu^2 + y_1^2) (\mu^2 - y_2^2) (\mu^2 + y_2^2),\nonumber \\
&&M=1: \qquad \mbox{aux. roots} \, \, \infty \qquad \mbox{eigenv.} \, \, \Delta'(\mu)(\mu^2 - y_1^2) (\mu^2 + y_1^2) (\mu^2 - y_2^2) (\mu^2 + y_2^2), \nonumber \\
&&M=1: \qquad \mbox{aux. roots} \, \, y_1 \qquad \mbox{eigenv.} \, \, \Delta'(\mu) (\mu^2 + y_1^2)^2 (\mu^2 - y_2^2) (\mu^2 + y_2^2), \nonumber \\
&&M=1: \qquad \mbox{aux. roots} \, \, - i y_1 \qquad \mbox{eigenv.} \, \, \Delta'(\mu)(\mu^2 - y_1^2)^2 (\mu^2 - y_2^2) (\mu^2 + y_2^2), \nonumber \\
&& etcetera\nonumber \\
&&M=3: \qquad \mbox{aux. roots} \, \, (y_1, - i y_1, y_2) \qquad \mbox{eigenv.} \, \, \Delta'(\mu)(\mu^2 - y_1^2) (\mu^2 + y_1^2)  (\mu^2 + y_2^2)^2, \nonumber \\
&&M=3: \qquad \mbox{aux. roots} \, \, (y_1, - i y_1, - i y_2) \qquad \mbox{eigenv.} \, \, \Delta'(\mu)(\mu^2 - y_1^2) (\mu^2 + y_1^2)  (\mu^2 - y_2^2)^2, \nonumber \\
&&etcetera.
\end{eqnarray}

\section{Derivation of the TBA equations}
\label{app:TBA}

First, let us take the logarithm of equations (\ref{betherel1})-(\ref{betherel3}), divide them by $i$ and
define the \emph{counting functions} so that they give integer multiples of $2\pi/L$ when evaluated at the Bethe roots or holes: 
\begin{equation}
Z(\theta_k)=\frac{2\pi n_k}{L}\,,\quad Z_{\pm1}(\beta_{\pm1,k})=\frac{2\pi m_{\pm1,k}}{L}\,,\quad Z_{\pm3}(\beta_{\pm3,k})=\frac{2\pi m_{\pm3,k}}{L}\,.
\end{equation}
Moreover, the counting functions should be conventionally defined in a way to be monotonically increasing functions; then we try to define them as follows
\begin{eqnarray}
 Z(\theta) \equiv e^{\theta} + \frac{2}{iL}\sum_{i=1}^N \log S(\theta-\theta_i) + \frac{1}{iL}\sum_{i=1}^{M_1^+}\log\coth\frac{z_{1,i}+i\pi/2-\theta}{2} + \frac{1}{iL}\sum_{i=1}^{M_1^-}\log\coth\frac{z_{1,i}-i\pi/2-\theta}{2}  \label{Zdef1}\nonumber\\
 + \frac{1}{iL}\sum_{i=1}^{M_3^+}\log\coth\frac{z_{3,i}+i\pi/2-\theta}{2} + \frac{1}{iL}\sum_{i=1}^{M_3^-}\log\coth\frac{z_{3,i}-i\pi/2-\theta}{2}\\ 
 Z_{\pm n}(\beta) \equiv \mp\frac{1}{iL}\sum_{i=1}^N\log\tanh\frac{z_{n,i}\pm i\pi/2-\theta}{2}\,;\ n=1,3\,.~~~~~~~~~~~~~~~~~~~~~~~~~~~~~~~~~~~~~~~~~~~~~~~~~~~~~~~~~~\label{Zdef2}
\end{eqnarray}
Therefore
\begin{eqnarray}
 &&L(Z(\theta_k)-Z(\theta_j))=2\pi(n_k-n_j)\,, \\
 &&L(Z_{\pm1}(\beta_{\pm1,k})-Z_{\pm1}(\beta_{\pm1,j}))=2\pi(m_{\pm1,k}-m_{\pm1,j})\,, \\
 &&L(Z_{\pm3}(\beta_{\pm3,k})-Z_{\pm3}(\beta_{\pm3,j}))=2\pi(m_{\pm3,k}-m_{\pm3,j})\,,
\end{eqnarray}
and the numbers of roots (holes) contained in the infinitesimal intervals $d\theta, d\beta_{\pm1},  d\beta_{\pm3}$ are given by $L\rho_0^r(\theta)d\theta$ ($L\rho_0^h(\theta)d\theta$) and $L\rho_{\pm1}^r(\beta)d\beta$ ($L\rho_{\pm1}^h(\beta)d\beta$), $L\rho_{\pm3}^r(\beta)d\beta$ ($L\rho_{\pm3}^h(\beta)d\beta$) respectively, where the densities are defined as
\begin{eqnarray}
&& \rho_0(\theta)=(\rho_0^r(\theta)+\rho_0^h(\theta))\equiv\frac{1}{2\pi}\frac{dZ(\theta)}{d\theta}\,,\label{rhodef1} \\
&& \rho_{\pm1}(\beta)=(\rho_{\pm1}^r(\beta)+\rho_{\pm1}^h(\beta))\equiv\frac{1}{2\pi}\frac{dZ_{\pm1}(\beta)}{d\beta}\,, \\
&& \rho_{\pm3}(\beta)=(\rho_{\pm3}^r(\beta)+\rho_{\pm3}^h(\beta))\equiv\frac{1}{2\pi}\frac{dZ_{\pm3}(\beta)}{d\beta}\,.\label{rhodef3}
\end{eqnarray}
In the thermodynamic limit the sums become integrals as $\frac{1}{L}\sum_i\rightarrow \int d\theta\rho^r(\theta)$, then definitions (\ref{Zdef1})-(\ref{Zdef2}) become
\begin{eqnarray}
 Z(\theta) = e^{\theta} +\frac{2}{i}\int d\theta' \log S(\theta-\theta') \rho_0^r(\theta')+\frac{1}{i} \sum_{\pm,n=1,3}\int d\beta\log\coth\frac{\beta-\theta\pm i\pi/2}{2}\rho_{\pm n}^r(\beta)\,,\\
 Z_{\pm n}(\beta) = \mp\frac{1}{i}\int d\theta\log\tanh\frac{\beta-\theta\pm i\pi/2}{2}\rho_{0}^r(\theta)\,;\ n=1,3\,.~~~~~~~~~~~~~~~~~~~~~~~~~~~~~~~~~~~~~~~~~
\end{eqnarray}
Let us then take the derivatives of the counting functions in their respective arguments: because of (\ref{rhodef1})-(\ref{rhodef3}), we get the nonlinear integral equations (\ref{rho1})-(\ref{rho2}) for the densities.

In order to derive the TBA equations, we start writing a generic expression of the free energy $F$
\begin{equation}
 F(T)=\widetilde E-T\mathcal{S}\,;\quad \widetilde E=\int d\theta\,\epsilon(\theta)\,\rho_0^r(\theta)=M\int d\theta\, e^{\theta}\rho_0^r(\theta)\,,
\end{equation}
where $\epsilon(\theta)$ is the energy density and $\mathcal{S}$ is the entropy, defined as
\begin{equation} 
\mathcal{S}=\sum_A \int d\theta \rho_A(\theta)\log\rho_A(\theta)-\rho_A^r(\theta)\log\rho_A^r(\theta)-\rho_A^h(\theta)\log\rho_A^h(\theta)\,.
\end{equation}
Now, taking the variation of $F$ with respect to the densities and using the following variations of the densities equations (\ref{rho0simple})-(\ref{rho1simple})
\begin{eqnarray}
 &&\delta\rho^h_0(\theta)=-\delta\rho^r_0(\theta) +\sum_{n=1,3}\phi*(\delta\rho_{-n}^r+\delta\rho_{+n}^h)\,,\\
 &&\delta\rho_{- n}^h(\beta)=-\delta\rho_{- n}^r(\beta) + \phi*\delta\rho_{0}^r\,;\ n=1,3\,,\\
&&\delta\rho_{+ n}^r(\beta)=-\delta\rho_{+ n}^h(\beta) + \phi*\delta\rho_{0}^r\,;\ n=1,3\,,
 \end{eqnarray}
we get
\begin{eqnarray}
 \delta F=\int d\theta \left\{\widetilde E(\theta)\delta\rho_0^r(\theta)-T\left[\log\frac{\rho_0(\theta)}{\rho_0^h(\theta)}\left(\sum_{n=1,3}\phi*(\delta\rho_{-n}^r+\delta\rho_{+n}^h)\right)(\theta)+\log\frac{\rho_0^h(\theta)}{\rho_0^r(\theta)}\delta\rho_0^r(\theta)\right.\right.\nonumber\\
 +\sum_{n=1,3}\log\frac{\rho_{- n}(\theta)}{\rho_{- n}^h(\theta)}(\phi*\delta\rho_0^r)(\theta)+\log\frac{\rho_{- n}^h(\theta)}{\rho_{- n}^r(\theta)}\delta\rho_{- n}^r(\theta)\nonumber\\
 \left.\left.+\sum_{n=1,3}\log\frac{\rho_{+ n}(\theta)}{\rho_{+ n}^r(\theta)}(\phi*\delta\rho_0^r)(\theta)+\log\frac{\rho_{+ n}^r(\theta)}{\rho_{+ n}^h(\theta)}\delta\rho_{+ n}^h(\theta)\right]\right\}\,.~~~~\label{deltaf}
\end{eqnarray}
Exchanging $\theta$ and $\theta'$ in the terms involving the convolution $\phi*\delta\rho_0^r=\int\phi(\theta-\theta')\delta\rho_0(\theta')$ and setting to zero the part of $\delta F$ proportional to $\delta\rho_0^r$, we get the TBA equation for $\varepsilon_0\equiv\log\frac{\rho_0^h}{\rho_0^r}$:
\begin{equation}
 \varepsilon_0(\theta) = R e^{\theta}  - \sum_{n=1,3}\phi*\left[\log(1+e^{-\varepsilon_{-n}})+\log(1+e^{\varepsilon_{+n}})\right]\,,\label{tba1}
 \end{equation}
 where we introduced also the pseudo-energies $\varepsilon_{\pm n}\equiv\log\frac{\rho_{\pm n}^h}{\rho_{\pm n}^r}$.
 Analogously, exchanging $\theta$ with $\theta'$ also in the convolutions $\phi*(\delta\rho_{-n}^r+\delta\rho_{+n}^h)=\int\phi(\theta-\theta')(\delta\rho_{-n}^r+\delta\rho_{+n}^h)(\theta')$ and imposing that the terms of (\ref{deltaf}) proportional to $\delta\rho_{- n}^r$ and $\delta\rho_{+ n}^h$ vanish, implies
 \begin{eqnarray}
  &&\varepsilon_{+ n}(\beta) = \phi*\log(1+e^{-\varepsilon_0})\,;\ n=1,3\,,\\
  &&\varepsilon_{- n}(\beta) = -\phi*\log(1+e^{-\varepsilon_0})\,;\ n=1,3\,.\label{tba3}
 \end{eqnarray}
Changing sign to $\varepsilon_{+n}\rightarrow -\varepsilon_{+n}$ and defining $L_A\equiv\log(1+e^{-\varepsilon_A}) $, the TBA equations (\ref{tba1})-(\ref{tba3}) can be compactly rewritten as equations (\ref{tba}).

\section{Derivation of the excited states' energy formula}
\label{app:excited}

Basically, in order to derive a closed formula for the energies of the excited states, we have to solve the system of equations given by (\ref{ex1}), (\ref{ex2}), (\ref{ex3}) and (\ref{ex4}),
and plug the solutions for $\varepsilon_0$ and $y_k$ into the excited states' energy formula
\beq
RE_{\mathrm{left}}(R,\gamma)=\sum_{k=1}^K e^{y_k}-\frac{1}{2\pi}\int d\theta\,e^{\theta} \log(1+e^{-\varepsilon_0(\theta)})\,.
\label{Emodk}
\eeq
As done in section \ref{sec:central} for the ground state, the starting trick consists in taking the first derivative of (\ref{ex1}) and solving it for $e^{\theta}$, so that we can plug
\beq
e^{\theta}=\varepsilon'_0(\theta)-4i\pi\sum_{j=J_{\infty}+1}^J\phi\left(x_j-\theta+\frac{i\pi}{2}\right)+2\left[\phi*(L_1^{\gamma})'\right](\theta)
\eeq
into (\ref{Emodk}), where we defined $L_1^{\gamma}(\theta)\equiv \log[1+e^{i\gamma}Y_1(\theta)][1+e^{-i\gamma}Y_1(\theta)]$ and took into account that the $x_j$'s at $\infty$ do not contribute.
Next, we can replace $\phi*L_0$ by using (\ref{ex2}), so that the second term of (\ref{Emodk}) becomes
\beqa
&&-\frac{1}{2\pi}\int d\theta\,e^{\theta} \log(1+e^{-\varepsilon_0(\theta)})=-\frac{1}{2\pi}\int_{\varepsilon_{0,min}}^{\infty} d \varepsilon_0 L_0 - \frac{1}{\pi}\int_{\varepsilon_{1,min}}^{0} d \varepsilon_1 \left(\frac{e^{i\gamma-\varepsilon_1}}{1+ e^{i\gamma-\varepsilon_1}}+\frac{e^{-i\gamma-\varepsilon_1}}{1+ e^{-i\gamma-\varepsilon_1}}\right)\nonumber\\
&&+2i\sum_{j=J_{\infty}+1}^J(\phi*L_0)\left(x_j+\frac{i\pi}{2}\right)+\frac{1}{\pi} \sum_{k=1}^K\int d\theta \left[\log\tanh\left(\frac{\theta-y_k}{2}\right)(L_1^{\gamma})'(\theta)\right]\,.
\label{Emodk2}
\eeqa
Integrating by parts the last term, we get
\beq
\frac{1}{\pi} \sum_{k=1}^K\int d\theta \left[\log\tanh\left(\frac{\theta-y_k}{2}\right)(L_1^{\gamma})'(\theta)\right]=i\sum_{k=1}^K\left[\phi*(L_1^{\gamma})\right]\left(y_k+\frac{i\pi}{2}\right)-\frac{1}{2\pi}\log[(-1)^K]\log[Y_0^2(-\infty)]\,.
\label{parts}
\eeq 
Now, using (\ref{ex4}) and knowing that the second term in its r.h.s. can be written as $i(\phi*L_0)(x_j+i\pi/2)$, we can replace the third term in (\ref{Emodk2})  by
\beq
2i\sum_{j=J_{\infty}+1}^J(\phi*L_0)\left(x_j+\frac{i\pi}{2}\right)=2\sum_{j=J_{\infty}+1}^J(\gamma-(2M_j+1)\pi)-2i\sum_{j=J_{\infty}+1}^J\sum_{k=1}^K\log\tanh\left(\frac{x_j-y_k+\frac{i\pi}{2}}{2}\right)\,.
\eeq
Similarly, we use (\ref{ex3}) to write the first term in the r.h.s. of (\ref{parts}) as
\beq
i\sum_{k=1}^K(\phi*L_1^{\gamma})\left(x_k+\frac{i\pi}{2}\right)=-\sum_{k=1}^K (2N_k+1)\pi-2i\sum_{k=1}^K\sum_{j=1}^J\log\tanh\left(\frac{y_k-x_j+\frac{i\pi}{2}}{2}\right)-\sum_{k=1}^K e^{y_k}\,.
\label{iphiL}
\eeq
Now, taking into account that
\beq
-2i\sum_{j=J_{\infty}+1}^J\sum_{k=1}^K\log\tanh\left(\frac{x_j-y_k+\frac{i\pi}{2}}{2}\right)-2i\sum_{k=1}^K\sum_{j=1}^J\log\tanh\left(\frac{y_k-x_j+\frac{i\pi}{2}}{2}\right)=2JK\pi
\eeq
and that for $\gamma>\pi$ the addition of $-i\pi$ to the r.h.s. of (\ref{ex1}) implies the replacements $2N_k+1\rightarrow2N_k$   and $J\rightarrow J-1$ in (\ref{iphiL}), the expression (\ref{Emodk2}) simplifies to (\ref{Efin}). 

Formula (\ref{Efin}) can be then rewritten in a form more similar to equation (43) of \cite{Fendley:1997ys}, by changing the integration variables as $e^{-\varepsilon_0}\rightarrow u$, $e^{-\varepsilon_1}\rightarrow v$ and integrating by parts the second integral in the r.h.s. of (\ref{Efin}):
\beqa
&&RE_{\mathrm{left}}(R,\gamma)=\frac{1}{2\pi}\int_{Y_0(-\infty)}^{\infty} du \frac{\log(1+u)}{u} + \frac{1}{2\pi}\sum_a\int_{Y_1(-\infty)}^{0} dv \frac{\log(1+\lambda_a v)}{v}\\
&&+\frac{1}{2\pi}\log[Y_0^2(-\infty)]\log[(-1)^KY_1(-\infty)]+2(J-1)K\pi+2\sum_{j=J_{\infty}+1}^J(\gamma-(2M_j+1)\pi)-\sum_{k=1}^K2N_k\pi\,.\nonumber
\label{Efin1b}
\eeqa
As we saw in section \ref{sec:exc}, it is possible to evaluate the first two integrals in terms of dilogarithms as in (\ref{Efin1}), while it is common in the TBA literature to write the energy in terms of Rogers dilogarithms, defined as 
\beq
L(x)\equiv-\frac{1}{2}\int_0^x dt \left[\frac{\log(1-t)}{t}+\frac{\log(t)}{1-t}\right].
\eeq
Formula (\ref{Efin1b}) assumes then the form
\beqa
RE_{\mathrm{left}}(R,\gamma)&=&\frac{1}{2\pi}\left\{L[-Y_0(-\infty)]+2L[-e^{i\gamma}Y_1(-\infty)]+2L[-e^{-i\gamma}Y_1(-\infty)]+\frac{5\pi^2}{24}+\log^2(e^{i\gamma})\right\}\nonumber\\
&&-\frac{\log^2(-e^{4i\gamma})}{16\pi}+2(J-1)K\pi+2\sum_{j=J_{\infty}+1}^J(\gamma-(2M_j+1)\pi)-\sum_{k=1}^K2N_k\pi\,,
\label{Efin2}
\eeqa
that can be also compactly rewritten as
\beqa
RE_{\mathrm{left}}(R,\gamma)&=&\frac{1}{2\pi}\left\{L[-Y_0(-\infty)]+2L[-e^{i\gamma}Y_1(-\infty)]+2L[-e^{-i\gamma}Y_1(-\infty)]\right\}\nonumber\\
&&+\frac{\gamma^2}{2\pi}+(2J-2K-1)\frac{\gamma}{2}+(6K-6J+1)\frac{\pi}{6}\,.
\label{Efin3}
\eeqa
Analogously, for $\gamma\in(\pi/2,\pi)$ we get
\beqa
RE_{\mathrm{left}}(R,\gamma)&=&\frac{1}{2\pi}\left\{L[-Y_0(-\infty)]+2L[-e^{i\gamma}Y_1(-\infty)]+2L[-e^{-i\gamma}Y_1(-\infty)]
+\frac{\pi^2}{3}+\log^2(e^{i\gamma})\right\}\nonumber\\
&&+\frac{\gamma-\pi}{2}-\frac{i}{4}\log(Y_1(-\infty)^2)\,,
\eeqa
where we recall that $Y_0(-\infty)=\tan(\gamma)^2$ and $Y_1(\*-\infty)=-1/\cos(\gamma)$, while for $\gamma\in(0,\pi/2)$
\beqa
RE_{\mathrm{left}}(R,\gamma)&=&\frac{1}{2\pi}\left\{L[-Y_0(-\infty)]+2L[-e^{i\gamma}Y_1(-\infty)]+2L[-e^{-i\gamma}Y_1(-\infty)]
+\frac{\pi^2}{3}-\log^2(e^{-i\gamma})\right\}\nonumber\\
&&-\gamma-\frac{i}{4}\log(Y_1(-\infty)^2)\,,
\eeqa
where $Y_0(-\infty)=Y_1(-\infty)^2=\infty$.

\subsection{Numerics}
\label{app:num}

Numerically, we started by solving the ground state TBA (\ref{twistTBA}) for $\gamma\in(0,\pi)$ and $R=1$: the result confirms nicely the analytic results (\ref{E0pi/2}) and (\ref{E0pi}), as one can see in Figure \ref{fig:Er}. \footnote{Our numerical results turned out to be in agreement with the expected values of $E(R=1,\gamma)$, calculated by (\ref{E0pi/2}), (\ref{E0pi}) and (\ref{par}), at least up to the second decimal digit.}
\begin{itemize}
\item For  $\gamma\in(\pi,3\pi/2)$, we adopted the prescription discussed in section \ref{sec:exc} (the additional $-i\pi$ in the equation for $\varepsilon_0$ (\ref{eps0ex})) and got perfect matching with the analytic prediction (\ref{E13/2}), except for $\gamma$ close to $3\pi/2$, where the numerical algorithm becomes sensitive to the approaching of a new zero.
\item For  $\gamma\in(3\pi/2,2\pi)$, we took then into account the first zero $y_1$ of $Y_1$ and solved iteratively the following equation for $y_1$: \footnote{We were always considering large negative real values for the initial conditions $y_k^{(0)}$ and $x_j^{(0)}$, even though we verified that the numerical algorithm remained stable by using other choices.}
\beq
y_1^{(n)}=\log\left\{-\int\frac{d\theta}{\pi}\frac{\ln\left[(1+e^{i\gamma}Y_1(\theta))(1+e^{-i\gamma}Y_1(\theta))\right]}{\sinh\left(y_1^{(n-1)}-\theta\right)}\right\}\,,
\label{y1}
\eeq
together with the TBA equations (\ref{eps0ex}) and (\ref{ex2}) for $K=1$. 
In this way we got $E_{\mathrm{left}}(R=1,\gamma=2\pi)=\pi/2$, for example, then a total energy $E(R=1,\gamma=2\pi)=2\pi$.
\item In order to push the numerics beyond $\gamma=2\pi$, we had to consider a second zero $x_2$ of $Y_0(\theta)$, with $M_2=1$, entering from $\theta=-\infty$ at $\gamma=2\pi$. Then we needed to add an equation for $x_2$, solving iteratively
\beq
i\log\tanh\left(\frac{x_2^{(n)}-y_1}{2}+\frac{i\pi}{4}\right)=\gamma-3\pi-\int\frac{d\theta}{2\pi}\frac{\ln\left(1+Y_0(\theta)\right)}{\sinh\left(x_2^{(n-1)}-\theta\right)}\,,
\label{x2}
\eeq
together with (\ref{ex2}) for $K=1$, (\ref{eps0ex}) with the additional term $-2\log\tanh\left[(\theta-x_2)/2\right]$ in the r.h.s., and modifying (\ref{y1}) as follows:
\beq
y_1^{(n)}=\log\left\{-2i\log\tanh\left(\frac{y_1^{(n)}-x_2}{2}+\frac{i\pi}{4}\right)-\int\frac{d\theta}{\pi}\frac{\ln\left[(1+e^{i\gamma}Y_1(\theta))(1+e^{-i\gamma}Y_1(\theta))\right]}{\sinh\left(y_1^{(n-1)}-\theta\right)}\right\}\,.
\label{y1mod}
\eeq
\item A second zero $y_2$ of $Y_1(\theta)$, with $N_2=1$, enters at $\gamma=5\pi/2$, then we need to add the following equation for $y_2$
\beq
y_2^{(n)}=\log\left\{-2\pi-2i\log\tanh\left(\frac{y_2^{(n-1)}-x_2}{2}+\frac{i\pi}{4}\right)-\int\frac{d\theta}{\pi}\frac{\ln\left[(1+e^{i\gamma}Y_1(\theta))(1+e^{-i\gamma}Y_1(\theta))\right]}{\sinh\left(y_2^{(n-1)}-\theta\right)}\right\}
\label{y2}
\eeq
to the iterative algorithm, to modify equation (\ref{x2}) by inverting
\beq
i\sum_{k=1}^2\log\tanh\left(\frac{x_2^{(n)}-y_k}{2}+\frac{i\pi}{4}\right)=\gamma-3\pi-\int\frac{d\theta}{2\pi}\frac{\ln\left(1+Y_0(\theta)\right)}{\sinh\left(x_2^{(n-1)}-\theta\right)}\,,
\label{x2mod}
\eeq
and (\ref{ex2}) by considering $K=2$.

We recall that the structure of the zeros are suggested by the behaviour of the zeros of $Y_0(\pm\infty)$ and $Y_1(\pm\infty)$ deduced by equations (\ref{minima}) and (\ref{maxima}), as discussed in section \ref{sec:exc}, see Figure \ref{fig:minmax}.
\item In particular, the zero of $Y_0(+\infty)$ at $\gamma=3\pi$ suggests us that $x_2$ goes to $+\infty$, but a zero of $Y_0(-\infty)$ for the same value of $\gamma$ implies that a new zero, $x_3$, enters from $\theta=-\infty$ with $M_3=2$. Then we have to solve the following equation for $x_3$
\beq
i\sum_{k=1}^2\log\tanh\left(\frac{x_3^{(n)}-y_k}{2}+\frac{i\pi}{4}\right)=\gamma-5\pi-\int\frac{d\theta}{2\pi}\frac{\ln\left(1+Y_0(\theta)\right)}{\sinh\left(x_3^{(n-1)}-\theta\right)}\,,
\label{x3}
\eeq
to add the term \footnote{The $-2i\pi$ is due to $x_2$ gone to $\infty$.} $-2i\pi-2\log\tanh\left[(\theta-x_3)/2\right]$ to the r.h.s. of (\ref{eps0ex}) and to modify accordingly equations (\ref{y1mod}) and (\ref{y2}):
\beqa
&&\hspace{-0.8cm}y_1^{(n)}=\log\left\{2\pi-2i\log\tanh\left(\frac{y_1^{(n-1)}-x_3}{2}+\frac{i\pi}{4}\right)-\int\frac{d\theta}{\pi}\frac{\ln\left[(1+e^{i\gamma}Y_1(\theta))(1+e^{-i\gamma}Y_1(\theta))\right]}{\sinh\left(y_1^{(n-1)}-\theta\right)}\right\}\,,\nonumber\\
&&\hspace{-0.8cm}y_2^{(n)}=\log\left\{-2i\log\tanh\left(\frac{y_2^{(n-1)}-x_3}{2}+\frac{i\pi}{4}\right)-\int\frac{d\theta}{\pi}\frac{\ln\left[(1+e^{i\gamma}Y_1(\theta))(1+e^{-i\gamma}Y_1(\theta))\right]}{\sinh\left(y_2^{(n-1)}-\theta\right)}\right\}\nonumber\,,
\eeqa
where $x_2$ at $\infty$ contributes with a $+2\pi$ w.r.t. (\ref{y1mod}) and (\ref{y2}).
\item Finally, the zero $y_3$ enters at $\gamma=7\pi/2$: then we have to use (\ref{ex2}) with $K=3$, add the term $i\log\tanh[(x_3-y_3)/2]$ to the l.h.s. of the equation for $x_3$ (\ref{x3}), and solve iteratively also
\beq
y_3^{(n)}=\log\left\{-2\pi-2i\log\tanh\left(\frac{y_3^{(n-1)}-x_3}{2}+\frac{i\pi}{4}\right)-\int\frac{d\theta}{\pi}\frac{\ln\left[(1+e^{i\gamma}Y_1(\theta))(1+e^{-i\gamma}Y_1(\theta))\right]}{\sinh\left(y_3^{(n-1)}-\theta\right)}\right\}\,.\nonumber
\eeq
\end{itemize}
In summary, we solved numerically the system of equations (\ref{ex1}), (\ref{ex2}), (\ref{ex3}) and (\ref{ex4}) for $R=1$ and up to $\gamma=4\pi$, taking into account the structure of zero discussed here and in section \ref{sec:exc}.
The corresponding results are plotted in Figure \ref{fig:Er}.

\bibliography{biblio}

\end{document}